\documentclass[a4paper,11pt]{article}
\usepackage{jheppub} 
\pdfoutput=1
\usepackage{graphicx,array}
\usepackage{color}
\usepackage{setspace}
\usepackage{amstext}
\usepackage{amssymb}
\usepackage{slashed}
\usepackage{makecell}
\usepackage{multirow}
\definecolor{orange}{rgb}{1,0.5,0}
\definecolor{brown}{rgb}{0.59, 0.29, 0.0}

\definecolor{note_fontcolor}{rgb}{0.80078125, 0.80078125, 0.80078125}

\def\beq{\begin{equation}}
\def\eeq{\end{equation}}
\def\bea{\begin{eqnarray}}
\def\eea{\end{eqnarray}}

\newcommand{\GeV}{\mathrm{\;GeV}}

\def\kk{{\rm KK}}

\usepackage{changepage}
\newenvironment{changemargin}[2]{%
\begin{list}{}{%
\setlength{\topsep}{0pt}%
\setlength{\leftmargin}{#1}%
\setlength{\rightmargin}{#2}%
\setlength{\listparindent}{\parindent}%
\setlength{\itemindent}{\parindent}%
\setlength{\parsep}{\parskip}%
}%
\item[]}{\end{list}}

\title{Dedicated Strategies for Triboson Signals from Cascade Decays of Vector Resonances}

\author[a]{Kaustubh Agashe,}
\author[a,b]{Jack H.~Collins,}
\author[a]{Peizhi Du,}
\author[a,c]{Sungwoo Hong,}
\author[d]{Doojin Kim,}
\author[a,e]{and Rashmish K.~Mishra}

\affiliation[a]{Maryland Center for Fundamental Physics, Department of Physics, University of Maryland, College Park, MD 20742, USA}
\affiliation[b]{Department of Physics and Astronomy, Johns Hopkins University, Baltimore, MD 21218, USA}
\affiliation[c]{Department of Physics, LEPP, Cornell University, Ithaca NY 14853, USA}
\affiliation[d]{Theory Department, CERN, CH-1211 Geneva 23, Switzerland}
\affiliation[e]{INFN, Pisa, Italy and Scuola Normale Superiore, Piazza dei Cavalieri 7, 56126 Pisa, Italy}
      
\emailAdd{kagashe@umd.edu}
\emailAdd{jhc296@umd.edu} 
\emailAdd{pdu@umd.edu} 
\emailAdd{sh768@cornell.edu} 
\emailAdd{doojin.kim@cern.ch} 
\emailAdd{rashmish@pi.infn.it}

\abstract{
New colorless electroweak (EW) charged spin-1 particles with mass of a few TeV arise in numerous extensions of the Standard Model (SM). Decays of such a vector into a pair of SM particles, either fermions or EW bosons, are well studied. Many of these models have an additional scalar, which can lead to (and even dominate in certain parameter regions) a novel decay channel for the heavy vector particles instead -- into a SM EW boson and the scalar, which subsequently decays into a SM EW boson pair. In this work, we focus on the scalar being relatively heavy, roughly factor of two lighter than the vector particles, rendering its decay products well separated. Such a cascade decay results in a final state with three isolated bosons. We argue that for this ``triboson'' signal the existing diboson searches are not quite optimal due to combinatorial ambiguity for three identical bosons, and in addition, due to a relatively small signal cross-section determined by the heaviness of the decaying vector particle. In order to isolate the signal, we demonstrate that tagging all three bosons, followed by use of the full triboson invariant mass distribution  as well as that of appropriate subsets of dibosons, is well motivated. We develop these general strategies in detail within the context of a specific class of models that are based on extensions of the standard warped extra-dimensional scenario. We also point out that a similar analysis would apply to models with an enlarged EW gauge sector in four dimensions, even if they involve a different Lorentz structure for the relevant couplings.
}

\begin{document} 

\begin{flushright}
UMD-PP-017-33 \\
CERN-TH-2017-233
\end{flushright}

\maketitle
\flushbottom

\section{Introduction}
\label{sec:introduction}
The presence of beyond the Standard Model (SM) physics at the TeV scale is highly motivated, most notably by attempts to address the Planck-weak hierarchy and dark matter problems. However, the direct searches performed for such new particles at the Large Hadron Collider (LHC) have so far found no compelling evidence for new physics beyond the SM. Perhaps the new particles, while still having a mass in the TeV ballpark, are just out of the kinematic reach of the LHC. Alternatively, the new particles are in principle accessible at the LHC, but are still missed by the existing experimental searches because these
particles decay into non-standard final state configurations, which the current strategies are not optimized for.

In this paper we consider the second possibility, in the context of new multi-TeV mass spin-1 particles. For simplicity, we assume that these particles are EW charged and color-neutral, and categorize them generically as $W^{ \prime }/Z^{ \prime }/\gamma^{ \prime }$. Such vector particles arise in a multitude of extensions of the SM and thus have been and continue to be the subject of intense study, both theoretically and experimentally. However, most of these analyses focus on the decay of the spin-1 particles into a pair of SM particles, either fermions or EW bosons (including $W/Z/\gamma$ and Higgs). Therefore, the discovery in this case is simply via a peak in the invariant mass distribution of the SM particles pair.

While the ``di-SM'' modes are the simplest modes available for the decays of such $W^{ \prime }/Z^{ \prime }/\gamma^{ \prime }$, we highlight that there are regions of parameter space (either in the same model, or in simple plausible variations and generalizations thereof) where a two-step cascade decay can be significant or even dominant. In this process, the heavy vector particle decays into a SM EW (gauge or Higgs) boson, plus a lighter scalar particle,\footnote{This intermediate on-shell particle can also be another spin-1 particle, but here we concentrate on the scalar option for simplicity.} and the scalar particle subsequently decays into a pair of SM EW bosons. The final state thus contains three SM EW bosons. Such a decay chain is schematically shown as 
\bea
V^{ \prime } & \rightarrow V + \phi , \nonumber \\
\phi & \rightarrow V + V,
\label{reaction}
\eea
where $V^{ \prime }$ denotes the heavy spin-1 particle, $\phi$ represents the intermediate scalar, and $V$ stands for the SM EW bosons $W/Z/\gamma$.
Note that, in general, the three ``$V$''s here could be all different.
The noteworthy feature of this new channel is the simultaneous presence of two resonances, resulting in a peak in the triboson invariant mass distribution
at the spin-1 mass, and a second peak in the invariant mass of decay products of the scalar particle at its mass.
Both resonances may be fully reconstructible, 
in particular for the hadronic decays of $W/Z/$Higgs bosons.
However, as we will see, the current experimental searches incorporate the diboson invariant mass distribution only,
as is sufficient for the corresponding vanilla two-body decay of these spin-1 particle. In this work we note that these searches might not be efficient for the signal with three bosons
(as is perhaps expected looking at the resonance structure outlined earlier, but we will argue explicitly in what follows, based on the analysis in \cite{Aguilar-Saavedra:2016xuc,Aguilar-Saavedra:2017iso}).
Hence, developing new dedicated search strategies for such signals is well motivated, which we will pursue in detail throughout this paper.

One illustration of this inadequacy of the standard diboson analysis for the final state with three bosons is the case where the intermediate scalar is lighter than the parent vector resonance but still relatively heavy, for example, $1-1.5$ TeV scalar vs.~$2-3$ TeV vector. Given the $\mathcal O$(TeV) mass gap between the scalar and the vector, the prompt SM boson which comes from the direct decay of the vector in the first line of eq.~(\ref{reaction}) will be highly boosted. So will be the other two bosons from the scalar decay due to its $\mathcal O$(TeV) mass. By contrast, the scalar itself will only have a mild boost so that the two bosons from the scalar decay will not be merged. Thus, the three SM bosons in the final state for this regime of scalar mass will be in general well-separated from each other 
and will all have roughly similar hardness so that the final state can be described as ``triboson''. Standard identification techniques can then be used to identify the three SM bosons and reconstruct the vector and scalar resonances.

Now, a crucial point is that current diboson analyses often select only the hardest one or two fat jets as hadronic $V$-candidates (depending on if the search is fully hadronic or semi-leptonic). This is true of all ATLAS diboson searches (e.g.~\cite{Aaboud:2017eta, Aaboud:2017fgj}) and some CMS diboson searches (e.g.~\cite{CMS-PAS-B2G-17-005}). For the vector and scalar mass choices above, this will typically select the prompt $V$ and the harder of the two decay products of $\phi$, and these do not form a resonance pair. This leads to a broad diboson invariant mass bump and a kinematic edge, rather than a resonance peak. Additionally, this broad bump will peak at masses significantly lower than the vector boson mass where the backgrounds are higher. These two effects (broadening and reduced invariant mass) lead to significantly diminished sensitivity in diboson searches compared to a diboson resonance of the same mass, despite the triboson signature being inherently more distinctive. Other searches (e.g.~\cite{CMS:2017skt}) allow subleading jets to form hadronic $V$-candidates. 
While these searches may more frequently select the two $\phi$ decay products as a resonance pair, the diboson invariant mass distribution of the signal will generically have multiple bumps corresponding to different possible diboson pairings, making even these diboson searches difficult to interpret in the context of this signal. Therefore, good sensitivity to these cascade decays requires a consideration of the full triboson invariant mass which should peak at the vector mass. Additionally, various diboson invariant masses would be required in order to uncover the entire structure of the signal process and observe the $\phi$ resonance peak.

The couplings involved in the cascade decays of eq.~(\ref{reaction}) may be of two types. First one involves a product of vector fields directly, requiring EW symmetry breaking (EWSB) spurions such as the Higgs vacuum expectation value (VEV), as can be found in models of Left-Right symmetry \cite{Mohapatra:2013cia}. The second one involves a products of field strength tensors, hence such a VEV is not needed. While we will comment on both possibilities, we focus on the latter case which can arise from a variation of the standard warped extra dimensional framework. In this case, the heavy spin-1 particle of the general triboson signal above corresponds to the extra-dimensional Kaluza-Klein (KK) excitations of the SM EW gauge bosons. The role of the scalar particle can be played by the radion, which is roughly the modulus corresponding to fluctuations in size of the extra dimension. The coupling of this scalar particle to the gauge fields (both SM and their KK modes), relevant to this work, is via the latter's field strength tensors. Such couplings treat all gauge bosons on a similar footing, up to the SM gauge couplings. Therefore this includes a coupling to photon as well. In minimal models this process has a small rate, but it can be significant in models of the class introduced in~\cite{Agashe:2016rle} which include an intermediate brane and an extended bulk. In this paper we consider two specific realization of this class of models, in which only a subset of the gauge fields propagate into the extended piece of the bulk.\footnote{The LHC signals for such a framework, but with {\em all} SM gauge fields propagating in the extended bulk were studied in \cite{Agashe:2016kfr}.}

In the first realization of such a framework, all the EW gauge fields (both $SU(2)$ and hypercharge) are allowed to propagate in the extended bulk, while the gluons do not. The largest signal channel in such a realization is tri-$W$, from KK $W$ production followed by its decay into a $W$ and a radion, and the subsequent radion decay to two $W$'s. 
There is also an interesting $W +$ diphoton channel from radion decay to diphotons which is sub-dominant in rate, but also has less SM background. This channel does not face the combinatorics issue in identifying the correct diboson (diphoton in this case) in order to form the resonant peak from the scalar decay. Nonetheless, for the existing diphoton search the background for such a signal is also set by the scalar mass. Thus, it is likely to overwhelm the signal whose production cross section is determined by the (larger) mass of the vector (i.e., the primary parent). In order to isolate this signal, and especially to pin down the complete model, we will show that tagging the $W$ boson as well, and using the diphoton and three boson invariant mass distributions is well motivated.%

The second realization of the framework, which is a slight variation of the first, allows only the hypercharge gauge boson to propagate in the extended bulk. This results in a dramatic triphoton signal from KK photon production and its decay into photon and radion, followed by a radion decay to diphoton.

Here is the outline of the rest of this paper. We begin in section~\ref{sec:Categorizing} with a rough classification of Lorentz structures relevant for the triboson signal. In section~\ref{warped}, we provide a more detailed description of the warped extra dimension model with extended bulk, which gives rise to the tensor-type coupling of the radion with a pair of vector bosons. Section~\ref{tools} is reserved for a brief description of the techniques that we use for studying the LHC signals. Section~\ref{sec:results} contains a detailed discussion of our main results in the $WWW$, $W\gamma\gamma$ and $\gamma \gamma \gamma$ channels. We provide a summary of our results in section~\ref{conclusion}, and touch on possible future directions. We provide brief discussions on same sign dilepton constraints for tri-$W$ signal and the choice of jet tagging method in the appendices.

\section{Categorizing Models for Triboson Signals}\label{sec:Categorizing}
We begin by giving a broad classification of the models with EW triboson signals that we will study in the rest of this paper. As outlined in the introduction, the basic event topology is shown in eq.~(\ref{reaction}), where $V^{ \prime }$ is a heavy, EW (only)-charged spin-1 particle, $\phi$ is a 
new, lighter than $V^{ \prime }$, scalar particle, and $V$'s (which could be all different) are SM gauge or Higgs bosons. It is clear that each step of this cascade decay involves a coupling between two vector particles (SM or BSM) and one scalar (new) particle. Based on the Lorentz structure of such a coupling, we consider two representative class of models, vector-type and tensor-type, which will be discussed in order. For simplicity, we assume that the same type of coupling is involved at each step of eq.~(\ref{reaction}), although a ``mixed'' decay chain is also allowed.

\subsection{Models with Vector-type Coupling}
In this class of models, the coupling between the scalar $\phi$ and the gauge bosons $V$ and $V'$ has the form
\bea
{\cal L } & \ni & v_{\text{EW}} \left(
a_{VV}\: g^2 V_\mu V^\mu + a_{VV'}\:g_\star g V_\mu V'^{\mu}
\right)\phi\:,
\label{vector1}
\eea
where $v_{\text{EW}}$ is the EW-scale vacuum expectation value, $g$ ($g_\star$) is the gauge coupling corresponding to $V_\mu$ ($V'_\mu$), and $a_{VV}$ ($a_{VV'}$) are dimensionless constants. We dub this type of models ``vector'' due to the coupling involving the vector fields directly.
Since $V^{ \mu } V_{ \mu }$ and $V_\mu V'^\mu$ are not gauge-invariant operators by itself, we need an EW gauge symmetry breaking spurion, thus accounting for the EW-scale vacuum expectation value $v_{\rm EW}$. Gauge invariance also requires the corresponding factors of the gauge couplings. 
With this structure, it is then clear that {\em no} such coupling is allowed between $\phi$ and two SM photons $\gamma$, i.e. $a_{\gamma\gamma} = 0$.
This kind of operators  may arise, for example, by mixing between $\phi$ and the SM Higgs. In fact the first term in eq.~(\ref{vector1}) is present in the SM with $\phi$ identified as the SM Higgs.

An attractive realization of the above structure 
is provided by the Left-Right symmetric model \cite{Mohapatra:2016twe}. Here the SM EW symmetry is extended to 
$SU(2)_R \times SU(2)_L \times U(1)_{ B - L }$, with $SU(2)_R \times U(1)_{ B - L }$ spontaneously broken down to $U(1)_Y$ at around TeV scale, giving massive (electrically charged) $W_R$ and (electrically neutral) $Z^{ \prime }$, the latter being the combination of $U(1)$ contained in $SU(2)_R$ and $U(1)_{ B - L }$ which is orthogonal to $U(1)_Y$. 
These extra EW gauge bosons then play the role of $V^{ \prime }$ in eq.~(\ref{reaction}), while the (natural) 
origin of $\phi$ is as follows. The SM Higgs field, which is an $SU(2)_L$ doublet, is promoted to be a bi-doublet of $SU(2)_L \times SU(2)_R$. This is similar to the construction in the two Higgs doublet model (2HDM). 
The role of $\phi$ is then be taken by the second physical/CP-even neutral scalar arising from this bi-doublet field. The couplings in eq.~(\ref{vector1}) follow in a rather straightforward manner, given that $\phi$ is charged under {\em both} $SU(2)_R$ and $SU(2)_L$.

\subsection{Models with Tensor-type Coupling}
In this class of models, scalar $\phi$ can couple to the field strength of the gauge bosons $V$ and $V'$ as
\bea
{\cal L} 
& \ni & 
\left( b_{VV} 
\; g^2 \; 
V^{ \mu \nu } V_{ \mu \nu } 
+ b_{VV'} \;  g_{ \star } \; g \;
V^{ \mu \nu } V^{ \prime }_{ \mu \nu } 
\right) 
\frac{ \phi }{ \Lambda }\,,
\label{tensor}
\eea
where $\Lambda$ parameterizes a relevant new physics scale, $g$ ($g_\star$) is the gauge coupling corresponding to $V$ ($V'$) and $b_{VV}$ ($b_{VV'}$) is a dimensionless constant. The name of this type of models follows from the coupling to the field strength tensors. Unlike the previous case, gauge invariance allows $\phi$ to also couple to two SM photons $\gamma$.
 
Note that the first term is analogous to the leading SM Higgs coupling to gluons and photons.
For concreteness, we have assumed $\phi$ to be CP-even here. The generalization to CP-odd case would obviously involve 
$\widetilde{V}^{ \mu \nu }$ instead, where $\widetilde{V}^{ \mu \nu }$ is the corresponding dual field strength tensor.
\begin{figure}[tbp]
\center
\includegraphics[width=0.6\linewidth]{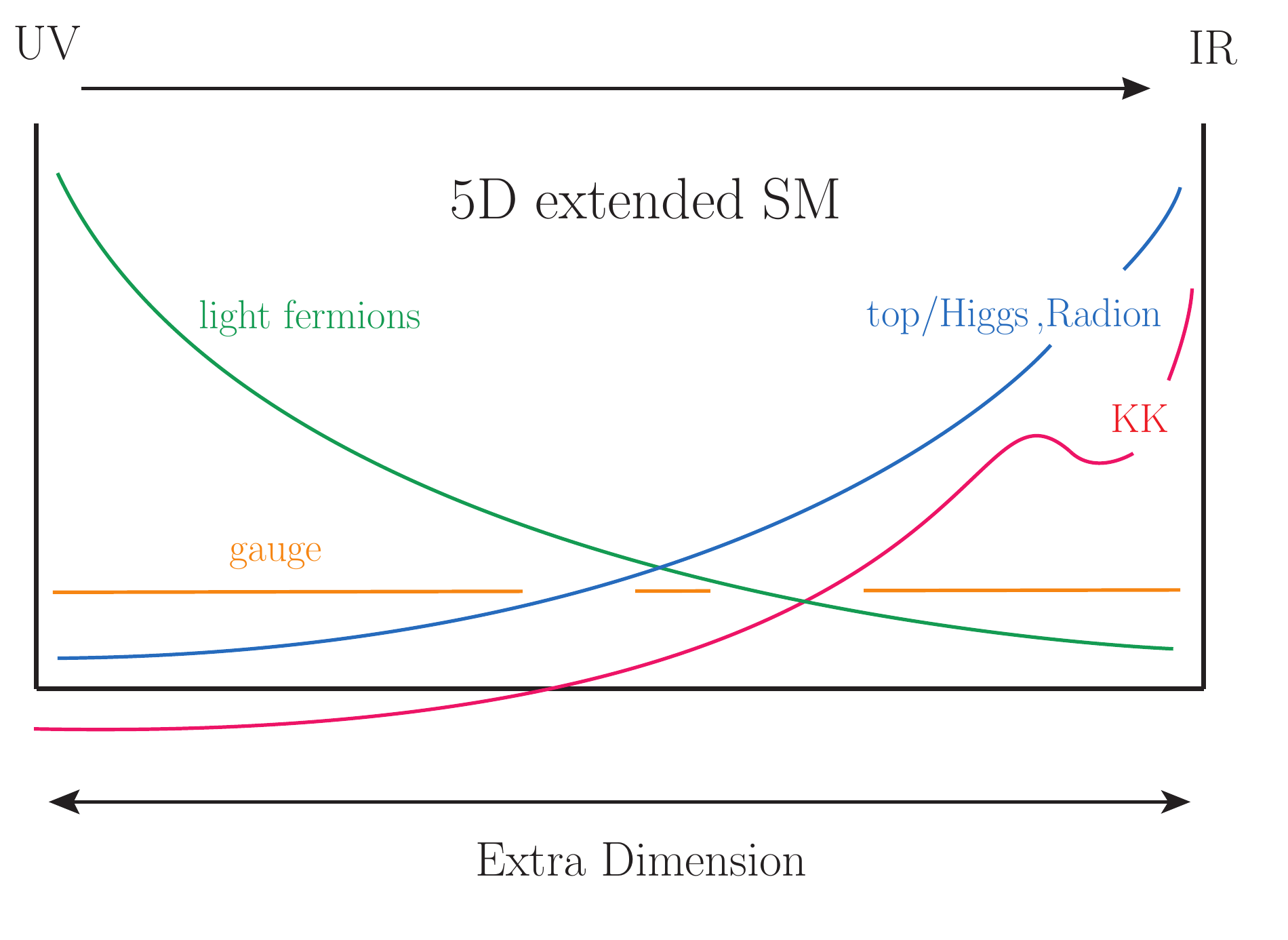}
\caption{Warped extra dimensional model with SM fields in the same bulk (standard framework). Schematic shapes of extra-dimensional wavefunctions for various particles (zero mode SM fermions and gauge particles, a radion, and a generic KK mode) are shown. }
\label{fig:standard}
\end{figure}

As we outline below, a well-motivated example where the above couplings provide the leading signals is a framework with the SM fields
propagating to {\em varying} degrees in a  
warped extra dimension. We will content ourselves with a quick review here in order to mostly explain the context. For more details, the reader is referred to previous papers~\cite{Agashe:2016rle,Agashe:2016kfr}. First, we summarize the standard set-up with two branes, i.e., UV/Planck and IR/TeV branes (see figure~\ref{fig:standard}).
The $\phi$ of the general case above corresponds here to the radion, namely, the scalar modulus which is roughly associated with the 
fluctuations in the size of the extra dimension.
It is well known that the radion couples to (transverse polarizations of) SM gauge bosons (including gluon) as in the first term in eq.~(\ref{tensor}), even before a mechanism for radius stabilization is included, i.e., for a massless radion. 
The KK excitations of the SM gauge fields play the role of $V^{ \prime }$ etc. Finally, the coupling of KK gauge bosons to SM gauge bosons and the radion [the second term in eq.~(\ref{tensor})] is subtler in that it arises only 
after radius stabilization \cite{Agashe:2016rle}: see section \ref{warped} for an explicit parametrization of these couplings.

However, in spite of the {\em existence} of the couplings in eq.~(\ref{tensor}), hence allowing the process in eq.~(\ref{reaction}), the crucial observation is that {\em both} the two-body decays in this chain are swamped by decays into top quark/Higgs (including longitudinal $W$ and $Z$ gauge bosons). Let us sketch the reason for this, which will also suggest a possible way to suppress this contribution.

Schematically, in extra-dimensional models, couplings between 4D fields are proportional to the overlap of corresponding profiles in the extra dimension. Therefore, the couplings in eq.~(\ref{tensor}) originate from overlap of the gauge KK/radion profile, which is peaked near the IR brane, with the SM gauge modes which are delocalized in the extra dimension (see for example figure~\ref{fig:standard}). The overlap is therefore between one (two) IR localized profile and two (one) delocalized profiles for the first (second) term in eq.~(\ref{tensor}).
By contrast, the gauge KK/radion coupling to top quark/Higgs particles involves {\em all} (three) profiles being peaked near the IR brane. Thus, the latter couplings dominate so that ditop and diboson final states have the highest rate, and therefore have been the staple searches in this context.\footnote{As far as {\em production} of these gauge KK modes is concerned, it turns out that dominant channel is via light quark-antiquark annihilation. These quark profiles are peaked near the UV brane; nonetheless the overlap with gauge KK is non-negligible, albeit somewhat suppressed compared to that with zero-mode SM gauge fields.}

\begin{figure}[tbp]
\centering
\includegraphics[width=0.6\linewidth]{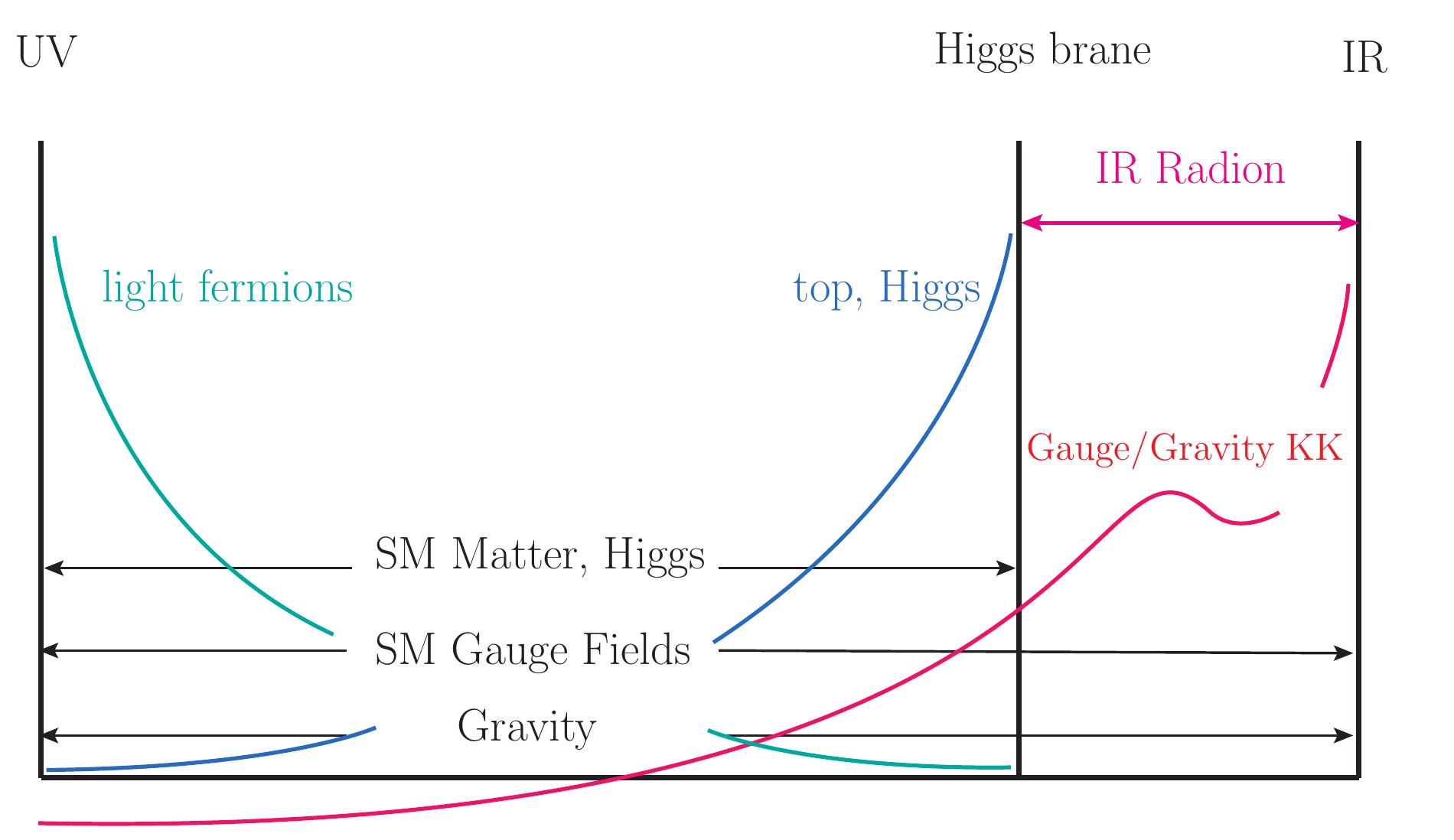}
\caption{Warped extra dimensional model with {\em all} SM gauge fields (and gravity) in the full bulk, but matter and Higgs fields in subspace (extended framework). Schematic shapes of extra-dimensional wavefunctions for various particles (zero mode SM fermions and gauge bosons, an (IR) radion, and a generic KK mode) are shown.  }
\label{fig:extended}
\end{figure}

Remarkably, in an extension of the above scenario of the type proposed in \cite{Agashe:2016rle}, the scales get tilted, as far as the dominant couplings and signals are concerned.
The central theme here is that only SM gauge and gravity could be living in the {\em full} bulk, while the matter/Higgs fields propagate only down to an intermediate brane with scale $\mathcal O(10)$ TeV (i.e., still in the ``IR'' region), which we call ``Higgs brane". The SM Higgs and top quark are peaked at the Higgs brane, but (modestly) ``split'' from the lightest gauge/graviton KK peaked at the (final) IR brane at a few TeV (see for example figure~\ref{fig:extended}).
Note that this situation necessarily leads to {\em two} radions, tied to the location of the intermediate and the IR branes, respectively. The lighter radion (which will be the relevant one here) is peaked near the IR brane, like the lightest gauge KK.
Such a modification of space occupancy implies that the couplings of top quark/Higgs particles to gauge/graviton KK and the lighter radion are suppressed due to small overlap of profiles, as seen in figure~\ref{fig:extended}.
Thus, signals of the type in eq.~(\ref{reaction}) can become highly relevant.
In a recent work~\cite{Agashe:2016kfr}, we initiated a study of such LHC signals of this novel idea with the focus being on the case where {\em all} SM gauge fields propagate in the full space, i.e., KK gluon and KK EW gauge bosons are {\em both} relevant for phenomenology.
The radion dominantly decays into SM gluons so that the cascade decay of the type in eq.~(\ref{reaction}) typically features a gluon initiated jet, either at the first stage (i.e., if we start with a KK gluon) or/and the second (i.e., radion decay).
Thus, even though cascade decays of the form in eq.~(\ref{reaction}) are uncovered by such an extension, the triboson signal therein is still negligible.

\begin{figure}[tbp]
\centering
\includegraphics[width=0.48\linewidth]{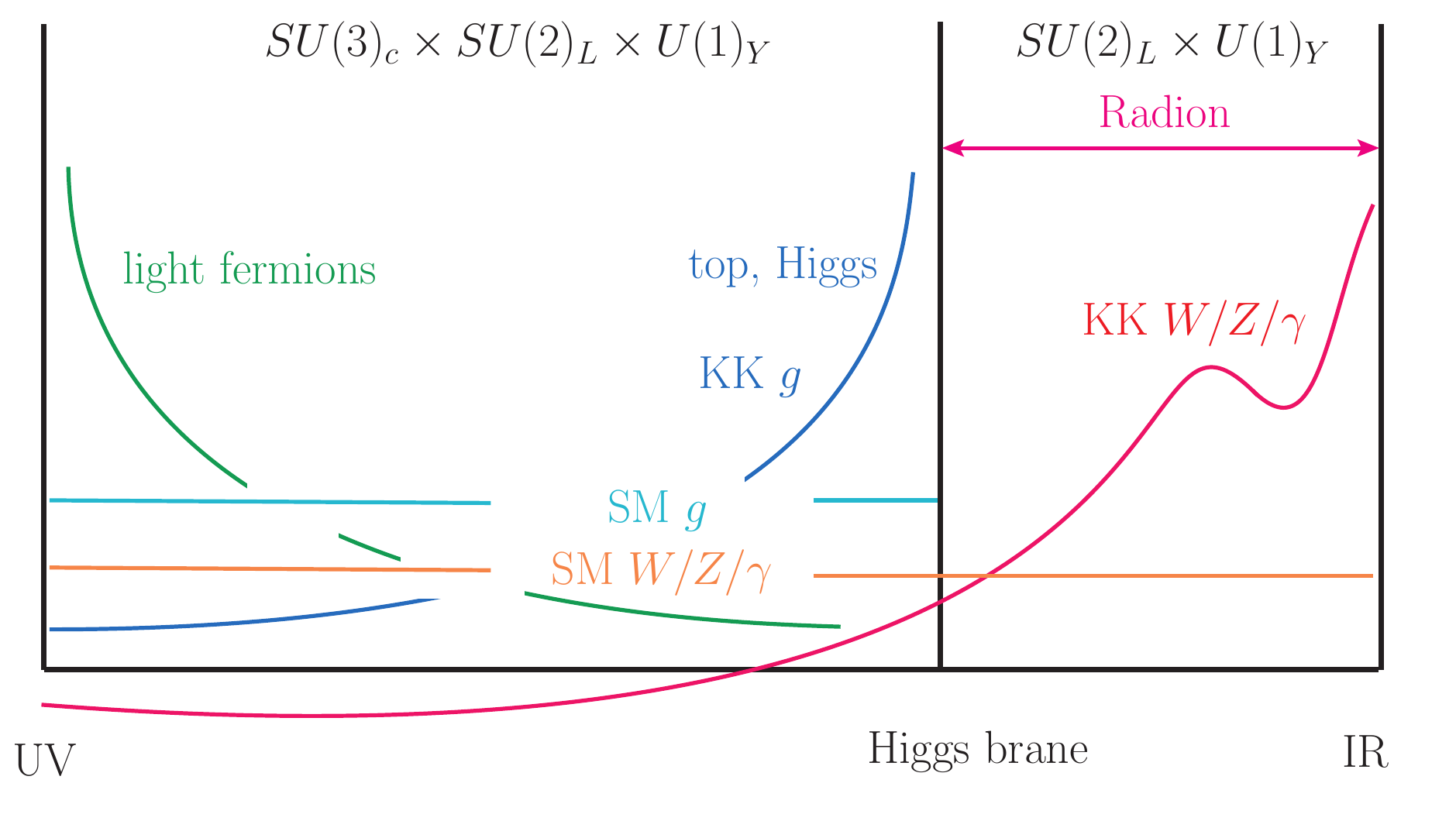}
\includegraphics[width=0.48\linewidth]{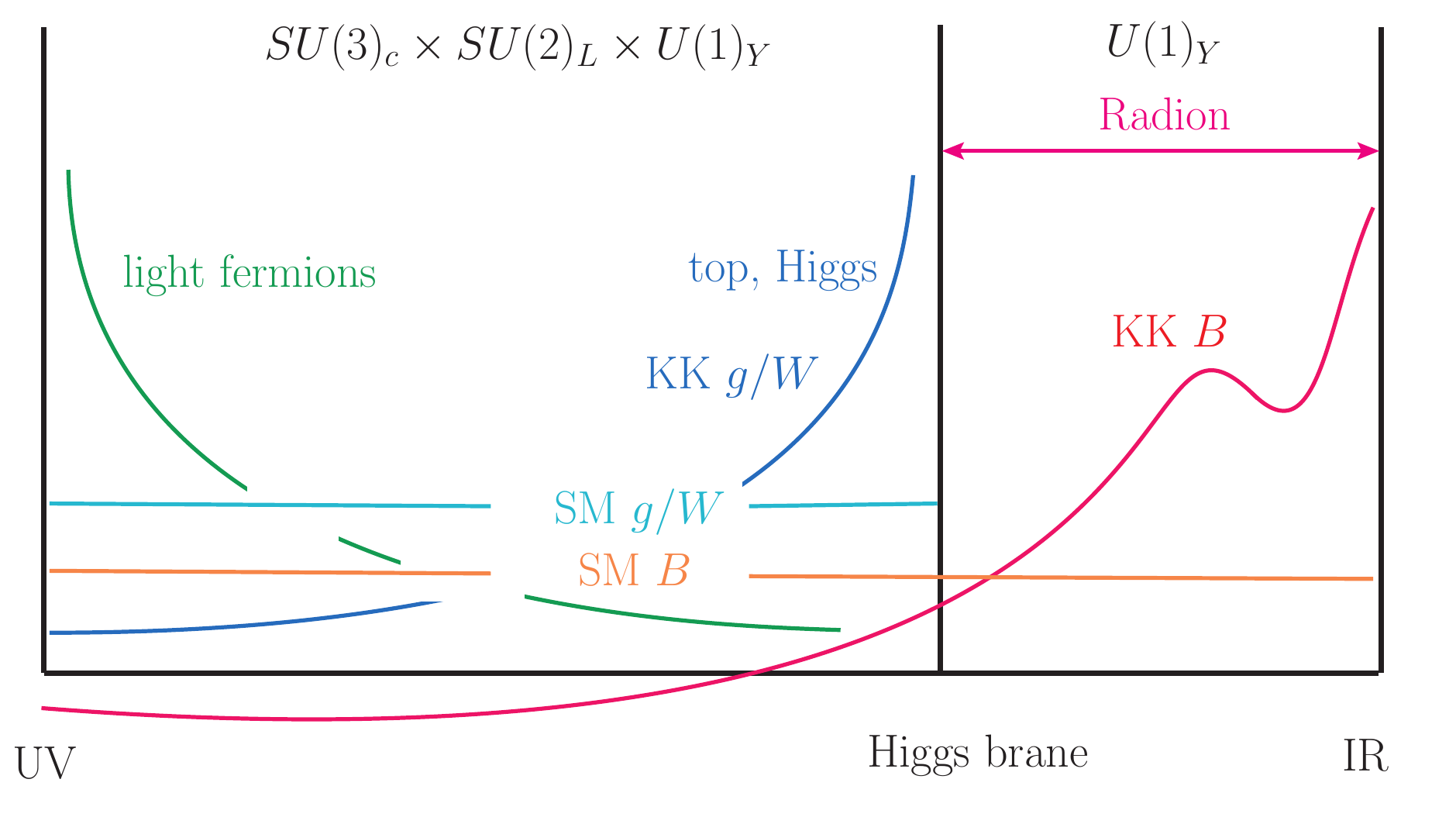}
\caption{Warped extra dimensional model with {\em only} SM EW gauge fields (on the left) and only SM hypercharge field (on the right) in the full bulk, but matter, Higgs and other gauge fields
in subspace (extended framework). Schematic shapes of extra-dimensional wavefunctions for various particles (zero mode SM fermions and gauge bosons, an (IR) radion, and a generic KK mode) are shown. }
\label{fig:EWextended}
\end{figure}

As a further modification, in this paper, we consider the possibility that only the EW gauge fields, or only the hypercharge field propagate down to the final IR brane, i.e., other gauge fields like gluon field (similar to matter/Higgs fields) stops at the intermediate brane itself (see for example figure~\ref{fig:EWextended}). In such a realization, the KK gluon is too massive to be produced at the LHC, whereas the KK EW gauge bosons, or KK hypercharge gauge boson are {\it still} accessible.
Crucially, the radion coupling to SM gluons is rendered negligible (as is the case for radion coupling to the Higgs/top). The rest of the couplings, i.e., involving EW gauge bosons (whether SM or heavy ones), are unchanged with respect to those studied (and given in a detailed form) in the earlier references. In particular, radion decay is then dominated by $W$, $Z$, and $\gamma$, leading to the triboson signal of eq.~(\ref{reaction}) as the dominant channel.

\section{Overview of Triboson Signals at the LHC for Warped Models}
\label{warped}
We are now in a position to discuss the tensor-type model in detail, arising within the framework of general extensions of standard warped extra dimensional models, as schematically displayed in figure~\ref{fig:EWextended}.
We first introduce two models under this extended framework; the first one allows the gauge bosons of $SU(2)_L\times U(1)_Y$ to propagate in the extended bulk while the second one allows only the gauge bosons associated with $U(1)_Y$ to do so. We henceforth denote them by ``EW model'' and ``hypercharge model'', respectively.
As discussed in the introductory section, different field propagation in the extended bulk results in different triboson signals in the two models: the $WWW$ and $W\gamma\gamma$ final states in the EW model, which result from the production of KK $W$ gauge boson and its subsequent decay, and the $\gamma\gamma\gamma$ final state in the hypercharge model arising from the decay of KK hypercharge gauge boson.

We begin with the discussion on the EW model in section~\ref{subsec:EW}, followed by the hypercharge model in section~\ref{subsec:hypercharge}. 
For each model, a list of relevant couplings are shown first, followed by analytic expressions for the decay widths for KK gauge bosons/radion. 
We then translate various experimental constraints to bounds on the couplings and mass parameters. Finally, we present several benchmark points for each model, which are safe from all bounds, in section~\ref{subsec:benchmark}.

\subsection{Model with EW Gauge Fields in the Extended Bulk}\label{subsec:EW}
In this model, all $SU(2)_L\times U(1)_Y$ gauge fields are allowed to propagate all the way down to the final IR brane. 
$SU(3)_c$ gauge fields, together with Higgs and the other fermion fields,  live only between the UV and intermediate Higgs branes. 
Therefore, the lightest KK modes are first KK modes of EW gauge bosons.\footnote{Although KK gravitons have the same KK scale as KK EW gauge bosons, they are typically heavier than KK gauge bosons.} In addition, the IR radion can be lighter than KK EW gauge bosons (see \cite{Agashe:2016rle} and references therein). Therefore, the three KK EW gauge bosons, the radion and the SM particles are phenomenologically relevant to our study. Note that due to the shape of the KK profiles (see e.g. figure~\ref{fig:EWextended}), the IR radion has sizable couplings to EW gauge bosons, whereas couplings to gluons and top quark/Higgs are highly suppressed. From all these considerations, the relevant couplings between KK EW gauge bosons, SM EW gauge bosons, and the radion can be written as 
\bea\label{eq:LVKK}
\mathcal L^{\rm EW}_{\rm warped} &\ni& \frac{g^2_V}{g_{V_{\rm KK}}} V_{\rm KK} ^\mu J_{V\mu} \nonumber \\
&+& \left(  -\frac{1}{4}\frac{g_{\rm grav}}{g_{V_{\rm KK}}^2} g_{V}^2  V_{\mu\nu}V^{\mu\nu}+ \epsilon \frac{g_{\rm grav}}{g_{V_{\rm KK}}^2} g_{V_{\rm KK}} g_{V} V_{\mu\nu}V^{\mu\nu}_{\rm KK} \right) \frac{\varphi}{m_{\rm KK}}\,,
\eea
where we denote KK EW gauge bosons and SM EW gauge bosons collectively as $V_{\rm KK}$ ($=W_{\rm KK},\,Z_{\rm KK},\,\gamma_{\rm KK}$) and $V$ ($=W,\,Z,\,\gamma$) respectively. 
Also, $V^{\mu\nu}$ and $V_{\rm KK}^{\mu\nu}$ are the usual field strength tensors for SM and gauge KK bosons respectively. The gauge couplings $g_V$ and $g_{V_{\rm KK}}$ are specific to SM and KK gauge fields, correspondingly, while KK gravity coupling is denoted by $g_{\rm grav}$. 
$m_{\rm KK}$ stands for the mass of KK gauge boson which can be identified as the new physics scale $\Lambda$ in eq.~\eqref{tensor}.
Finally, $\varphi$ symbolizes the radion field and $J_{V\mu}$ is the current made of SM fields (fermions, gauge bosons and the Higgs) associated with the SM gauge boson $V$. The KK EW gauge bosons $V_\text{KK}$ can be identified with $V'$ and the radion $\varphi$ can be identified with $\phi$ in eq.~\eqref{tensor}. 

The first term in eq.~\eqref{eq:LVKK} contains the interaction between the gauge KK bosons and the SM fermion currents, which is the same as that in standard warped models, thus responsible for the production of $V_{{\rm KK}}$ from light quarks inside protons at the LHC. 
The second term is the coupling of radion to SM EW gauge bosons with the prefactor $-\frac{1}{4} \frac{g_{\rm grav}}{g^2_{W_{\rm KK}}}$ identified as $b_{VV}$ in eq.~\eqref{tensor}. Radions in warped models specifically couple to field strength tensors of SM gauge bosons~\cite{Agashe:2016rle,Agashe:2016kfr}, 
and this operator governs the decay of $\varphi$ to a SM gauge boson pair. 
The last term is the coupling among the radion, SM gauge, and KK gauge bosons, and only arises after radius stabilization with $\epsilon \frac{g_{\rm grav}}{g^2_{W_{\rm KK}}}$ identified as $b_{VV'}$ in eq.~\eqref{tensor}. The parameter $\epsilon$ is roughly $1/\log (\Lambda_{\rm Higgs}/\Lambda_{\rm IR})$ where $\Lambda_{\rm Higgs}$ and $\Lambda_{\rm IR}$ are the scales of intermediate Higgs and IR branes, respectively. This last term describes the decay of $V_{{\rm KK}}$ to $V$ and $\varphi$, and as we show below, in the parameter space of interest, it is the dominant decay mode.

\begin{figure}
\centering
\includegraphics[width=0.6\linewidth]{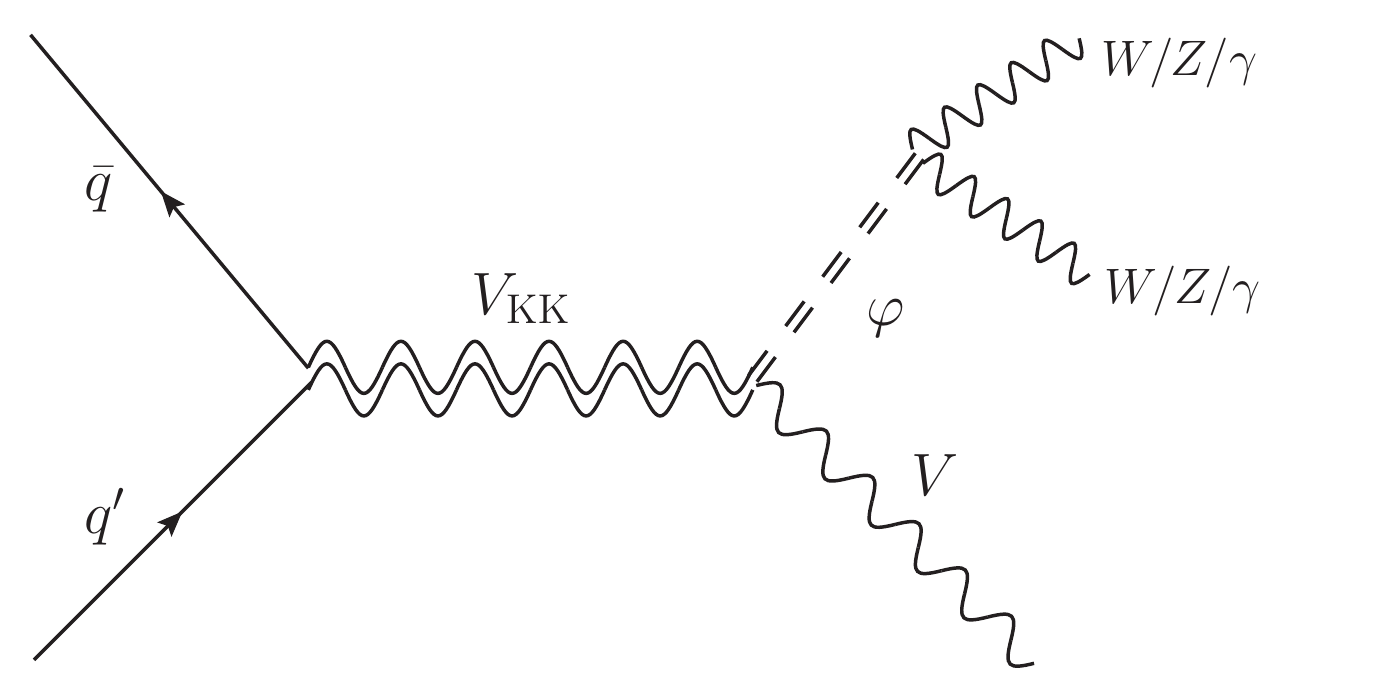}
\caption{Feynman diagram for the signal process from production to decay. 
Double (single) lines represent KK (SM) particles and $q / q'$ denote light quarks inside the proton.
The signal process is characterized by two resonance bumps given by $V_{{\rm KK}}$ and $\varphi$.}
\label{fig:triboson}
\end{figure}

From all these considerations, we have ended up with our triboson signal channels as follows: KK EW gauge bosons are produced on shell via light quark annihilation and then decay to a corresponding SM gauge boson and a radion, the latter of which subsequently disintegrating to a pair of SM EW gauge bosons. The generic event topology is displayed in figure~\ref{fig:triboson}. As stated before, we focus on two triboson signals as their production rates are expected to be large:
\begin{align}
W_\kk \rightarrow W \varphi \rightarrow WWW\:,\:\:\text{and}\:\: W_\kk \rightarrow W \varphi \rightarrow W\gamma\gamma\,.
\end{align}
We are now about to investigate allowed values for the model parameters appearing in eq.~\eqref{eq:LVKK}.

\medskip
\noindent \textbf{KK gauge and KK gravity couplings}: There are only two free KK gauge couplings in this model: $g_{W_{\rm KK}}$ and $g_{B_{\rm KK}}$ corresponding to couplings of $SU(2)_L$ and $U(1)_Y$. The KK gauge couplings of $\gamma_{\rm KK}$ and $Z_{\rm KK}$ are obtained, similarly as in the SM, by the following relations: 
\bea
g_{\gamma_{\rm KK}} = \frac{g_{W_{\rm KK}}g_{B_{\rm KK}}}{\sqrt{g_{W_{\rm KK}}^2+g_{B_{\rm KK}}^2}}\,,~~~~ g_{Z_{\rm KK}} = \sqrt{g_{W_{\rm KK}}^2+g_{B_{\rm KK}}^2}.
\label{eq:g_gamma_g_Z_from_others}
\eea
In order for the weakly coupled 5D EFT to be valid (i.e. remain perturbative) up to a high enough mass scale, so as to include $\mathcal{O}(1)$ KK modes, $g_{g/W/B_{\rm KK}} $ should satisfy an upper bound of roughly 6 (see also refs.~\cite{Agashe:2016rle,Agashe:2016kfr} for a more detailed argument).
On the other hand, a lower limit of roughly 3 for gauge KK coupling comes from the requirement that the Landau pole scale is higher than the GUT scale. Therefore, the allowed ranges for KK gauge couplings are
\bea
3 \: \lesssim g_{W_\text{KK}/B_\text{KK}} \: \lesssim 6, \label{eq:paramranges}
\eea
from which we deduce the limits for $g_{\gamma_{\rm KK}}$ and $g_{Z_{\rm KK}}$ 
from eq.~(\ref{eq:g_gamma_g_Z_from_others}).
Similar to the case of KK gauge couplings, the KK gravity coupling has the upper limit around 6. However, since there is no Landau pole issue in the gravity sector, the KK gravity coupling can naturally go down to $\mathcal O(1)$. Hence, the allowed range for the KK gravity coupling is given by 
\bea
\mathcal{O}(1) \lesssim g_{\rm grav} \lesssim 6\,.
\eea

\medskip

\noindent \textbf{KK gauge boson and radion masses:} 
Various resonance searches at the LHC constrain the masses for KK gauge bosons.  We shall discuss the associated bounds in section~\ref{subsec:EWbound} in detail.
We choose $m_{\rm KK}$ to be slightly heavier than the current bound: in most channels $m_\kk = 3$ TeV. Radion masses are chosen to be $m_\varphi = 1$ TeV and $1.5$ TeV in this study. 

\medskip

\noindent \textbf{Parameter $\epsilon$:} Generically, $\epsilon$ needs to be $\mathcal{O}$(1/ a few) in order for the hierarchy $\Lambda_{\rm Higgs} / \Lambda_{\rm IR}$ to be stabilized.  As evident from eq.~(\ref{eq:LVKK}), large values of $\epsilon$ allow enhanced signal cross section. In this regards, for our benchmark points, we take $\epsilon=0.5$.
\subsubsection{Relevant Particles and Current Bounds}\label{subsec:EWbound}
In this section, we briefly discuss the mass bounds on new particles involved in heavy KK gauge boson decay processes. To this end, we first review the tree level decay widths of relevant particles, followed by discussions on current bounds.

\subsubsection*{(a) Radion}
In this EW model, the direct production of radion is available through vector boson fusion (VBF) encoded in the second coupling in eq.~(\ref{eq:LVKK}). The same interaction vertices are responsible for its dominant decays to SM EW gauge boson pairs such as $WW$, $ZZ$, and $\gamma \gamma$. To leading order, the radion decay width is given by
\bea
\Gamma(\varphi \to V V )\approx N_V g^2_{\rm grav} \left( \frac{g_{V}}{ g_{V_{\rm KK}}} \right )^4\left(\frac{m_{\varphi}}{m_{\rm KK}}\right )^2 \frac{m_\varphi}{64 \pi} \,,
\label{eq:Radion_BR}
\eea
where $N_V$ is the degrees of freedom of SM gauge boson: 2 for $W$ and 1 for $\gamma$ and $Z$. Note that the approximation symbols in some of the partial decay width formulae in this section originate from taking the massless limit of SM particles.

\paragraph*{Current bounds on radion:} 
The bounds for radion mass come from the radion decay to diboson or diphoton. The only direct production channel for radion is via VBF, and the production rate is typically smaller than those of quark pair-annihilation and gluon fusion. Therefore, radion mass can take values in a wide range of parameter space, which are consistent with current bounds. This is to be contrasted with the situation in 2 brane RS models, where the radion production rate from quark pair annihilation and gluon fusion is large, and constraints the radion mass to be heavier. 

As mentioned in section~\ref{sec:introduction}, we are focusing on heavy radion, say $m_\varphi\gtrsim 1$ TeV, even though lighter radion mass is allowed. This is in order to study three well-separated bosons.\footnote{For a light radion, the bosons from its decay can get merged, requiring dedicated strategies for isolating the signal. We will study this in a follow-up paper~\cite{triboson_lightRad}.} We have explicitly checked that $m_\varphi\gtrsim 1$ TeV is safe from diboson~\cite{Aaboud:2017eta,Aaboud:2017fgj,CMS-PAS-B2G-17-005,CMS-PAS-B2G-16-023} and diphoton~\cite{Aaboud:2017yyg} bounds for direct production of $\varphi$. 

Another production mechanism for the radion is from the KK gauge boson decay, which tends to be the dominant production channel. For $W_\kk \to W \varphi \to W\gamma\gamma$ channel, the invariant mass distribution of two photons will  show a bump at the radion mass, and therefore diphoton bounds are relevant. However, as we choose KK gauge bosons to be as heavy as 3 TeV, the radion production rate is small enough to be safe from current bounds. 
On the contrary, $W_\kk \to W \varphi \to WWW$ channel is sensitive to diboson searches in which the first two hardest $W$-jets are taken. Given our mass spectrum, it is very unlikely that two hardest-$p_T$ $W$-jets are from the radion decay. Therefore, the resulting diboson invariant mass would develop quite a broad distribution rather than a sharp bump at the radion mass. In summary, the radion mass bound is rather weak in this case not only due to a small production rate but also due to the combinatorial ambiguity discussed above. We will justify this point in the actual analysis in section~\ref{subsubsec:WWW}.

\subsubsection*{(b) KK EW gauge bosons}
KK gauge bosons are produced by light quarks inside proton via the first coupling in eq. (\ref{eq:LVKK}). In the parameter space of interest, the dominant decay channel for KK gauge bosons is $V_\kk \to V \varphi$, which we call ``radion channel'' for short. The structure of radion channel is encoded in the last coupling in eq. (\ref{eq:LVKK}). Also, KK gauge bosons have subdominant decay channels to two SM particles via flavor-universal coupling, i.e.~the first term in eq. (\ref{eq:LVKK}). 

\paragraph*{KK $W$:} 
The KK $W$ boson decay has radion channel and diboson, dijet, and dilepton channels: 
\bea
 &&\Gamma  (W_{\rm KK} \to \varphi\  W)\approx  \left( \epsilon  g_{\rm grav} \frac{g_{W}}{ g_{W_{\rm KK}}} \right )^2\left(1-\left(\frac{m_{\varphi}}{m_{\rm KK}}\right)^2\right)^3 \frac{m_{\rm KK}}{24 \pi}\,, \\
 && \Gamma  (W_{\rm KK} \to WZ)\approx\Gamma  (W_{\rm KK} \to Wh) \approx \left(\frac{g^2_{W}}{ g_{W_{\rm KK}}} \right )^2\frac{m_{\rm KK}}{192 \pi}\,, \\
 &&\Gamma  (W_{\rm KK} \to \psi \psi') \approx N_c \left(\frac{g^2_{W}}{ g_{W_{\rm KK}}} \right )^2\frac{m_{\rm KK}}{48 \pi}\,, 
\eea
where $W_{\rm KK} \to \psi \psi'$ represents KK $W$ decay into a pair of (different-flavored) SM fermions. $N_c$ denotes the color degrees of freedom of SM fermions (e.g., 3 for quarks and 1 for leptons).

\paragraph*{KK photon:} 
The KK photon decay has radion channel as well as $WW$, dilepton, dijet, and ditop channels:
\bea
 &&\Gamma  (\gamma_{\rm KK} \to \varphi \gamma) =  \left(  \epsilon g_{\rm grav} \frac{g_{ \gamma }}{ g_{\gamma_{\rm KK}}} \right )^2\left(1-\left(\frac{m_{\varphi}}{m_{\rm KK}}\right)^2\right)^3 \frac{m_{\rm KK}}{24 \pi}\,, \\
 && \Gamma  (\gamma_{\rm KK} \to W W ) \approx \left(\frac{g_{ \gamma }^2}{ g_{\gamma_{\rm KK}}} \right )^2\frac{m_{\rm KK}}{48 \pi}\,, \\
 && \Gamma  (\gamma_{\rm KK} \to \psi \psi ) \approx N_c Q_\gamma^2 \left(\frac{g_{ \gamma }^2}{ g_{\gamma_{\rm KK}}} \right )^2\frac{m_{\rm KK}}{12 \pi}\,, 
\eea
where $\gamma_{\rm KK} \to \psi \psi $ represents the KK photon decay into a pair of SM fermions. $Q_\gamma$ stands for the electric charge of the associated fermion $\psi$. 
\paragraph*{KK $Z$:}
The KK $Z$ boson decay has radion channel and diboson, dijet, ditop, and dilepton channels:
\bea
&&\Gamma (Z_{\rm KK} \to \varphi\  Z)\approx  \left( \epsilon  g_{\rm grav} \frac{g_{Z}}{ g_{Z_{\rm KK}}} \right )^2\left(1-\left(\frac{m_{\varphi}}{m_{\rm KK}}\right)^2\right)^3 \frac{m_{\rm KK}}{24 \pi}\,, \\
&&\Gamma (Z_{\rm KK} \to WW/Zh) \approx Q_Z^2 \left(\frac{g^2_{Z}}{ g_{Z_{\rm KK}}} \right )^2\frac{m_{\rm KK}}{48 \pi}\,,\label{eq: ZWW} \\
&&\Gamma  (Z_{\rm KK} \to \psi \psi) \approx  N_c Q_Z^2\left(\frac{g^2_{Z}}{ g_{Z_{\rm KK}}} \right )^2\frac{m_{\rm KK}}{24 \pi}.\label{eq: Zff}
\eea
Here $Q_Z$ for the $WW$ channel and $Zh$ channel are $\frac{1}{2}-\sin^2\theta_W$ and $\frac{1}{2}$, respectively, where $\theta_W$ is the usual Weinberg angle. $Q_Z$ in eq.~(\ref{eq: Zff}) is simply given by the SM $Z$ charge of the associated fermion $\psi$.

\paragraph*{Current bounds on KK gauge boson:} 
We tabulate allowed mass values for each KK gauge boson in table~\ref{tab:Bounds}. As the mass parameters of different KK gauge bosons are, in general, constrained by different searches at the LHC,\footnote{We studied in detail the potential constraint from same sign dilepton searches in appendix \ref{SSDL}.} we first identify LHC searches setting the most stringent bound on each of KK bosons. They appear in the third column of table~\ref{tab:Bounds} along with associated search channels.  
Since the production rate for KK gauge bosons in our models solely depends on two parameters, $g_{V_\kk}$ and $m_{V_\kk}$ (see also decay widths to fermion pairs), we provide the allowed parameter space in terms of them (see the last column of table~\ref{tab:Bounds}). In order to find a bound for each gauge boson mass, we simply carry out simulation at the leading order with \textsc{MG5@aMC}~\cite{Alwall:2014hca} and compare the output with existing data under the assumption of BR$(V_{{\rm KK}} \rightarrow V\varphi)=50\%$.   

\begin{table}[t]
\centering
\begin{tabular}{c | c c l}
\hline
 Model & Name & Current search & ~Allowed mass values [TeV]~  \\
\hline \hline
 \multirow{5}{*}{EW model}
 &\multirow{2}{*}{~KK $W$~}   &  \multirow{2}{*}{$\ell+E_T^{{\rm miss}}$  \cite{Aaboud:2017efa}}
  &  ~$m_{W_\kk}\gtrsim 2.5$ for $g_{W_\kk}\sim 4$\\
 & & & ~$m_{W_\kk}\gtrsim 3$ ~~for $g_{W_\kk}\sim 3$\\
\cline{2-4}
 &\multirow{2}{*}{KK $Z$}  & \multirow{2}{*}{~$\ell\ell$ resonance~\cite{Aaboud:2017buh}~}
 & ~$m_{Z_\kk}\gtrsim 2$~~ for $g_{Z_\kk}\sim 5$\\
& & &~$m_{Z_\kk}\gtrsim 2.5$ for $g_{Z_\kk}\sim 3$\\
\cline{2-4}
 &KK $\gamma$ & $\ell\ell$ resonance \cite{Aaboud:2017buh} & ~$m_{\gamma_\kk}\gtrsim 2$~~~ for $g_{\gamma_\kk}\sim 4$\\
 \hline
 ~Hypercharge model~& KK $B$& $\ell\ell$ resonance \cite{Aaboud:2017buh}& ~$m_{\gamma_\kk}\gtrsim 2$~~~ for $g_{\gamma_\kk}\sim 3$\\
 \hline
\end{tabular}
\caption{\label{tab:Bounds} A list of current searches which constrain KK gauge boson masses in the two models. The third column shows the current search which provides the most stringent bound on the associated KK gauge boson mass. All mass quantities are in TeV.}
\end{table}

\begin{figure}
\centering
\includegraphics[width=0.5\linewidth]{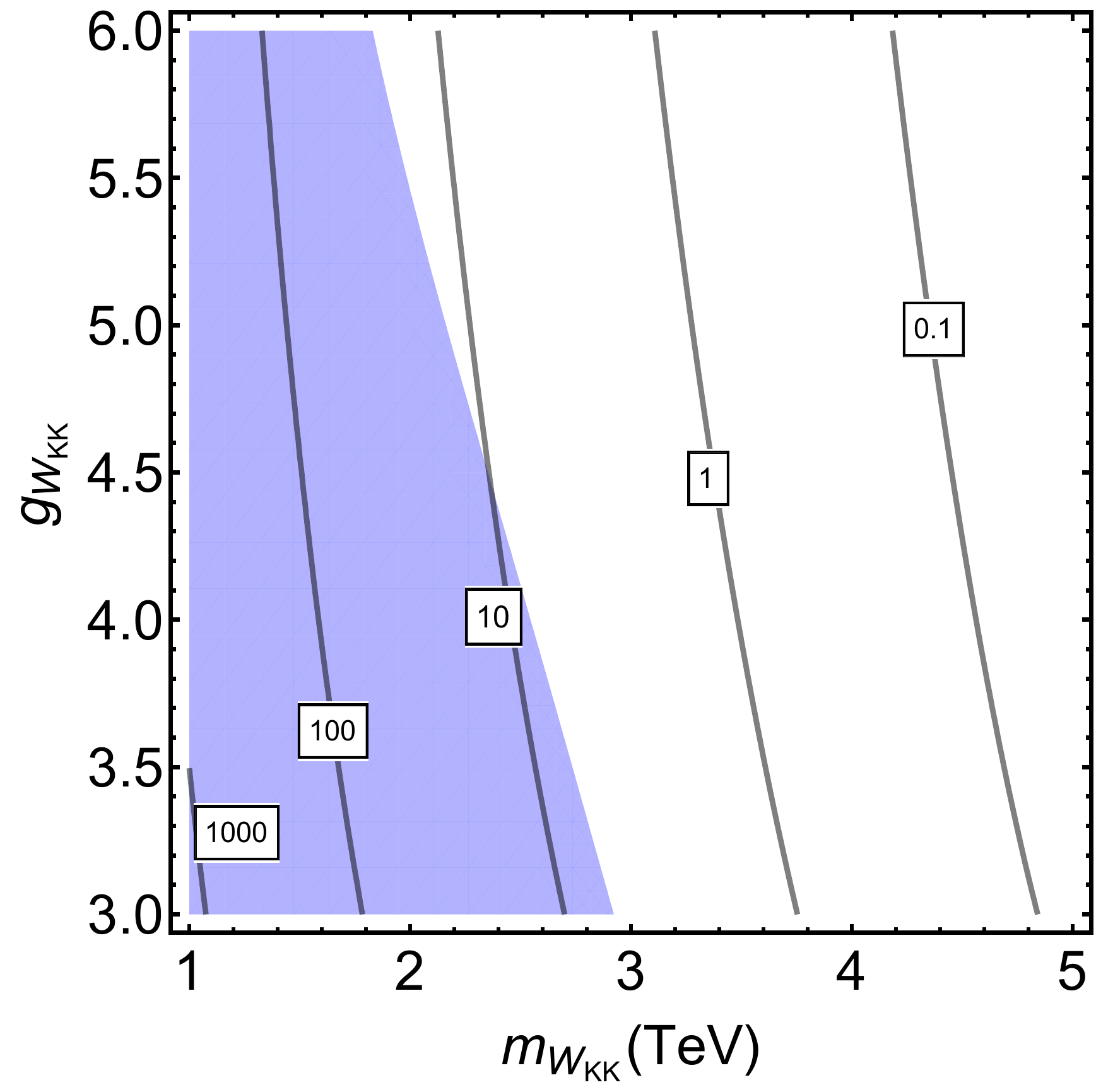}
\caption{Contours of the KK $W$ production at $\sqrt{s}=13$ TeV of the LHC as a function of $g_{W_\kk}$ and $m_{W_\kk}$. The numbers in boxes are the cross sections of producing KK $W$ in fb. The shaded region is ruled out by the search in ref.~\cite{Aaboud:2017efa} where $W'$ decays to a single lepton associated with a large $E_T^{{\rm miss}}$. See the text for more detailed information.}
\label{fig:WKK_prod}
\end{figure}

As mentioned earlier, we shall concentrate on the production of KK $W$ boson in our actual analysis since it often comes with the largest production cross-section.\footnote{Our analysis schemes and techniques are straightforwardly applied to the other KK boson searches arising in our model.} In this context, we exhibit the contours of KK $W$ production cross-section at the LHC-13 TeV as a function of $g_{W_\kk}$ and $m_{W_\kk}$, and shade parameter space disfavored by the current bounds, in figure~\ref{fig:WKK_prod}. 
Since the most stringent bound comes from the leptonic $W'$ search in conjunction with a leptonic transverse mass distribution~\cite{Aaboud:2017efa}, the leptonic decay mode of $W_{{\rm KK}}$, i.e., $W_{{\rm KK}} \to \ell \nu$, is most relevant. 
Of course, the other channels may yield the same final state (in an inclusive manner), but we expect that their contributions should be subdominant because corresponding transverse mass distribution either develops a singular peak at $m_W$ or populates more in the softer regime (i.e., no singular structure at all) so that most of events are rejected by 
the selection criteria in~\cite{Aaboud:2017efa}. 
To estimate the bound, we again assume that the radion channel of the KK $W$ decay takes over 50\%, resulting in BR$(W_{{\rm KK}} \to \ell \nu)\approx 4\%$ for each flavor of $\ell$.

\subsection{Model with Hypercharge Gauge Field in the Extended Bulk}\label{subsec:hypercharge}
In the hypercharge model, only $U(1)_Y$ gauge field propagates in the whole bulk up to the final IR brane, while the other gauge fields propagate only up to the intermediate Higgs brane. 
Therefore, the relevant particles in this model are the first KK hypercharge gauge boson, the radion and the SM particles. 
Just like the EW model discussed before, the relevant couplings related to KK hypercharge gauge boson and radion can be written as 
\bea\label{eq:LBKK}
\mathcal L^{\rm hypercharge}_{\rm warped} &\ni& \frac{g^2_B}{g_{B_{\rm KK}}} B_{\rm KK} ^\mu J_{B\mu} \nonumber \\
&+& \left( -\frac{1}{4}\frac{g_{\rm grav}}{g_{B_{\rm KK}}^2} g_{B}^2 B_{\mu\nu}B^{\mu\nu} 
+ \epsilon \frac{g_{\rm grav}}{g_{B_{\rm KK}}^2} g_{B_{\rm KK}} g_{B} B_{\mu\nu}B^{\mu\nu}_{\rm KK} \right)\frac{\varphi}{m_{{\rm KK}}}\,.
\eea
Again the operators in the second line are arranged in such a way as to make an easy comparison with eq.~\eqref{tensor}.
Here $B_{\kk}$ and $B$ denote first KK and SM hypercharge fields whose hypercharge couplings are represented by $g_{B_\kk}$ and $g_B$, respectively. 
As before, $J_{B\mu}$ represents current made of SM fields associated with hypercharge gauge boson.
Note that since $B$ is not the mass eigenstate after EW symmetry breaking, one should think of $B$ as a linear combination of $Z$ and $\gamma$. A similar gauge eigenstate mixing would go through for the $B_{{\rm KK}}$. However, considering the setup that in this hypercharge model, $SU(2)_L$ gauge fields propagate only up to the intermediate brane, we expect that the neutral components of $W_{{\rm KK}}$ are much heavier than $B_{{\rm KK}}$. Therefore, the size of mixing will be negligible.

The triboson signal processes in this model arise in a fashion analogous to those in the EW model. That is, a signal process is initiated by the production of $B_{{\rm KK}}$ via light quark annihilation [first term in eq.~(\ref{eq:LBKK})], followed by the $B_{{\rm KK}}$ decay to a photon and a radion [third term in eq.~(\ref{eq:LBKK})], the latter of which further decaying to $\gamma\gamma$, $Z\gamma$, or $ZZ$ [second term in eq.~(\ref{eq:LBKK})].
In this study, we focus on the triphoton mode:
\bea
B_\kk \to \gamma \varphi \to \gamma\gamma\gamma\,.
\eea
We next examine allowed values for the model parameters shown in eq.~\eqref{eq:LBKK}.

\subsubsection{Relevant Particles and Current Bounds}
\subsubsection*{(a) Radion}
The direct production of radion at the LHC is proceeded via either photon fusion or $Z$ boson fusion, which is encapsulated in the second operator in eq.~(\ref{eq:LBKK}). 
The same coupling is responsible for its (dominant) decays to  $\gamma\gamma$, $Z\gamma$, and $ZZ$. Using the standard parametrization of $B=\cos \theta_W \gamma-\sin\theta_W Z$, 
with $\theta_W$ being the usual Weinberg angle, we find the following decay widths of the radion:
\bea
&&\Gamma(\varphi \to \gamma \gamma )= \cos^4\theta_Wg^2_{\rm grav} \left( \frac{g_{B}}{ g_{B_{\rm KK}}} \right )^4\left(\frac{m_{\varphi}}{m_{\rm KK}}\right )^2 \frac{m_\varphi}{64 \pi}\,, \label{eq:Radion_Decay_0} \\
&&\Gamma(\varphi \to Z \gamma )\approx 2\sin^2\theta_W \cos^2\theta_Wg^2_{\rm grav} \left( \frac{g_{B}}{ g_{B_{\rm KK}}} \right )^4\left(\frac{m_{\varphi}}{m_{\rm KK}}\right )^2 \frac{m_\varphi}{64 \pi}\,, \\
&&\Gamma(\varphi \to Z Z)\approx \sin^4\theta_Wg^2_{\rm grav} \left( \frac{g_{B}}{ g_{B_{\rm KK}}} \right )^4\left(\frac{m_{\varphi}}{m_{\rm KK}}\right )^2 \frac{m_\varphi}{64 \pi}\,.
\label{eq:Radion_Decay}
\eea
It is clear from eqs.~\eqref{eq:Radion_Decay_0}-(\ref{eq:Radion_Decay}) that the BR to each pair of decay products is independent of detailed couplings since they appear as common factors. Setting $\sin^2\theta_W=0.23$, we obtain the BR's as
\bea
\textrm{BR}(\varphi \to \gamma \gamma)= 59\%,~~\textrm{BR}(\varphi \to Z \gamma)= 35\%,~~\textrm{BR}(\varphi \to Z Z)= 5.3\%.
\eea
There are two main channels for radion production in this model:
photon fusion and $B_\kk$ decay. The strongest constraint comes from diphoton searches. We have checked that radions produced through direct photon fusion with $m_\varphi\gtrsim 1$ TeV are not constrained for $g_{B_\kk}=3$. Further, their production from $B_\kk$ decay in the channel of $B_\kk \to \gamma \varphi \to \gamma\gamma\gamma$ is also safe from the diphoton bound, due to the large value of $m_{B_\kk}$ and the combinatorial ambiguity discussed in subsections for the EW model.

\subsubsection*{(b) KK $B$} 
The KK $B$ boson decay has radion channels (i.e., both $\varphi Z$ and $\varphi \gamma$) and diboson, dijet, ditop, and dilepton channels:
\bea
&&\Gamma (B_{\rm KK} \to \varphi\  \gamma)= \cos^2\theta_W\left( \epsilon  g_{\rm grav} \frac{g_{B}}{ g_{B_{\rm KK}}} \right )^2\left(1-\left(\frac{m_{\varphi}}{m_{\rm KK}}\right)^2\right)^3 \frac{m_{\rm KK}}{24 \pi}\,, \\
&&\Gamma (B_{\rm KK} \to \varphi\  Z)\approx \sin^2\theta_W\left( \epsilon  g_{\rm grav} \frac{g_{B}}{ g_{B_{\rm KK}}} \right )^2\left(1-\left(\frac{m_{\varphi}}{m_{\rm KK}}\right)^2\right)^3 \frac{m_{\rm KK}}{24 \pi}\,, \\
&&\Gamma (B_{\rm KK} \to WW)\approx\Gamma (B_{\rm KK} \to Zh) \approx \left(\frac{g^2_{B}}{ g_{B_{\rm KK}}} \right )^2\frac{m_{\rm KK}}{192 \pi}\,,\\
&&\Gamma  (B_{\rm KK} \to \psi \psi) \approx  N_c Q_B^2\left(\frac{g^2_{B}}{ g_{B_{\rm KK}}} \right )^2\frac{m_{\rm KK}}{24 \pi}\,,\label{eq: Bff}
\eea
where $Q_B$ in eq.~(\ref{eq: Bff}) denotes the SM hypercharge of the associated fermion $\psi$. Since KK $B$ is very similar to KK $Z$, the most stringent bound on its mass also comes from dilepton searches. The current bounds result in (see table~\ref{tab:Bounds}) 
\bea
m_{B_\kk}\gtrsim 2 \hbox{ TeV for } g_{B_\kk} \sim 3.
\eea

\subsection{Benchmark Points}\label{subsec:benchmark}
Table~\ref{tab:BPtable} shows a list of our benchmark points for various triboson channels in the EW and hypercharge models. Our benchmark points are chosen to be safe from all bounds discussed above, yet also have enough significance to be probed at the LHC. 
Since the production of KK gauge bosons is inversely proportional to the corresponding $g_{V_{\rm KK}}$, a smaller $g_{V_{\rm KK}}$ renders the bigger cross-section. However, as argued in section~\ref{subsec:EW}, all $g_{V_\kk}$ have a lower bound of roughly 3. Hence, we choose $g_{W_{\rm KK}}=3$ in all channels in the EW model, and $g_{B_\kk}=3$ in the hypercharge model. In order to enhance the branching ratio of KK gauge bosons to a radion, $g_{\rm grav}$ is set to its naturally allowed maximum value of 6, while the parameter $\epsilon$ is chosen to be 0.5. In the $W_\kk$ cascade decay to $WWW$, we choose $g_{B_\kk}=6$ to benefit by a boost of signal cross-section due to a larger branching ratio of radion to $WW$. Similarly, we choose $g_{B_\kk}= 3$ in the $W_\kk$ cascade decay to $W\gamma\gamma$ to obtain a larger BR($\varphi\to\gamma\gamma$).

\begin{table}[t]
\centering
\begin{tabular}{c|c c c c c c }
\hline
 Model & Process & Name & $m_{\rm KK}$ & $m_\varphi$ & $g_{B_{\rm KK}}$ & $g_{W_{\rm KK}}$  \\
\hline \hline
\multirow{4}{*}{ }
 & \multirow{2}{*}{} $W_{\rm KK}  \rightarrow W \varphi \to W W W$ 
   & $W$-$WWW$-BP1 & 3 & 1 &6 &3    \\
EW & (\ref{subsubsec:WWW}) & $W$-$WWW$-BP2 & 3 & 1.5 &6 &3\\ 
 \cline{2-7}
model & \multirow{2}{*}{} $W_{\rm KK} \rightarrow W \varphi \to W\gamma\gamma$ 
  & $W$-$W\gamma \gamma$-BP1 & 3 & 1 & 3 &3 \\
 & (\ref{subsubsec:Waa}) & $W$-$W\gamma \gamma$-BP2 & 3 & 1.5 &3&3  \\
\hline
  \multirow{2}{*}{}Hypercharge
 & \multirow{2}{*}{}$B_{\rm KK}  \rightarrow \gamma \varphi \to \gamma\gamma\gamma$ 
  & $B$-$\gamma\gamma\gamma$-BP1 & 3 & 1 & 3 &--   \\
model & (\ref{subsubsec:aaa}) & $B$-$\gamma\gamma\gamma$-BP2  & 3 & 1.5 &3 &--   \\
 \hline
\end{tabular}
\caption{\label{tab:BPtable} A list of benchmark points defined by their associated process and chosen parameter values in both models. All mass quantities are in TeV.
For all of them, $g_{\rm grav}$ and $\epsilon$ parameters are set to be 6 and 0.5, respectively. 
We assign the name of the channels in the following pattern: {\it the name of the KK gauge boson - final states - BP1 or BP2}. 
The numbers in the parentheses of the second column refer to the subsection elaborating the corresponding collider analysis. In the hypercharge model, there is only one KK gauge coupling $g_{B_\kk}$, and thus $g_{W_\kk}$ is left blank in the last row.  }
\end{table}

\section{Tools for Collider Study}
\label{tools}
We now briefly discuss the key techniques used in our collider study. Our Monte Carlo event simulation for both signal and background processes takes into account various realistic effects such as showering, hadronization, and detector response. 
Signal model UFO files are created using \textsc{FeynRules}~\cite{Alloul:2013bka}, and \textsc{MG5@aMC}~\cite{Alwall:2014hca} was used for parton-level event generation. 
Leading-order signal and background events are simulated under an environment of the LHC 14 TeV together with parton distribution functions parameterized by \textsc{NN23LO1}~\cite{Ball:2012cx}. 
Showering, hadronization and detector effects are incorporated using \textsc{Pythia6.4}~\cite{Sjostrand:2006za} and \textsc{Delphes3}~\cite{deFavereau:2013fsa}. 

Jets are reconstructed in \texttt{fastjet}~\cite{Cacciari:2011ma} using the anti-$k_t$ algorithm~\cite{Cacciari:2008gp}, with a jet radius $R = 1$ (we study the optimal choice in future sections). These jets are pruned using the Cambridge--Aachen algorithm \cite{Dokshitzer:1997in,Wobisch:1998wt} with $z_{cut} = 0.1$ and $R_{cut}$ factor $= 0.5$, and it is the $p_T$ and jet mass of the pruned jets that are used in further stages of the analysis. Fat jets are selected as $W$-candidates if they satisfy two tagging criteria. First, it is required that their pruned jet mass falls in the range $65 \leq m_J \text{ (GeV)} \leq 105$. The second tagging criterion is based on the $\tau_{21} = \tau_2 / \tau_1$ N-subjettiness ratio~\cite{Thaler:2010tr}, which is effective in separating two-pronged boosted $W$-jets from single-pronged QCD jets. The individual $\tau_N$ are defined by
\begin{align}
\tau_N = \left(\sum_k p_{T,k}\: R\:\right)^{-1}\sum_k\: p_{T,k}\: \min{\Big(\Delta R_{1,k} ,\Delta R_{2,k},\dots,\Delta R_{N,k} \Big)},
\end{align}
where the subscript $k$ runs over all jet constituents and the angles $\Delta R$ are measured with respect to $N$ subjet axes. These axes are selected using one-pass minimization from a starting seed determined by exclusive $k_t$ reclustering~\cite{Thaler:2011gf}. We require as a tagging criterion that a hadronic $W$-candidate passes a loose cut $\tau_{21} < 0.75$, where $\tau_{21}$ is measured on the ungroomed jet. We subsequently impose harder cuts in some of the analyses which will follow in later sections. These choices are motivated by the selections used in the CMS hadronic diboson search in~\cite{CMS:2015nmz}.

The tri-$W$ channel with one of the $W$ decaying leptonically and other two hadronically (henceforth referred as semileptonic channel or single lepton channel interchangeably) contains a single (invisible) neutrino in the final state, so its four momentum can be reconstructed event-by-event under the assumption that it comes from a $W$ decay. Two transverse components of the momentum are simply given by the missing transverse momentum, $\vec{P}_T^{\textrm{miss}}$:
\bea
\vec{p}_{T,\nu}=\vec{P}_T^{\textrm{miss}}=-\sum_i \vec{p}_{T,i}\,,
\eea
where $i$ runs over all visible particles in the final state. 
We then calculate the momentum component in the $z$ direction by requiring  an on-shell $W$ mass condition in combination with the lepton four momentum:
\bea
m_W^2 = (p_{\nu}+p_\ell)^2 = 2\left(E_\ell \sqrt{|\vec{P}_T^{\textrm{miss}}|^2+p_{z,\nu}^2}-\vec{p}_{T,\ell}\cdot \vec{P}_T^{\textrm{miss}} - p_{z,\ell}p_{z,\nu}\right)\,, 
\eea
where we assume that both neutrino and lepton are massless as usual. This quadratic equation leads to a two-fold ambiguity in the solution for $p_{z,\nu}$, and because of possible jet mismeasurements and other detector smearing effects this may sometimes be complex.
When the quadratic equation results in two real solutions we select the smaller one, and in the case of a complex solutions we choose the real part. This is the same strategy used, for example, in the ATLAS diboson resonance search in semi-leptonic final state~\cite{Aaboud:2017fgj}.

\section{Results for LHC Signals}
\label{sec:results}
In this section, we present the results for LHC reach of our signals, in various decay channels discussed in section \ref{warped}. In particular, we focus on the production and dominant decay channels of the lightest BSM particles -- the first KK partners of SM gauge bosons and the radion. We present the analysis for the representative benchmark points presented in table~\ref{tab:BPtable}. For each channel, we take two benchmark points, which correspond to two values of the radion mass: 1.0 TeV and 1.5 TeV. 
In section \ref{subsec:results_full_EW_model}, we consider EW model, where the full EW sector propagates in the entire bulk, while in section \ref{subsubsec:aaa}, we study hypercharge model, where only the gauge boson corresponding to hypercharge propagates in the full bulk. As we have seen in section \ref{warped}, both situations are viable from current experimental constraints. We will show in the following sections that both these scenarios can give rise to clear signals that can be observed at the LHC with a large significance.

\subsection{Full EW in Extended Bulk}
\label{subsec:results_full_EW_model}
The relevant couplings for the production and the decays of the KK EW gauge bosons are shown in eq.~(\ref{eq:LVKK}). In particular, KK EW gauge boson can be produced via its coupling to the SM fermion current [the first term in eq.~(\ref{eq:LVKK})]. For example, KK $W$ boson can be produced at the LHC by $q\bar{q}$ fusion, with the size of coupling being reduced compared to the SM gauge coupling, by a factor $g_V/g_{V_{\rm KK}}$. The coupling of the KK EW gauge boson to the SM currents allows it to decay into a pair of SM particles, resulting in difermion final state for fermion current and diboson final state for current made of gauge bosons and the Higgs. However, the main focus of our study is the cascade decay of KK EW gauge bosons. Couplings responsible for this are the second and the third terms in eq.~(\ref{eq:LVKK}). The third term enables the decay of KK gauge boson into the corresponding SM gauge boson and a scalar, radion, and the second term generates the subsequent decay of the scalar into a pair of SM gauge bosons. As can be clearly seen from eq.~(\ref{eq:LVKK}), the size of the decay rate of KK gauge boson into radion + SM gauge boson relative to a pair of SM states is governed by $\epsilon g_{\rm grav}$ and $m_{\varphi}$. In our study, we focus on part of parameter space where the above described cascade decay acquires significant rates or even become the dominant decay mode. As we will see, such scenarios require dedicated strategies in order to reveal the nature of the signals. 

For concreteness, we consider the production and the cascade decay of KK $W$ boson. Regarding the radion decay, we consider the possibility of its decay into either a pair of $W$'s or diphoton. Therefore, the primary signal has three SM gauge bosons in the final state -- either three $W$ bosons or a $W$ boson and two photons. 

\subsubsection{Tri-$W$ Signal}\label{subsubsec:WWW}
The tri-$W$ process leads to many possible final states, the largest of which are either fully hadronic ($6 q$) or have a single lepton ($4 q + \ell \nu$), with branching fractions 31\% and 30\% respectively.\footnote{Here $\ell$ refers to $e, \mu$; we neglect the final states involving $\tau$ leptons which are rare and experimentally challenging to detect.} These final states allow full reconstruction of the signal event (as described in section~\ref{tools} in the case where there is a single neutrino), hence have the potential to reconstruct both resonances involved in the cascade decay. In this section we discuss dedicated search strategies that would allow for simultaneous discoveries of the two resonances in these final states.

Of the decay modes accompanying two or more charged leptons, the most dangerous are those involving same-sign charged leptons, i.e. $q\bar{q}'\ell^\pm \ell^\pm \nu \nu$  whose branching fraction is 3.1\%. Although this final state does not allow to reconstruct any resonances due to the presence of two neutrinos, it is a clean and distinctive final state to which existing LHC searches for same-sign dileptons might be sensitive. We analyse the constraints from these searches in appendix~\ref{SSDL}, and find that they are less sensitive than the searches that we will describe for the dominant fully hadronic and single-lepton decays.

\subsubsection*{Bump-hunting for two resonances}
LHC searches for diboson resonances typically involve a bump-hunt on a diboson invariant mass distribution, $M_{VV}$, which would exhibit a peak at the mass of a diboson resonance signal. A simple extension of this search strategy for a triboson resonance is to execute a bump-hunt on a triboson invariant mass distribution, $M_{VVV}$, which would exhibit a resonance peak at $m_{W_\mathrm{KK}}$. A bump hunting shape-analysis can be roughly approximated with a cut-and-count analysis in a $M_{VVV}$ window around $m_{W_\mathrm{KK}}$, and we perform a study of such a strategy later in this section (calling it {\textbf{1D analysis}}). However, such a search would not make use of the information contained in the second radion resonance -- that there should also be, among the three possible diboson pairings, an $M_{VV}$ close to $m_\varphi$. The combinatorial ambiguity among the three $W$'s can be resolved by ordering them by $p_T$, and selecting the $p_T$-ordered pairing $M_{V_i V_j}$ which in simulation most frequently reconstructs the radion mass.
In our parton level simulation, we find that for $m_\varphi = 1 \; \text{TeV}$ this is $M_{V_2 V_3}$ which selects the correct pair in 56\% of events, and for $m_\varphi = 1.5 \; \text{TeV}$ it is $M_{V_1 V_3}$ which makes the correct selection in 53\% of events. A two-dimensional bump hunt can then be performed in the selected $M_{VVV}$-$M_{V_i V_j}$ plane, which will exhibit a resonance peak at $m_{W_\mathrm{KK}}$-$m_\varphi$. We will perform an approximate version of such a search using a cut-and-count analysis with a square window cut around $m_{W_\mathrm{KK}}$-$m_\varphi$, which we call {\textbf{2D analysis}}.

The advantage of the 2D analysis over the 1D analysis is the observation of the radion resonance and the rejection of a great deal of SM background. However, due to the combinatorial ambiguity, no single choice of $M_{V_i V_j}$ will correctly identify the two decay products of the radion in all events, and therefore the 2D analysis effectively throws away almost half of the signal events in which the ``wrong'' diboson selection was made. We therefore consider a third analysis (henceforth called {\textbf{3D analysis}}) in which we take, in addition to the $M_{V_i V_j}$ pairing which most frequently selects the radion decay products,  the $M_{V_i V_k}$ pairing which is second most frequent. Most events will have either $M_{V_i V_j}$ or $M_{V_i V_k}$ very close to $m_\varphi$, and therefore the signal will form a ``+" shape in this plane (see also left panels of figures~\ref{fig:WKK_WWW_Full_Had_2D_SG1} and \ref{fig:WKK_WWW_Full_Had_2D_SG2}). In parton level simulations, we find that the optimal pair for both benchmarks is $M_{V_1 V_3}$-$M_{V_2 V_3}$, which correctly selects the radion decay products in 94\% of events for the $m_\varphi = 1 \; \text{TeV}$ benchmark and 83\% of events for $m_\varphi = 1.5 \; \text{TeV}$. This approach may therefore recover the majority of signal events that would be rejected due to kinematics by the 2D analysis. It could be possible to perform a template shape analysis in this three-dimensional space $M_{VVV}$-$M_{V_1 V_3}$-$M_{V_2 V_3}$, which we approximate with a binned cut-and-count analysis in this ``+" shape.

\begin{figure}
\center
\includegraphics[width=7.5cm]{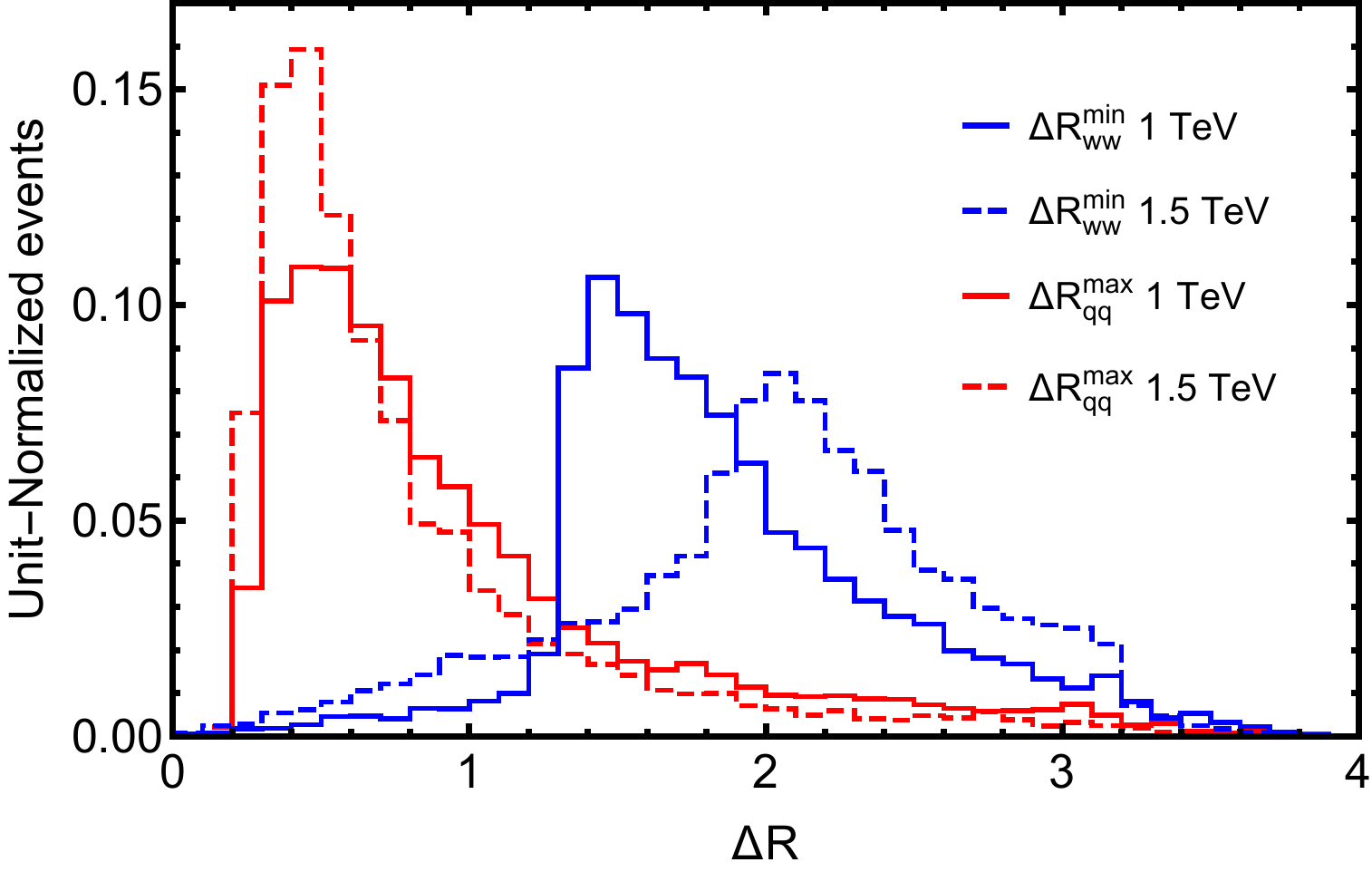}
\includegraphics[width=7.3cm]{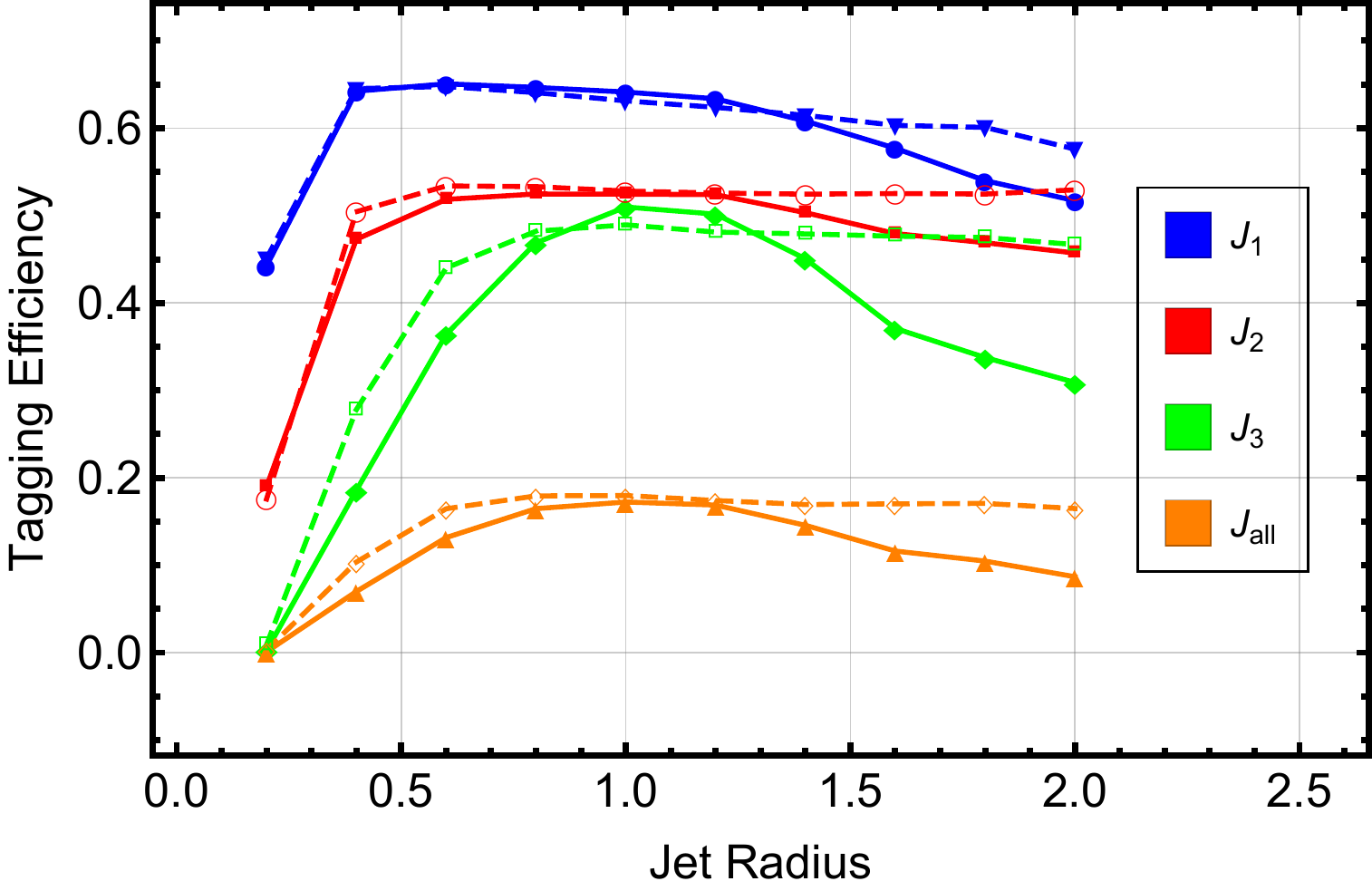}
\caption{The plot on the left shows the maximum of the $\Delta R_{qq}$ from each $W$ and minimum of $\Delta R_{WW}$. The plot on the right shows tagging efficiency as a function of jet radius. In both plots, the solid (dashed) lines show the result for the radion mass of 1 TeV (1.5 TeV). }
\label{fig:tag_eff_with_jet_radius_Deta_R}
\end{figure}

\subsubsection*{Fully hadronic channel}
The process of interest is
\bea
p p \to W_{\rm KK} \to W \varphi \to W W W,
\label{eq:process_WWW}
\eea
with all $W$'s decaying hadronically. 
As discussed in section \ref{sec:introduction}, our choice of $m_{W_{\rm KK}}$ and $m_\varphi$ leads to all three $W$'s in the final state typically being highly boosted and well separated, resulting in three fat jets $J$ each of which contains a pair of quarks from a $W$-decay. The typical separation between two quarks coming from a boosted $W$ decay is given by
\begin{equation}\label{equ:deltaR}
\Delta R \simeq \frac{2 m_W}{p_{T, W}}.
\end{equation}
Therefore, the softest $W$ will typically result in the $q\bar{q}$-pair with the greatest separation, and the radius $R$ chosen for fat-jet reconstruction should be sufficiently large to capture both of these quarks in a single jet. However, if this radius is too large, then multiple $W$'s will be clustered into the same fat jet, leading to contamination in event reconstruction. This effect is illustrated in figure~\ref{fig:tag_eff_with_jet_radius_Deta_R}, for both 1 and 1.5 TeV radion. In the left panel we plot the largest angle between a $q\bar{q}$ coming from a single $W$, and the smallest angle between two $W$'s (at parton level). There is relatively good separation between these distributions and we find that $R = 1$ results in a good compromise between these competing considerations. In the right panel we illustrate this also at the detector level, plotting the efficiency in signal events for tagging the $p_T$-ordered fat jet $J_i$ as a $W$-jet with the cuts on jet mass and $\tau_{21}$ described in section~\ref{tools}, as well as the overall efficiency for tagging all three jets. The latter efficiency peaks at 18\% at $R = 1$.

Because the characteristic $W$-jet radius has an inverse dependence on $p_T$, the three $W$'s in an event will typically exhibit some hierarchy in radii. This motivates a consideration of using a variable $p_T$-dependent jet radius as suggested in ref.~\cite{Krohn:2009zg}. We study this variable-$R$ approach in appendix~\ref{App:JetTaggingMethods} and compare the performance to that with constant $R$. We find only a small improvement for our benchmarks, and therefore consider only the simpler fixed-$R$ approach in the main body of the paper.

The primary background to the $JJJ$ signal is SM multijets in which at least three jets pass the $W$-jet tagging criteria. In order to produce sufficient Monte-Carlo statistics for the background in the signal region, we simulated SM trijet production with the following hard parton-level cuts which are weaker than the final cuts which will ultimately be imposed at the detector level.
\bea
&1.& \; p_{T j_1} > 500 \;{\rm GeV}, \;\; p_{T j_2} > 400 \;{\rm GeV}, \;\; p_{T j_3} > 200 \;{\rm GeV}, \; \; \left| \eta_j \right| < 5, \nonumber \\
&2.& \; M_{j_1 j_2} > 700 \;{\rm GeV}, \;\; M_{j_1 j_3} > 700 \;{\rm GeV}, \;\; M_{j_2 j_3} > 700 \;{\rm GeV}, \label{eq:parton_level_cuts_WWW_full_had} \nonumber \\
&3.& \; M_{j_1 j_2 j_3} > 2500 \;{\rm GeV}.
\eea  
At the detector level, we define a set of ``pre-selection'' cuts applied to both signal and background, requiring at least three reconstructed fat jets $N_J > 3$ which pass the same cuts as in eq.~(\ref{eq:parton_level_cuts_WWW_full_had}), in addition to harder $|\eta_j| < 2.4$ cut.

\begin{figure}
    \centering
    \includegraphics[width = 7.5 cm]{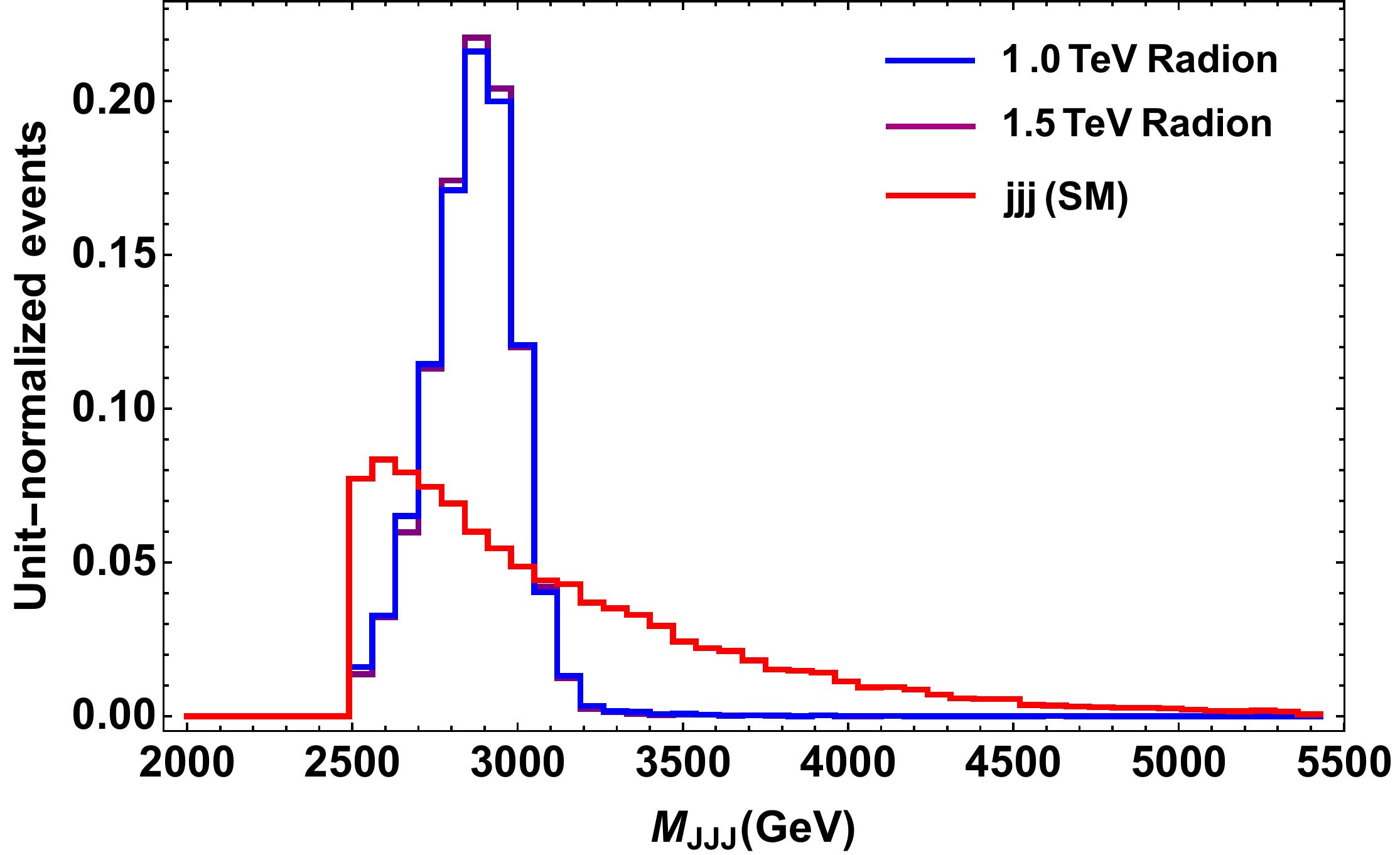}
    \includegraphics[width = 7.5 cm]{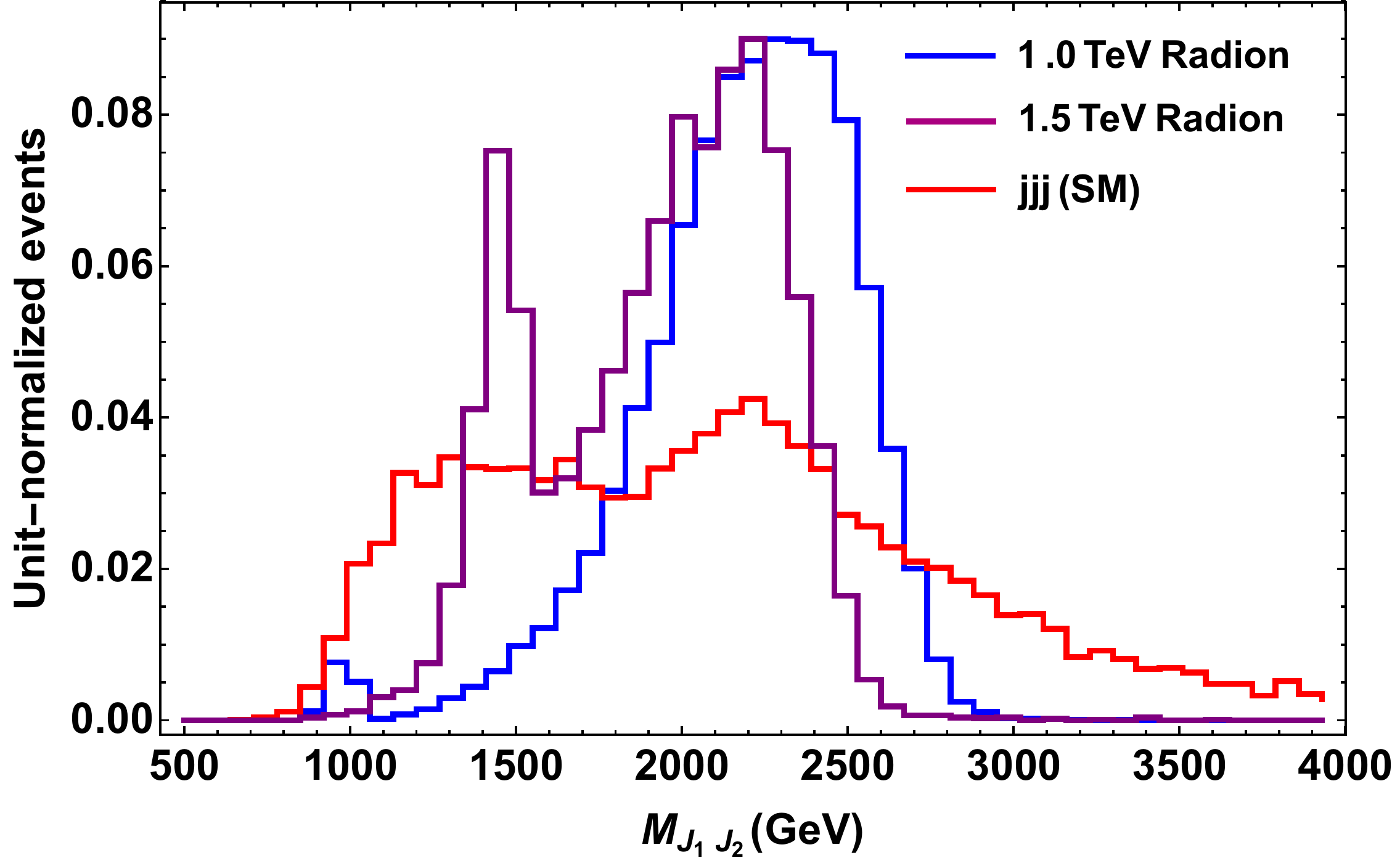}
    \includegraphics[width = 7.5 cm]{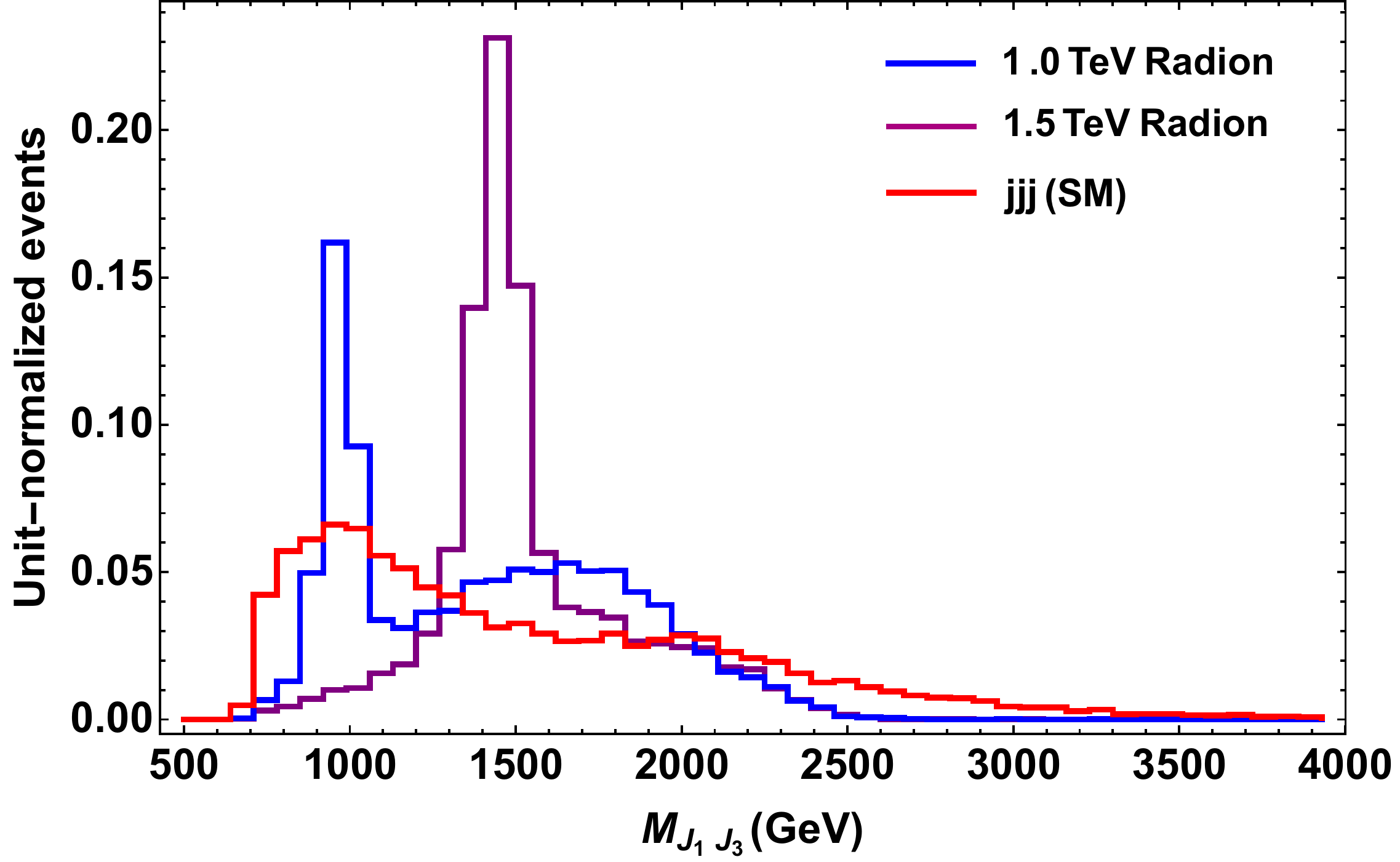}
    \includegraphics[width = 7.5 cm]{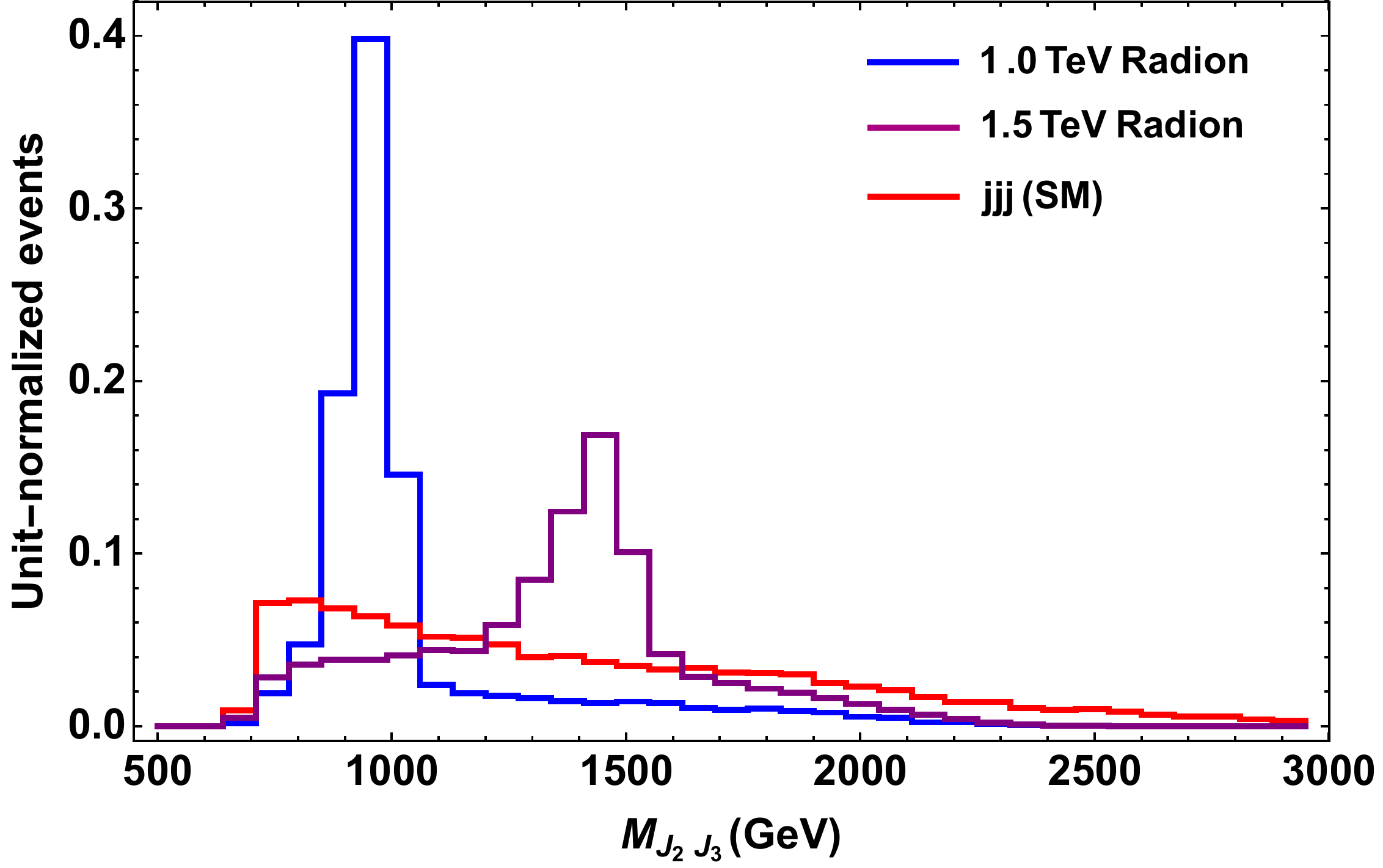}
    \includegraphics[width = 4.8 cm]{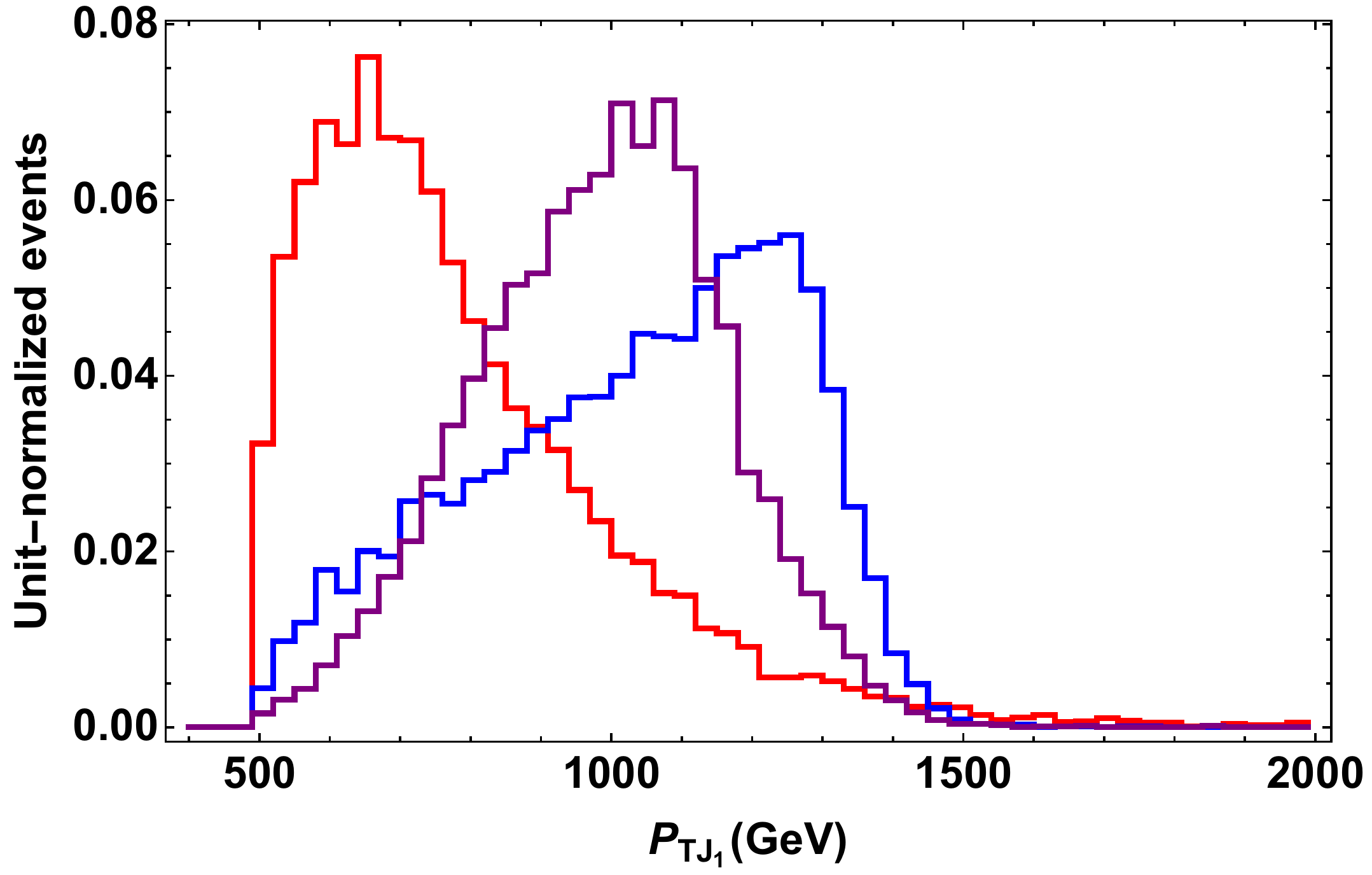}
    \includegraphics[width = 4.8 cm]{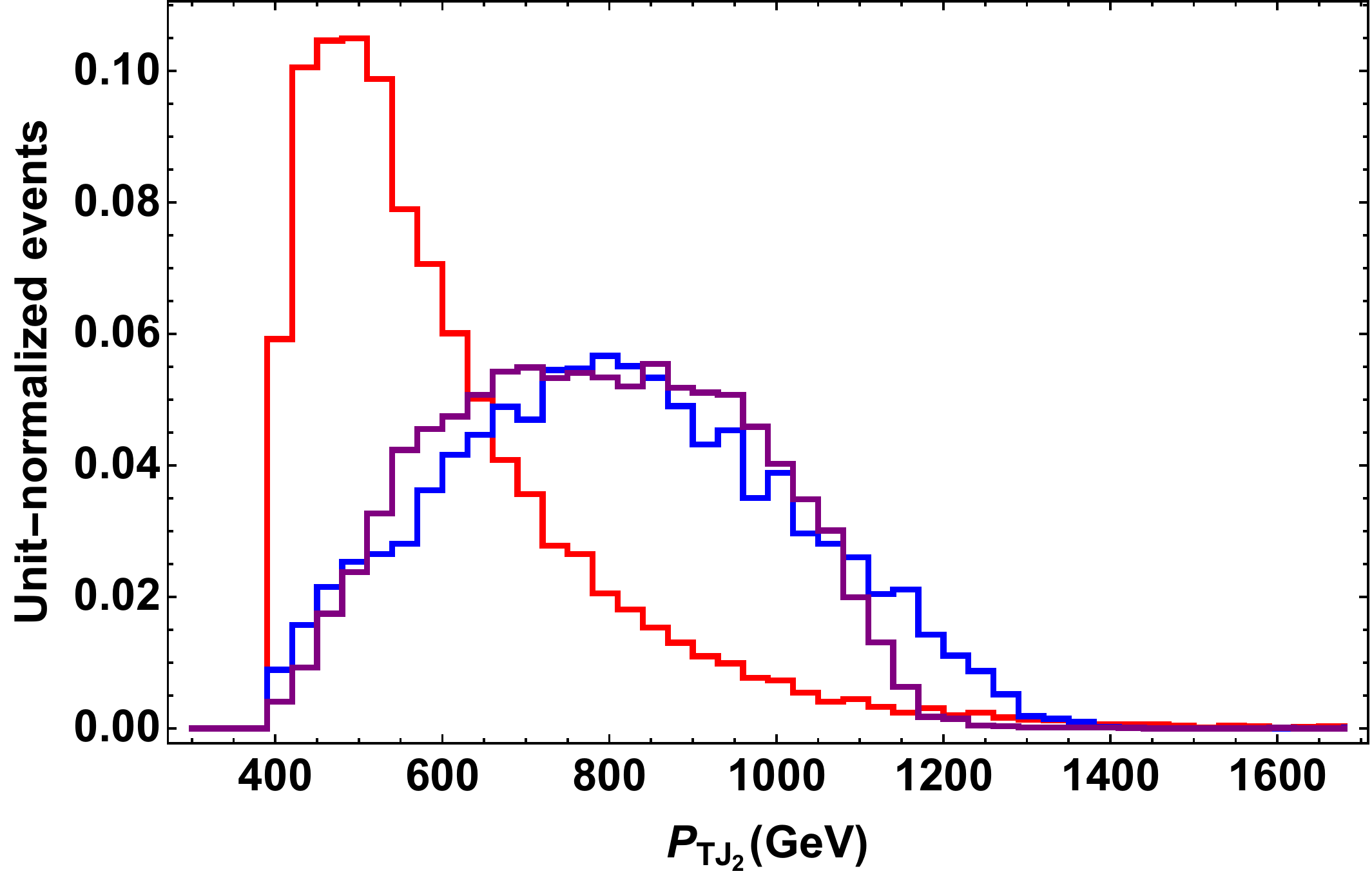}
    \includegraphics[width = 4.8 cm]{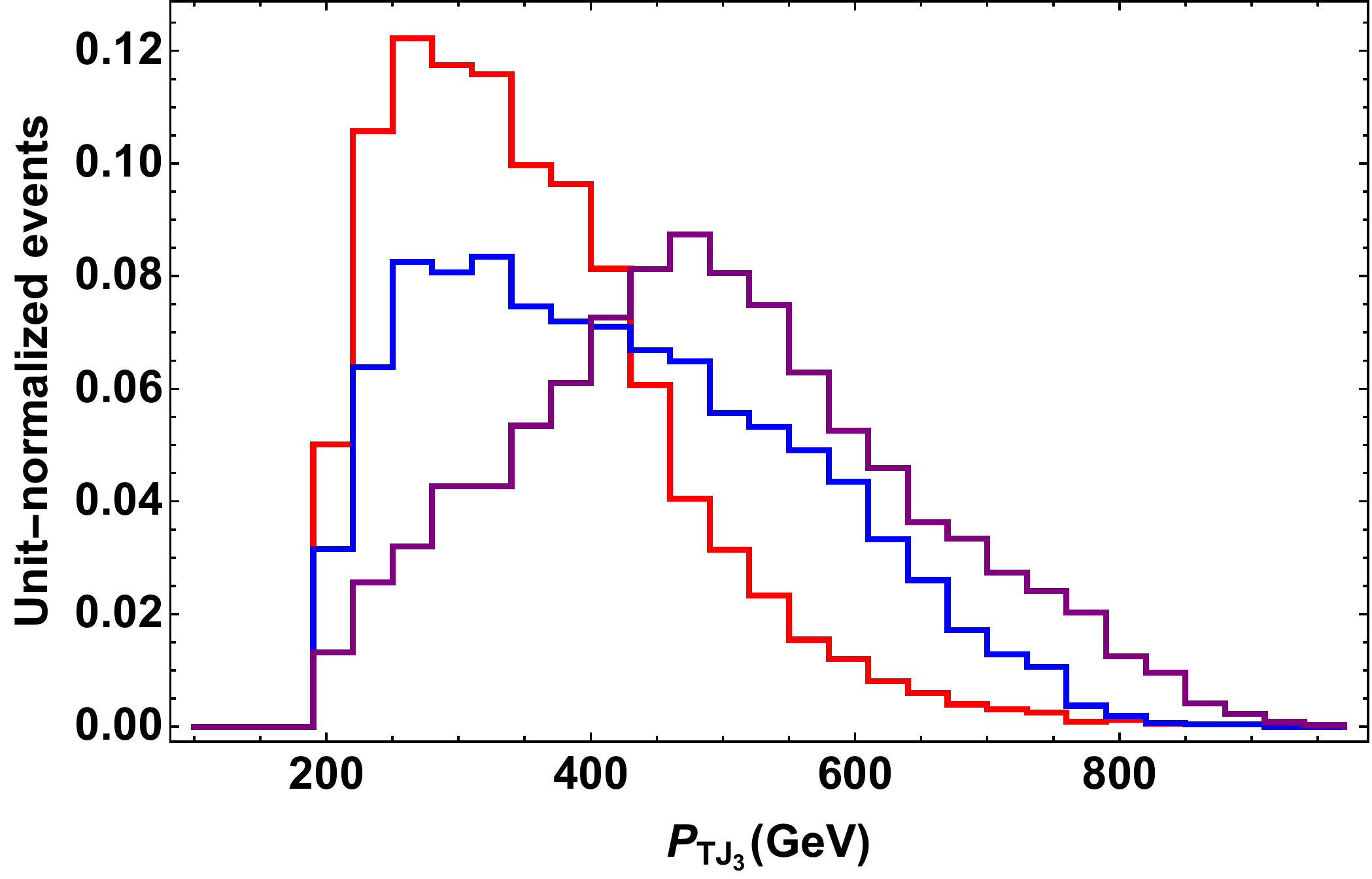}
    \includegraphics[width = 4.8 cm]{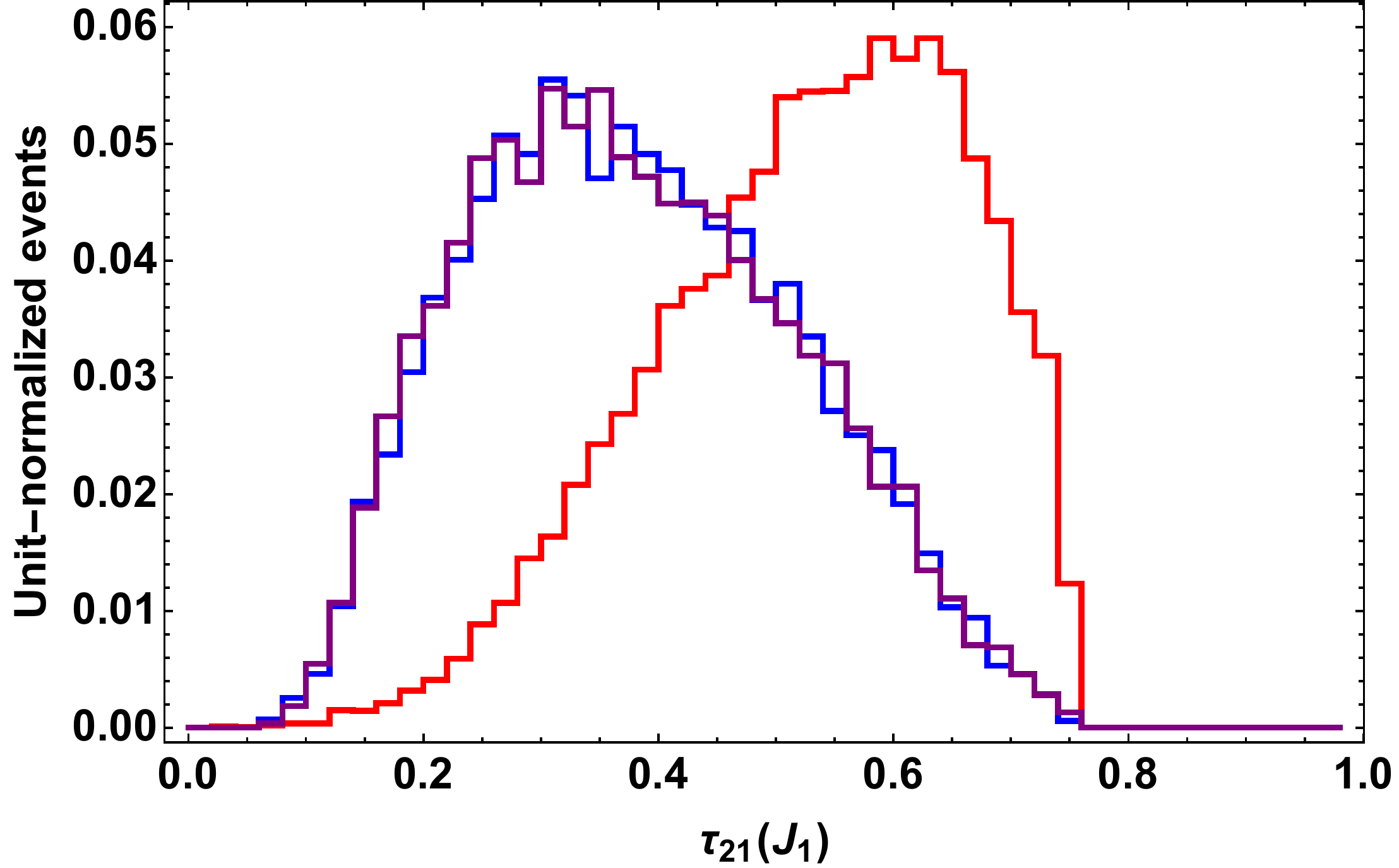}
    \includegraphics[width = 4.8 cm]{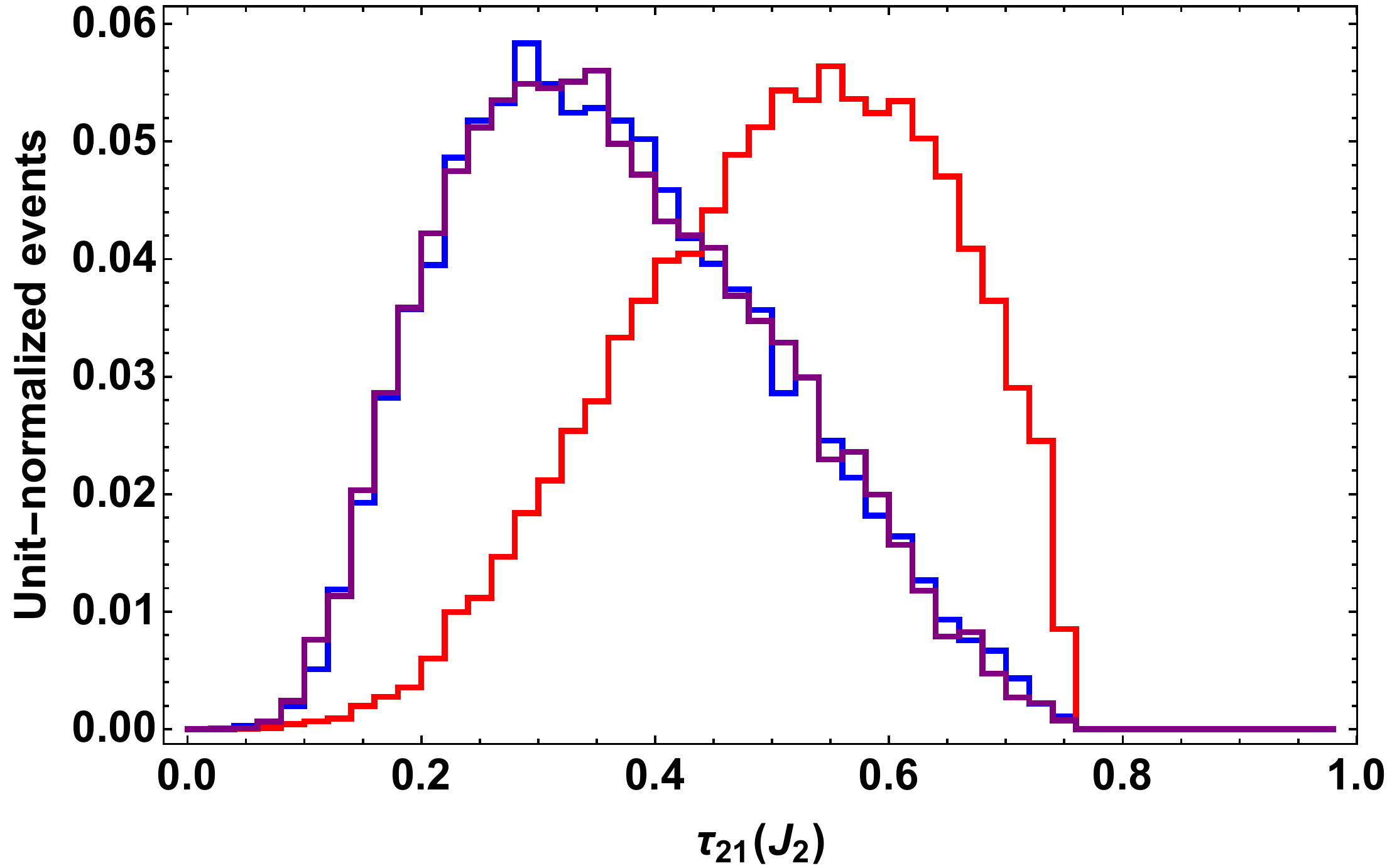}
    \includegraphics[width = 4.8 cm]{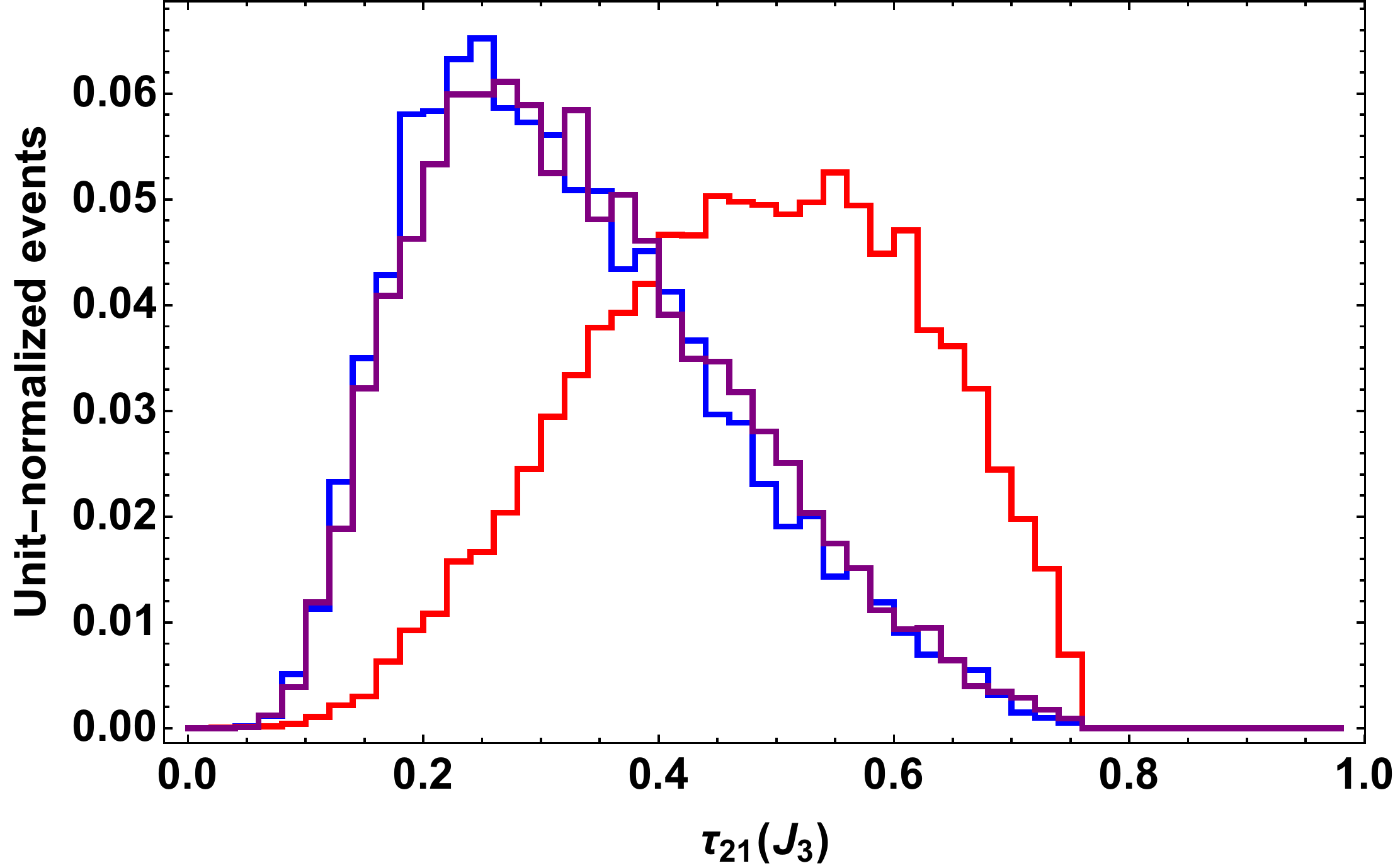}

    \caption{Distribution of kinematic variables for $W$-$WWW$-BP1 ($W$-$WWW$-BP2) fully hadronic channel: $M_{JJJ}$ (top row, left), $M_{J_1 J_2}$ (top row, right),
$M_{J_1 J_3}$ (second row, left), $M_{J_2 J_3}$ (second row, right), $p_{T J_1}$ (third row, left), $p_{T J_2}$ (third row, middle), $p_{T J_3}$ (third row, right), $\tau_{21 J_1}$ (bottom row, left), $\tau_{21 J_2}$ (bottom row, middle), $\tau_{21 J_3}$ (bottom row, right), for signal with 1 TeV radion (solid blue), signal with 1.5 TeV radion (solid purple) and backgrounds (solid red). We denote $p_T$-ordered jet as $J_{1,2,3}$, $J_1$ being the hardest jet. 
}
\label{fig:WKK_WWW_Full_Had_1D}
\end{figure}

Figure~\ref{fig:WKK_WWW_Full_Had_1D} shows distributions of various kinematic variables after pre-selection.
We see the sharp resonant peak at $m_{W_{\rm{KK}}}$ in  $M_{JJJ}$. The radion mass peak at $m_\varphi$ is divided mainly between $M_{J_2 J_3}$ and $M_{J_1 J_3}$, but there is also a small peak in  $M_{J_1 J_2}$ for the $1.5 \; \text{TeV}$ radion. In addition to mass window cuts in these distributions, further cuts on $p_T$ and $\tau_{21}$ will be useful for background rejection.

\begin{figure}
    \centering
    \includegraphics[width = 7 cm]{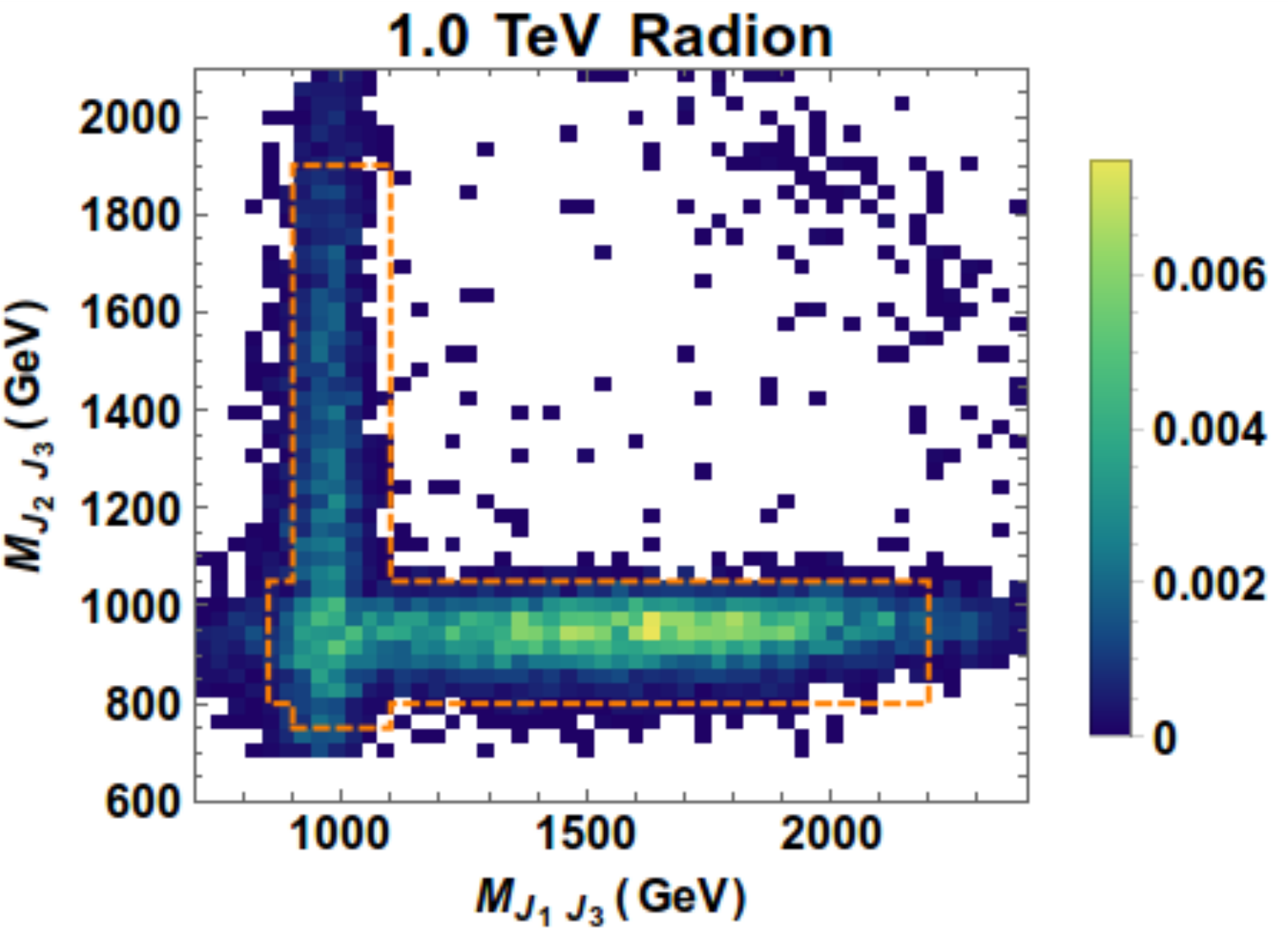}
    \includegraphics[width = 7.1 cm]{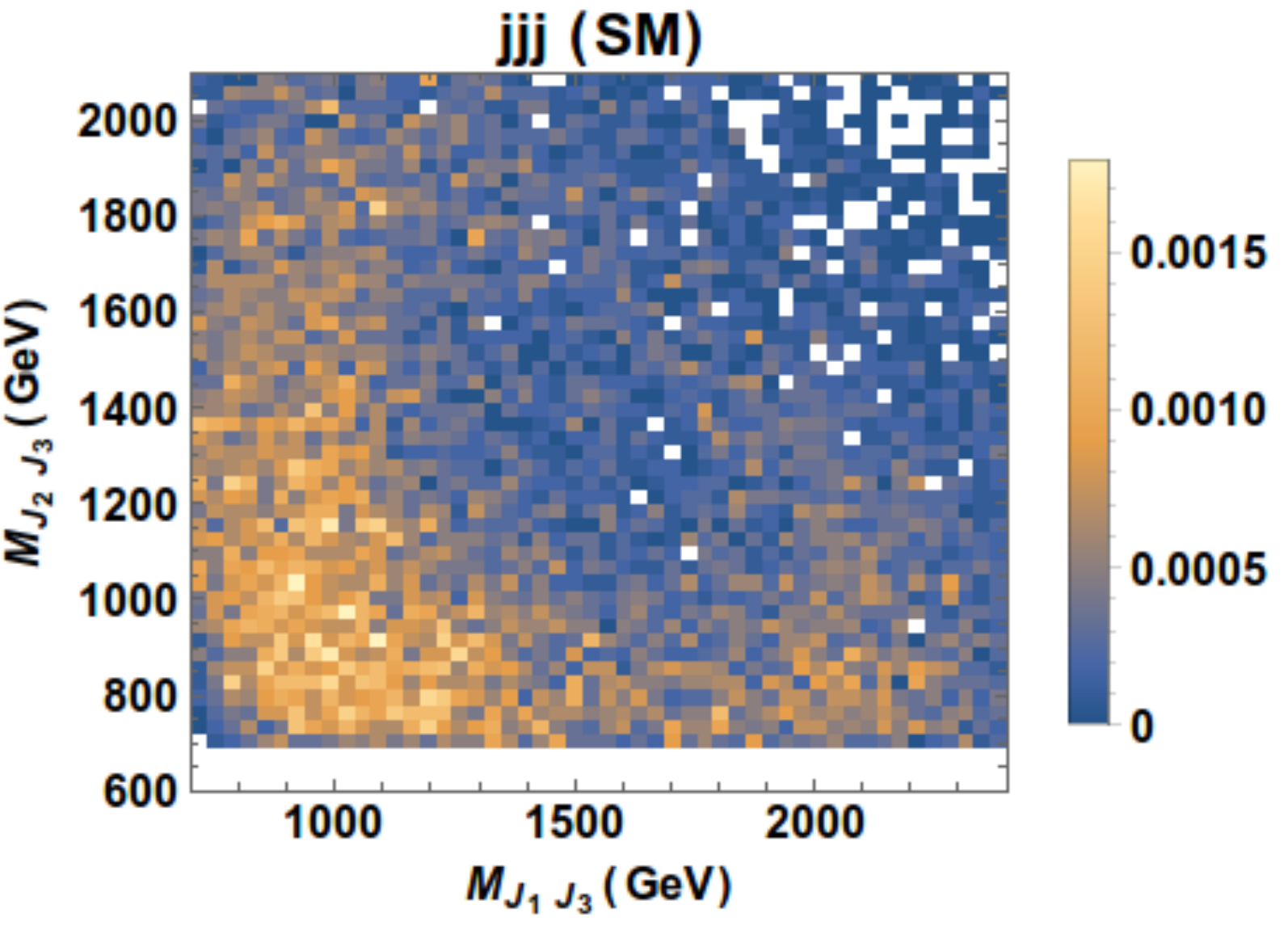}
    \includegraphics[width = 7 cm]{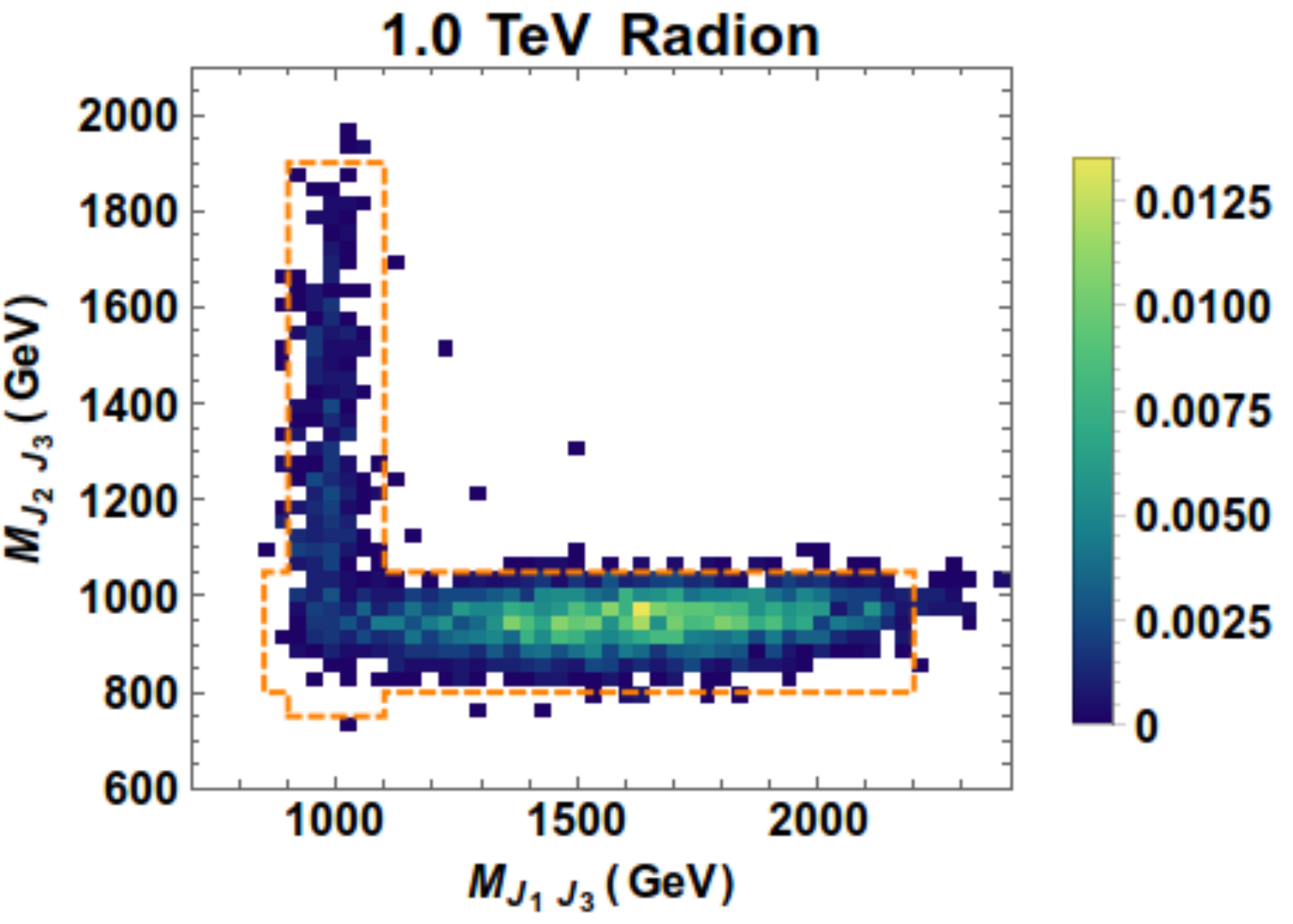}
    \includegraphics[width = 7 cm]{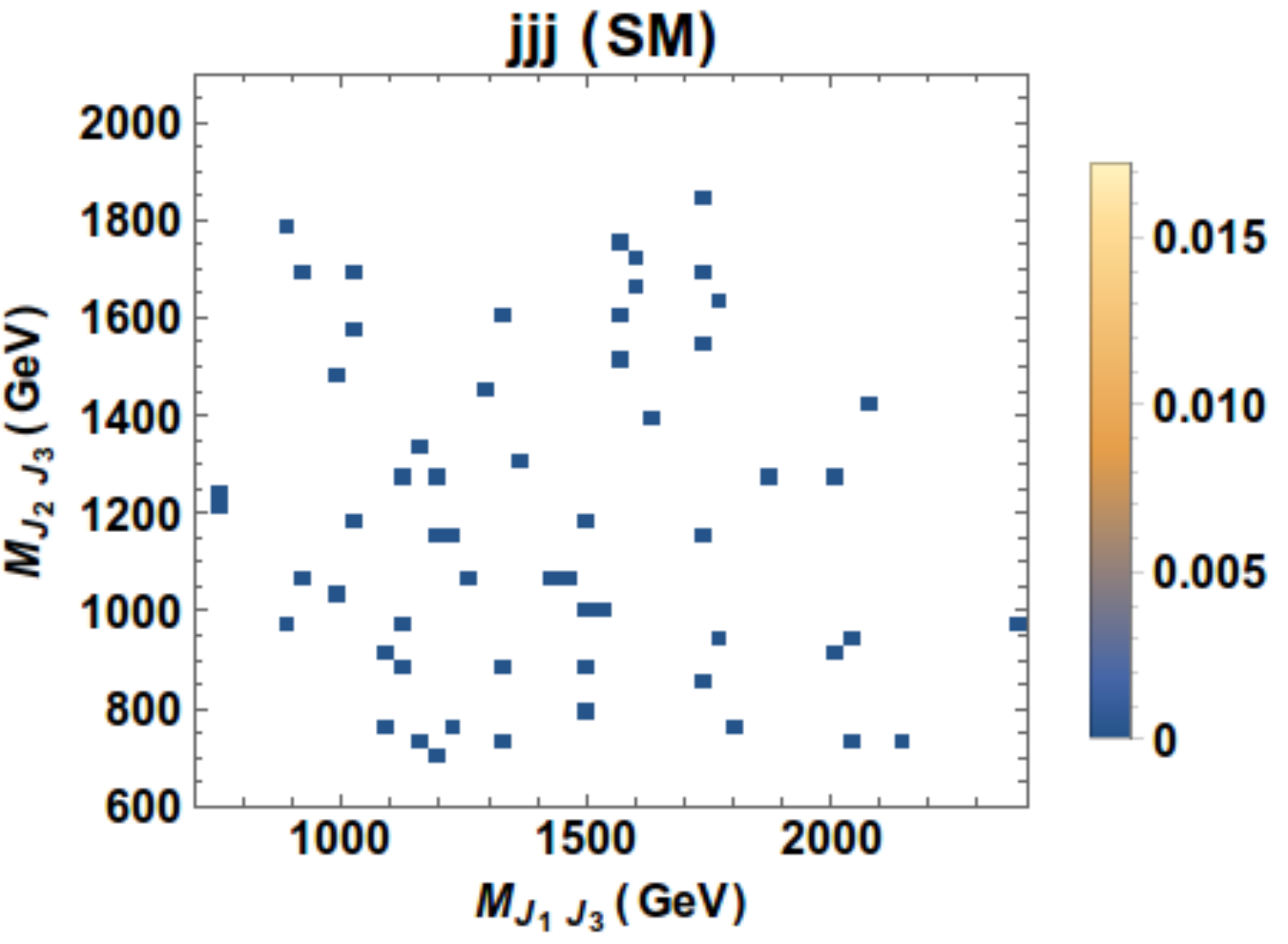}
    \caption{2D $M_{J_1 J_3}$-$M_{J_2 J_3}$ distributions for $W$-$WWW$-BP1 fully hadronic channel: SG after pre-selection but before any other cuts (top row, left), BK after pre-selection but before any other cuts (top row, right), SG after all other cuts but 2D $M_{J_1 J_3}$-$M_{J_2 J_3}$ cut (bottom row, left), BK after all other cuts but 2D $M_{J_1 J_3}$-$M_{J_2 J_3}$ cut (second row, right). The region surrounded by orange dashed line shows the signal selected by the ``+'' cut used in our 3D analysis. We refer to the main text for more detailed information.
}
\label{fig:WKK_WWW_Full_Had_2D_SG1}
\end{figure}
\begin{figure}
    \centering
    \includegraphics[width = 7 cm]{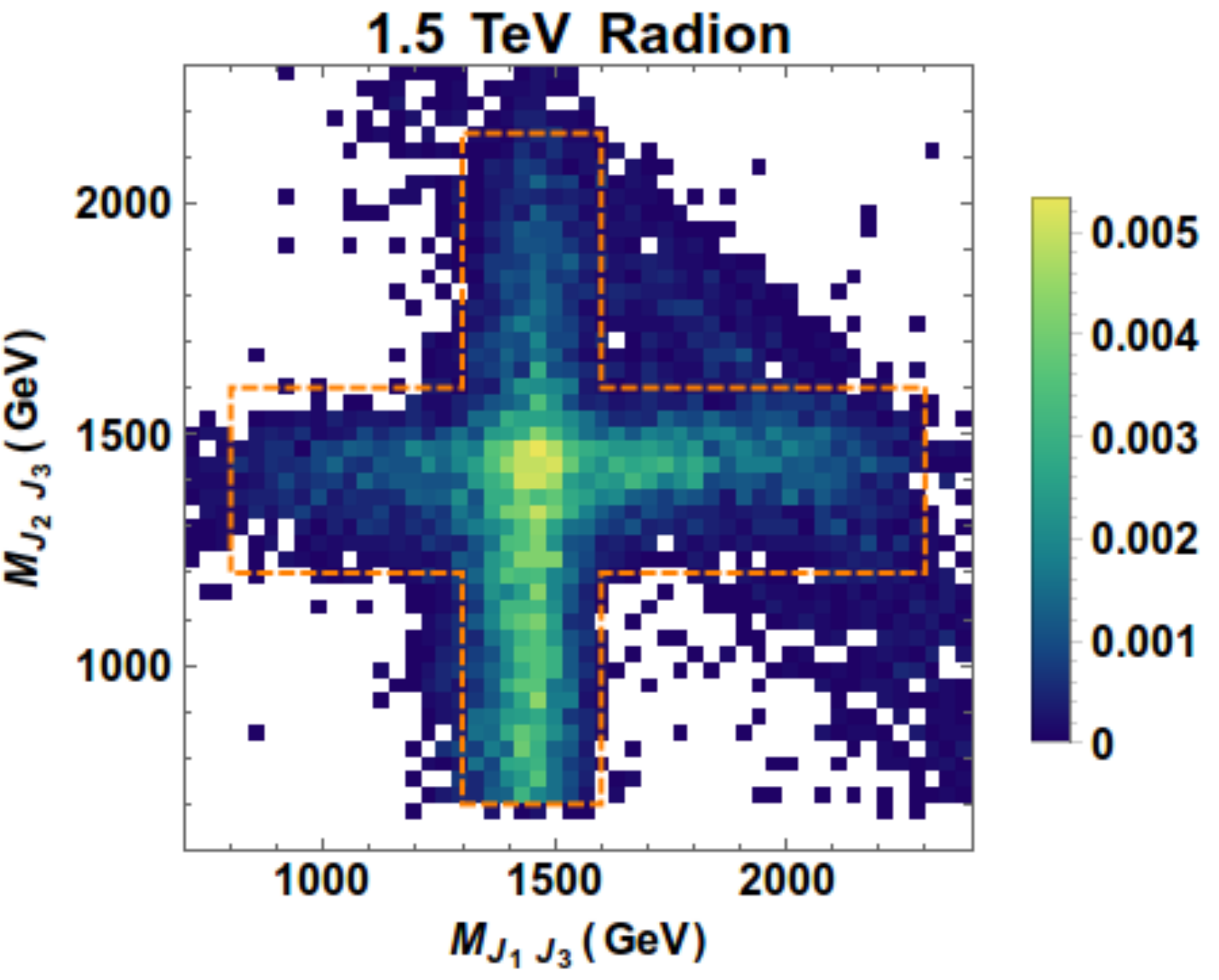}
    \includegraphics[width = 7.2 cm]{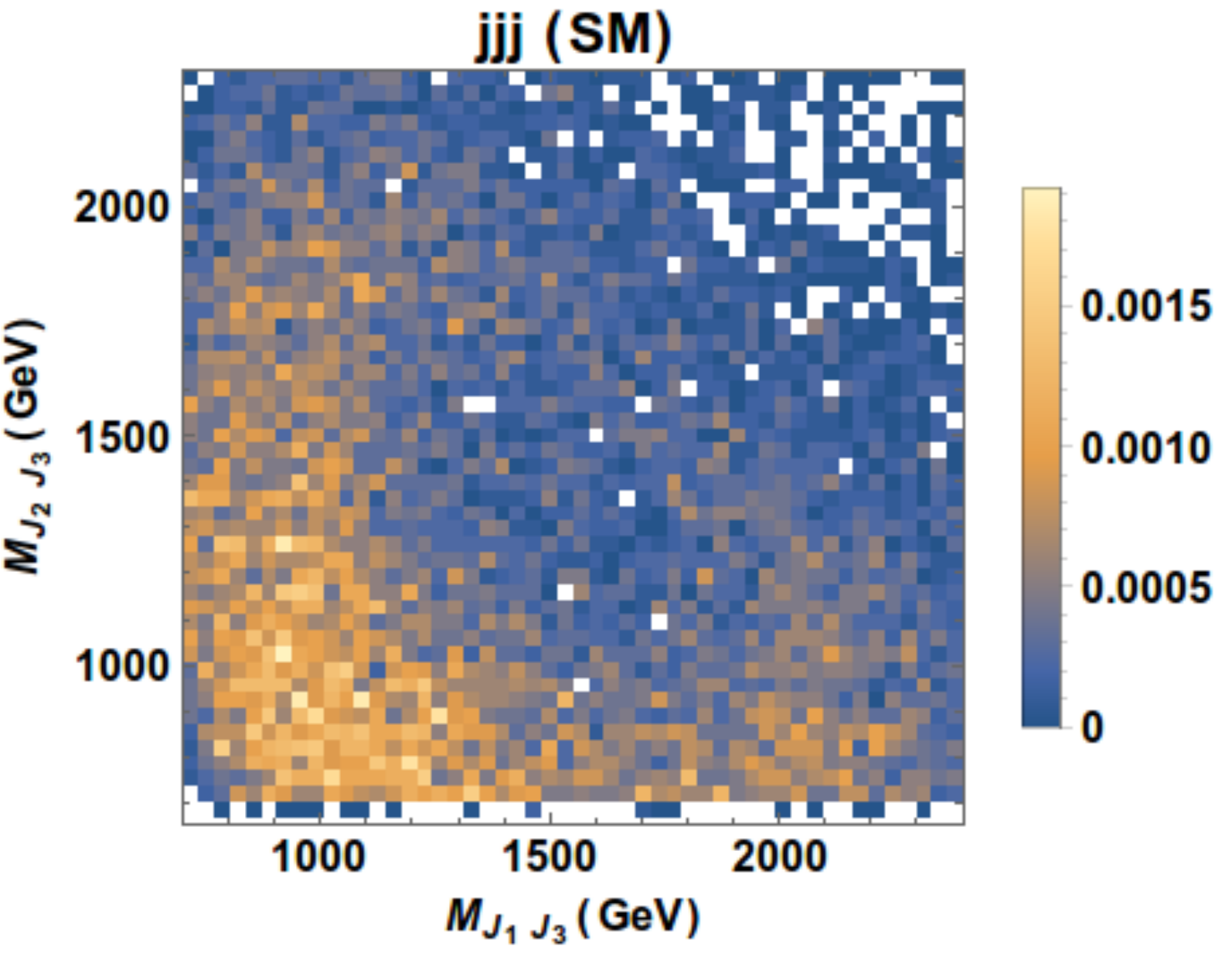}
    \includegraphics[width = 7 cm]{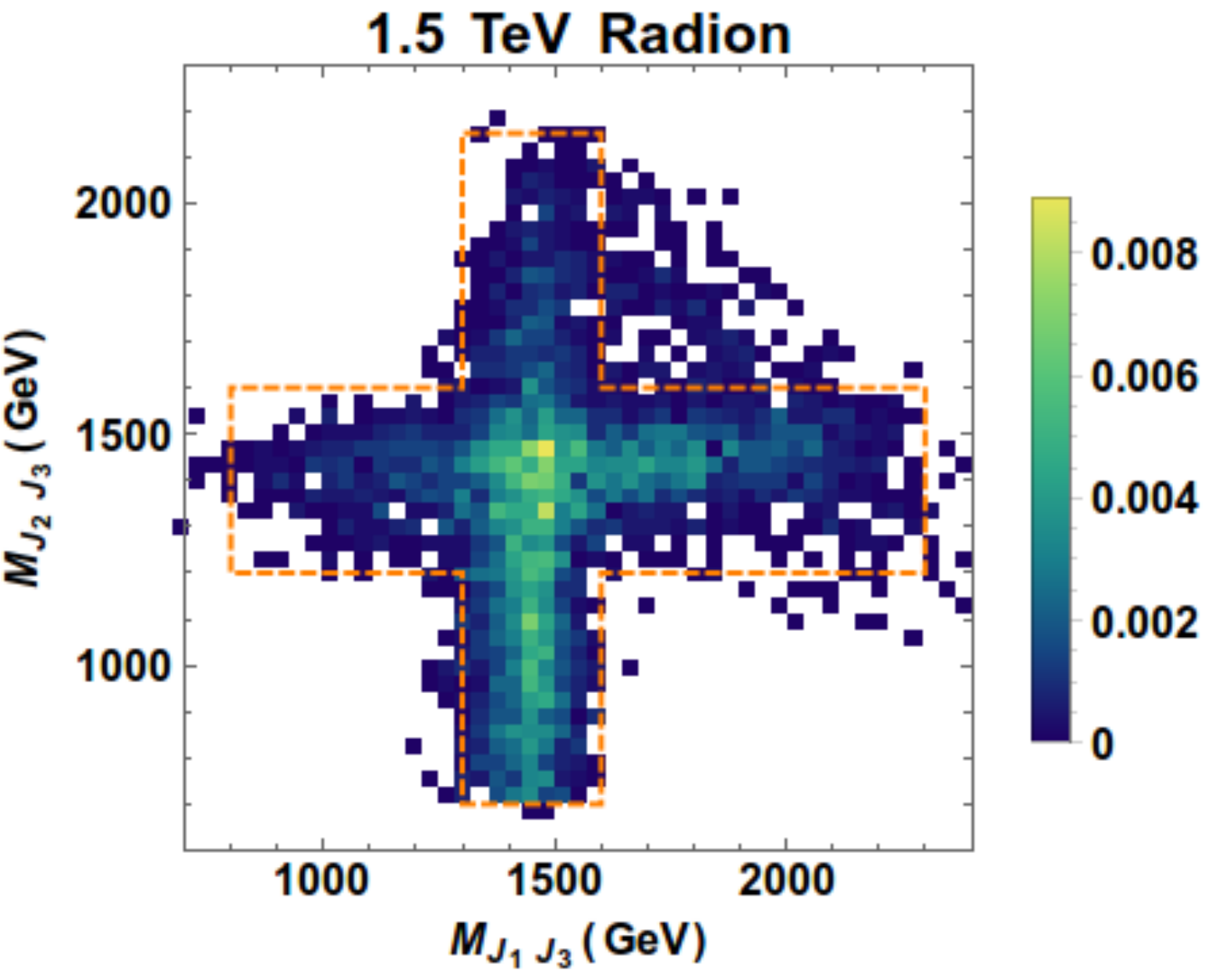}
    \includegraphics[width = 7 cm]{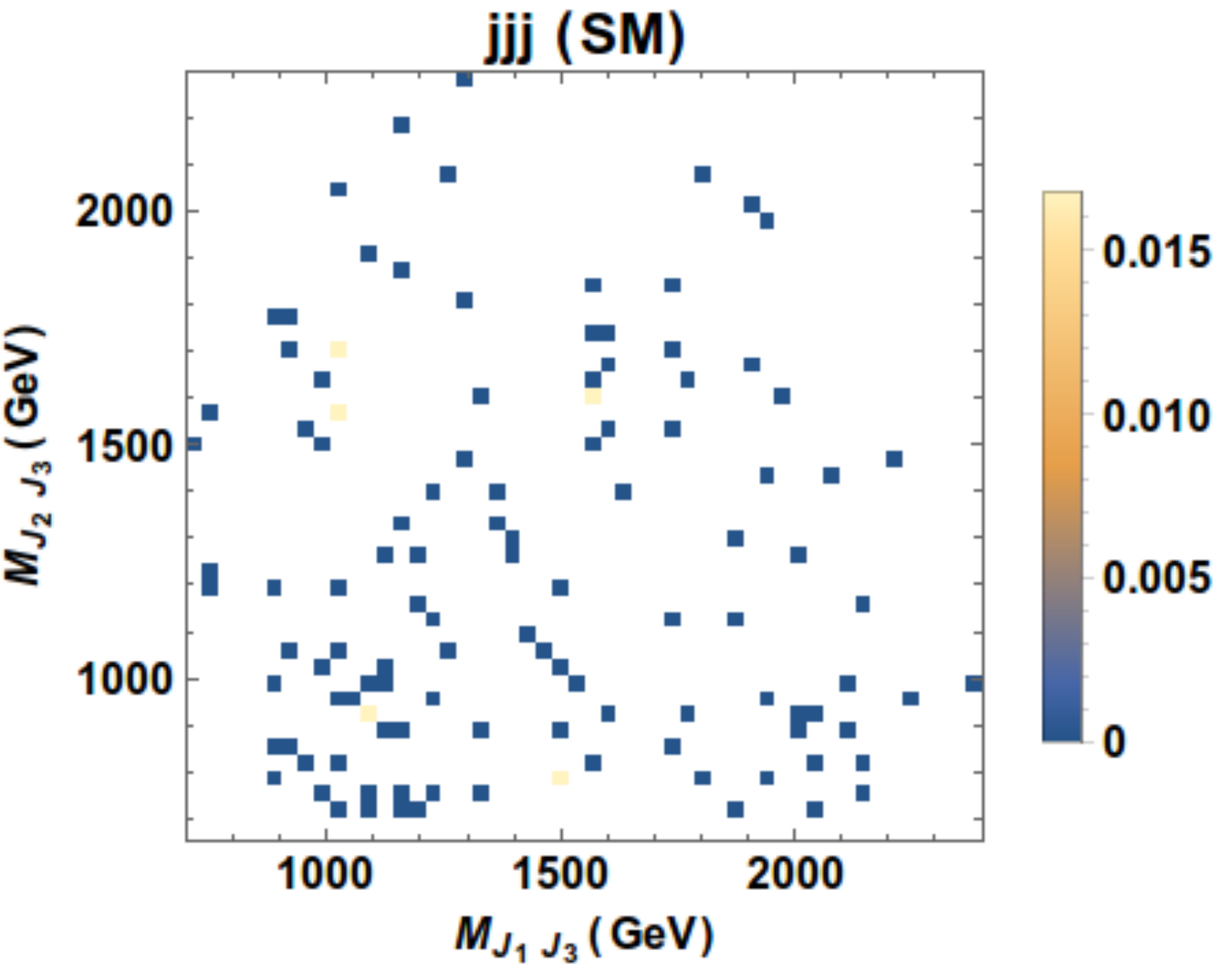}
    \caption{2D $M_{J_1 J_3}$-$M_{J_2 J_3}$ distributions for $W$-$WWW$-BP2 fully hadronic channel: SG after pre-selection but before any other cuts (top row, left), BK after pre-selection but before any other cuts (top row, right), SG after all other cuts but 2D $M_{J_1 J_3}$-$M_{J_2 J_3}$ cut (bottom row, left), BK after all other cuts but 2D $M_{J_1 J_3}$-$M_{J_2 J_3}$ cut (second row, right). The region surrounded by orange dashed line shows the signal selected by the ``+'' cut used in our 3D analysis. We refer to the main text for more detailed information.
}
\label{fig:WKK_WWW_Full_Had_2D_SG2}
\end{figure}

We first apply a set of $p_T$ and $\tau_{21}$ cuts common to the 1D, 2D, and 3D analyses, summarized at the top of table~\ref{tab:W-WWW-BP1-Full-Had-cutflow}.
We add a mass window cut in $M_{JJJ}$ for the 1D analysis to pick out the $W_{\rm{KK}}$ resonance. In the 2D analysis we additionally impose a single $M_{JJ}$ mass window cut for the $\varphi$ resonance. Also, we add mild $M_{JJ}$ cuts for other combinations of two fat jets to further reduce backgrounds.

For the 3D analysis, we desire to take maximum advantage of the distinctive signal regions shown in figures \ref{fig:WKK_WWW_Full_Had_2D_SG1} and \ref{fig:WKK_WWW_Full_Had_2D_SG2}. Plots in the top row show distributions after pre-selection and those in the bottom row represent distributions after all other cuts besides the 2D $M_{J_1 J_2}$-$M_{J_2 J_3}$ cut. We see that after the other cuts the signal is very well localized in a characteristic ``+'' shape, and the 2D analysis will discard all events falling into one of the two arms. In order to approximate a shape analysis we define two bins for each benchmark corresponding to the horizontal and vertical arms, with events falling into the intersection being assigned to the horizontal bin. We form a combined significance for the bins of each benchmark by adding individual bins in quadrature. This is accurate only in the limit when all systematic uncertainties can be neglected and only statistical uncertainties are relevant. This is not likely to correspond to a realistic search. However, it may give some general insight into the power that a full shape analysis might provide. For benchmark $W$-$WWW$-BP1, we define one bin corresponding to the horizontal arm as $850 \; \text{GeV} < M_{J_1 J_3} < 2200 \; \text{GeV}$ and $800 \; \text{GeV} < M_{J_2 J_3} < 1050 \; \text{GeV}$, and a second bin corresponding to the vertical arm as $900 \; \text{GeV} < M_{J_1 J_3} < 1100 \; \text{GeV}$ and $750 \; \text{GeV} < M_{J_2 J_3} < 1900 \; \text{GeV}$. The corresponding bins for benchmark $W$-$WWW$-BP2 are $800 \; \text{GeV} < M_{J_1 J_3} < 2300 \; \text{GeV}$ and $1200 \; \text{GeV} < M_{J_2 J_3} < 1600 \; \text{GeV}$ (horizontal arm), and $1300 \; \text{GeV} < M_{J_1 J_3} < 1600 \; \text{GeV}$ and $700 \; \text{GeV} < M_{J_2 J_3} < 2150 \; \text{GeV}$ (vertical arm). More details are summarized in table~\ref{tab:W-WWW-BP1-Full-Had-cutflow}.
\begin{table}[h]
\centering
\begin{tabular}{|c|c|c|c|c|c|c|}
\hline 
\multicolumn{7}{|c|}{$W$-$WWW$-BP1: Fully hadronic channel } \\
\hline
-- & \multicolumn{2}{c|}{1D} &  \multicolumn{2}{c|}{2D} & \multicolumn{2}{c|}{3D} \\
\hline
Cuts & SG & $jjj$ & SG & $jjj$ & SG & $jjj$ \\
\hline \hline
Parton-level cuts & 0.13  & 40 & 0.13  & 40 & 0.13 & 40 \\
\hline
$N_j \geq 3$, pre-selection cuts & 0.11 & 11 & 0.11 & 11 & 0.11 & 11 \\
\hline
$ p_{T, J_1} \in [800, \infty] \GeV$ & 0.092 & 4.1 & 0.092 & 4.1 & 0.092 & 4.1  \\
$ p_{T, J_2} \in [600, \infty] \GeV$ & 0.087 & 3.0 & 0.087 & 3.0 & 0.087 & 3.0 \\
$ p_{T, J_3} \in [300, \infty] \GeV$ & 0.066 & 2.0 & 0.066 & 2.0 & 0.066 & 2.0 \\
\hline
$ \tau_{21} \in [0, 0.5] $ & 0.037 & 0.12 & 0.037 & 0.12 & 0.037 & 0.12 \\
\hline
$ M_{JJJ} \in [2600, 3200] \GeV$ & 0.036 & 0.046 & 0.036 & 0.046 & 0.036 & 0.046 \\
$ M_{J_1 J_2} \in [1600, 2600] \GeV$ & -- & -- & 0.035 & 0.038 & 0.035 & 0.038 \\
\hline
$ M_{J_2 J_3} \in [850, 1050] \GeV$ & -- & -- & 0.030 & 0.0092 & -- & -- \\
$ M_{J_1 J_3} \in [800, \infty] \GeV$ & -- & -- & 0.030 & 0.0092 & -- & -- \\
\hline
``+" cut & -- & -- & -- & -- & 0.034 & 0.013 \\
\hline
$S/B$ & 0.77 & -- & 3.3 & -- & 3.6 & -- \\
$S/\sqrt{B}$ ($\mathcal{L}=300$ fb$^{-1}$) & 2.9 & -- & 5.4 & -- & 5.8 & -- \\
$S/\sqrt{S+B}$ ($\mathcal{L}=300$ fb$^{-1}$) & 2.2 & -- & 2.6 & -- & 2.7 & -- \\
\hline
\end{tabular}  \\
\vspace{1.0cm}
\begin{tabular}{|c|c|c|c|c|c|c|}
\hline 
\multicolumn{7}{|c|}{$W$-$WWW$-BP2: Fully hadronic channel } \\
\hline
-- & \multicolumn{2}{c|}{1D} &  \multicolumn{2}{c|}{2D} & \multicolumn{2}{c|}{3D} \\
\hline
Cuts & SG & $jjj$ & SG & $jjj$ & SG & $jjj$ \\
\hline \hline
Parton-level cuts & 0.11  & 40 & 0.11  & 40 & 0.11 & 40 \\
\hline
$N_j \geq 3$, pre-selection cuts & 0.10 & 11 & 0.10 & 11 & 0.10 & 11 \\
\hline
$ p_{T, J_1} \in [700, \infty] \GeV$ & 0.094 & 6.3 & 0.094 & 6.3 & 0.094 & 6.3  \\
$ p_{T, J_2} \in [600, \infty] \GeV$ & 0.081 & 3.8 & 0.081 & 3.8 & 0.081 & 3.8 \\
$ p_{T, J_3} \in [250, \infty] \GeV$ & 0.078 & 3.2 & 0.078 & 3.2 & 0.078 & 3.2 \\
\hline
$ \tau_{21} \in [0, 0.5] $ & 0.043 & 0.21 & 0.043 & 0.21 & 0.043 & 0.21 \\
\hline
$ M_{JJJ} \in [2600, 3200] \GeV$ & 0.042 & 0.086 & 0.042 & 0.086 & 0.042 & 0.086 \\
$ M_{J_1 J_2} \in [1200, 2600] \GeV$ & -- & -- & 0.042 & 0.078 & 0.042 & 0.078 \\
\hline
$ M_{J_1 J_3} \in [1300, 1600] \GeV$ & -- & -- & 0.028 & 0.016 & -- & -- \\
$ M_{J_2 J_3} \in [800, \infty] \GeV$ & -- & -- & 0.026 & \hspace{0.1cm}0.014\hspace{0.1cm} & -- & -- \\
\hline
``+" cut & -- & -- & -- & -- & 0.040 & 0.027  \\
\hline
$S/B$ & 0.49 & -- & 1.8 & -- & 2.1 & -- \\
$S/\sqrt{B}$ ($\mathcal{L}=300$ fb$^{-1}$) & 2.5 & -- & 3.8 & -- & 4.3 & -- \\
$S/\sqrt{S+B}$ ($\mathcal{L}=300$ fb$^{-1}$) & 2.0 & -- & 2.3 & -- & 2.7 & -- \\
\hline
\end{tabular}
\vspace{0.5cm}
\caption{Cut flows for $W$-$WWW$-BP1/$W$-$WWW$-BP2 (upper/lower table) fully hadronic channel and their major background in terms of cross sections (in fb). Parton-level cuts in~\eqref{eq:parton_level_cuts_WWW_full_had} are imposed only on the background events at the generation-level, while pre-selection cuts, which consist of the same cuts as in the parton-level cuts, are imposed on both signal and background events at the detector level after jet tagging. We refer to the main text for more detailed information on the ``+'' cut.}
\label{tab:W-WWW-BP1-Full-Had-cutflow}
\end{table} 

The comparison between the 1D and 2D analyses suggests that it is rather essential to utilize both two-body and three-body invariant mass cuts to achieve a better signal sensitivity. For example for $W$-$WWW$-BP1, we notice that the 2D analysis enables us to reduce background events to $\sim 3$ while retaining $\sim10$ signal events, with an integrated luminosity of $\mathcal{L} = 300 {\rm fb}^{-1}$. Promoting to the 3D analysis allows a more improved signal sensitivity, as it saves more signal events than background ones.

\subsubsection*{Potential improvement with alternative variables}

\begin{figure}
    \centering
    \includegraphics[width = 7 cm]{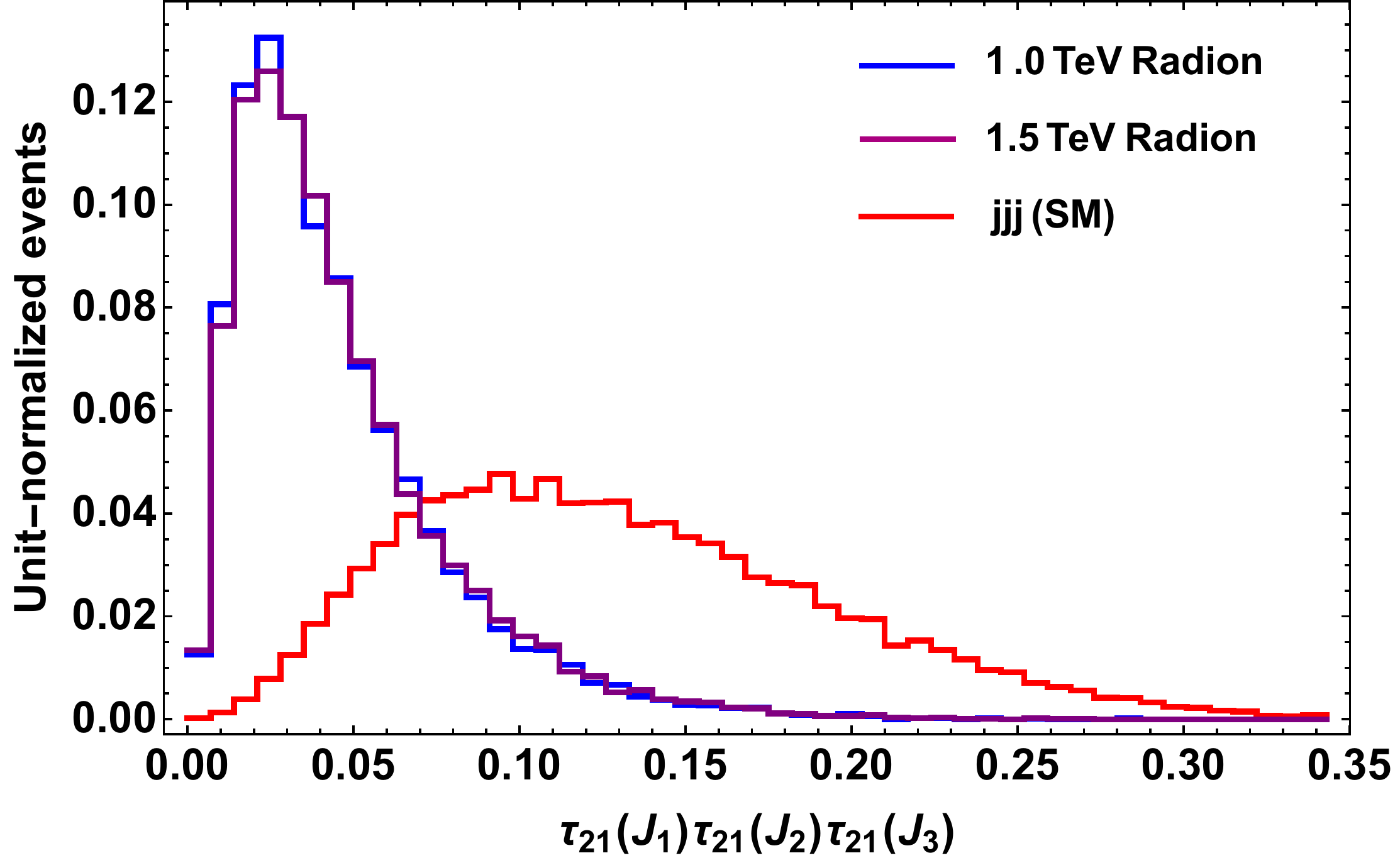}
    \includegraphics[width = 7 cm]{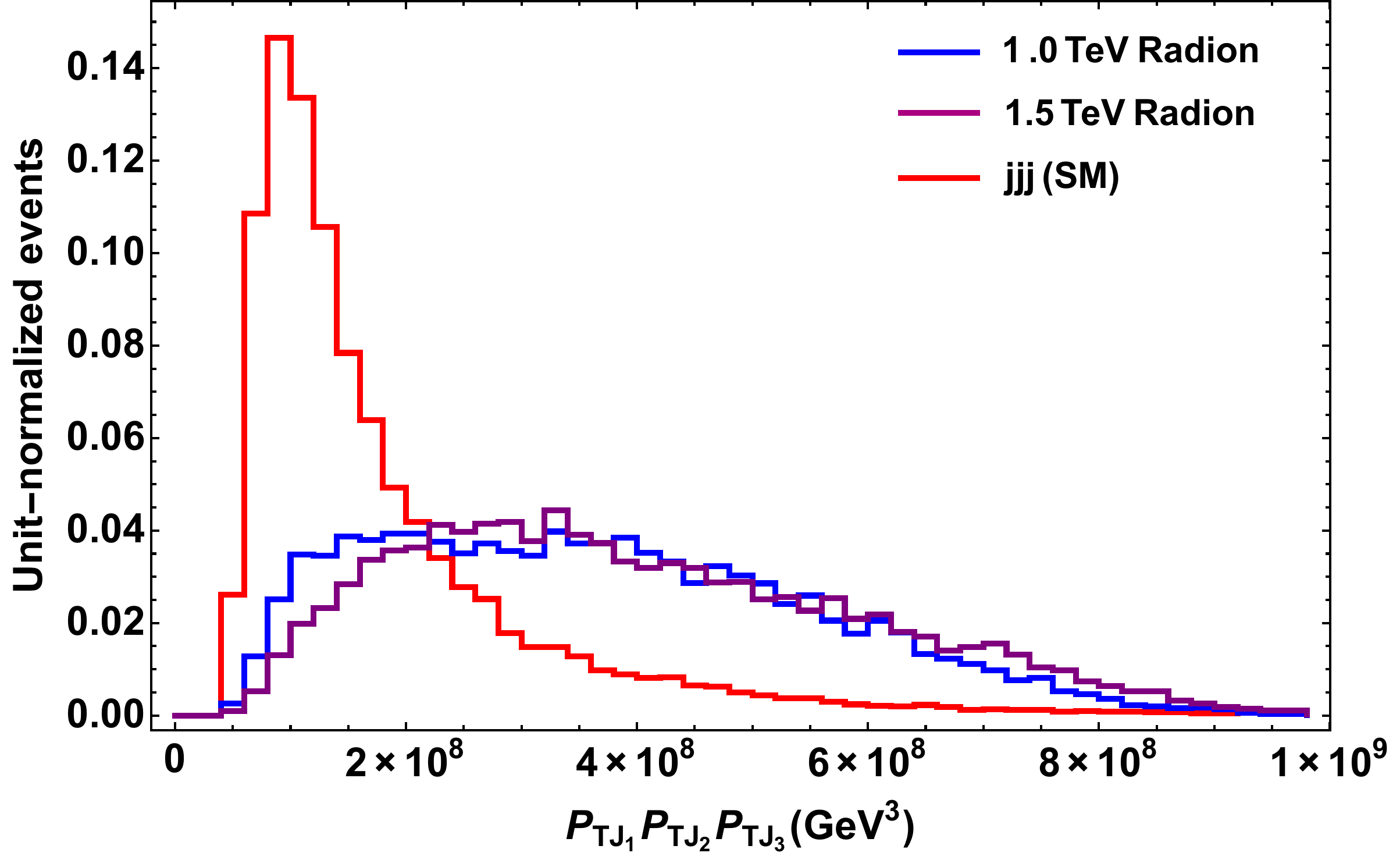}
    \caption{Distributions of new kinematic variables for the $W$-$WWW$-BP1($W$-$WWW$-BP2) fully hadronic channel: $\tau_{123}$ (left) and $p_{T, 123}$ (right) for signal with 1 TeV radion (solid blue), signal with 1.5 TeV radion (solid purple) and backgrounds (solid red). We denote $p_T$-ordered jet as $J_{1,2,3}$, $J_1$ being the hardest jet. These variables allow sharper distinction between the signal and the background compared with the standard $p_T$ and $\tau_{21}$ variables (see figure~\ref{fig:WKK_WWW_Full_Had_1D}).
}
\label{fig:WKK_WWW_Full_Had_1D_New_Variables}
\end{figure}

Finally, we briefly comment on a potential variation in the set of our selection variables, which can improve the signal significance.
We find that it is more efficient to place a single cut on the product of the three $\tau_{21}$'s, $\tau_{123} \equiv \tau_{12, J_1} \times \tau_{12, J_2}\times \tau_{12, J_3}$, and also on the product of the three $p_T$'s, $p_{T, 123} \equiv p_{T, 1}\times p_{T, 2}\times p_{T, 3}$. These alternative variables reveal better contrast between the signal and the background, as shown in figure~\ref{fig:WKK_WWW_Full_Had_1D_New_Variables}. Using them in conjunction with other conventional variables, we obtain a signal cross section of 0.044 (0.039) fb for $W$-$WWW$-BP1 ($W$-$WWW$-BP2) while having a background cross section of 0.013 (0.017) fb. With $\mathcal{L} = 300 \; {\rm fb}^{-1}$, we assess the $S/\sqrt{S+B}$ significance to be $3.2 \; (2.9)$ for $W$-$WWW$-BP1 ($W$-$WWW$-BP2) which is the largest compared to the results in table~\ref{tab:W-WWW-BP1-Full-Had-cutflow}. Usage of such novel variables may need a SM background modelling to quantify their potential more precisely. We leave further exploration of this for future.

\subsubsection*{Semi-leptonic channel}

\begin{figure}[h]
    \centering
    \includegraphics[width = 7.5 cm]{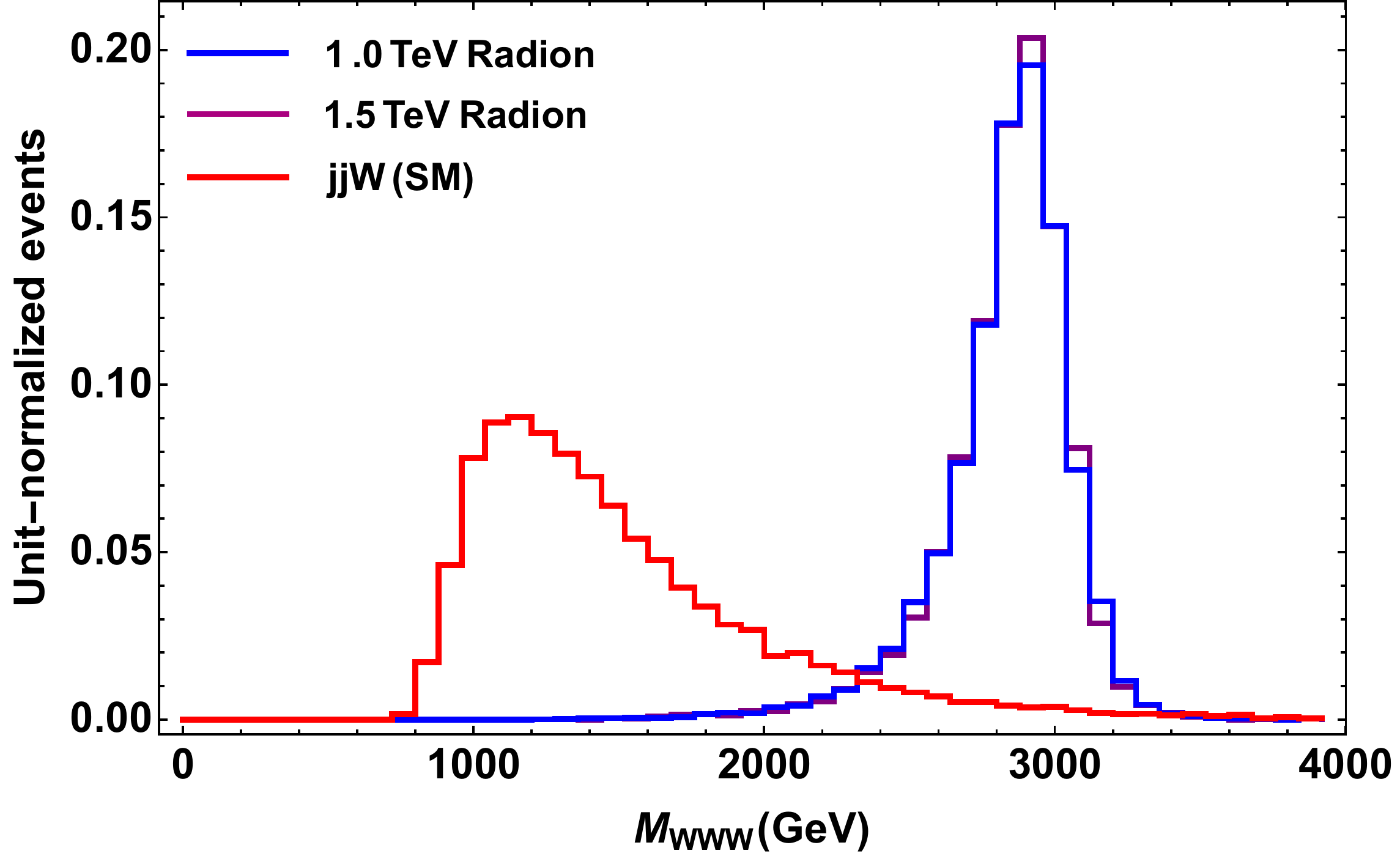}
    \includegraphics[width = 7.3 cm]{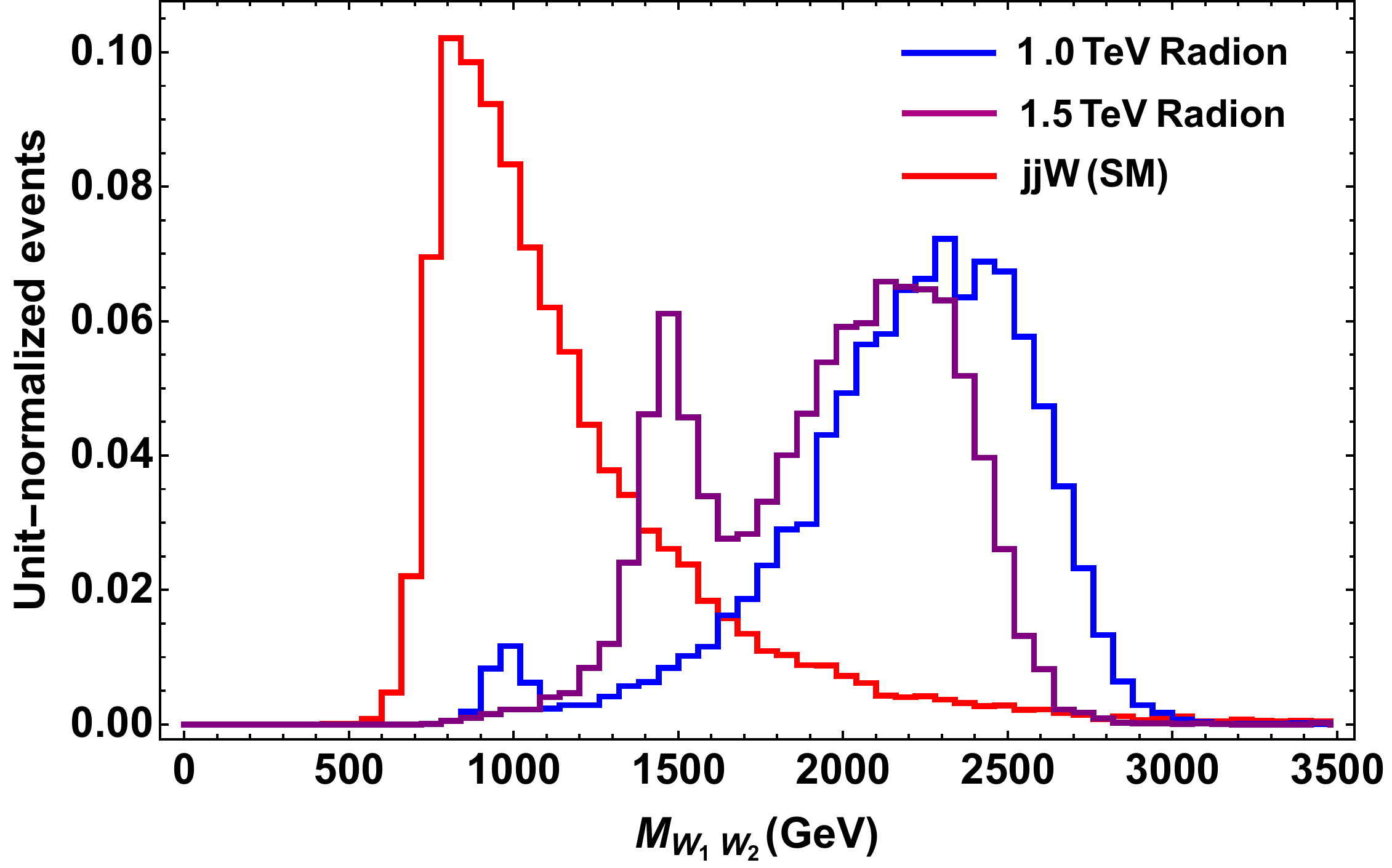}
    \includegraphics[width = 7.5 cm]{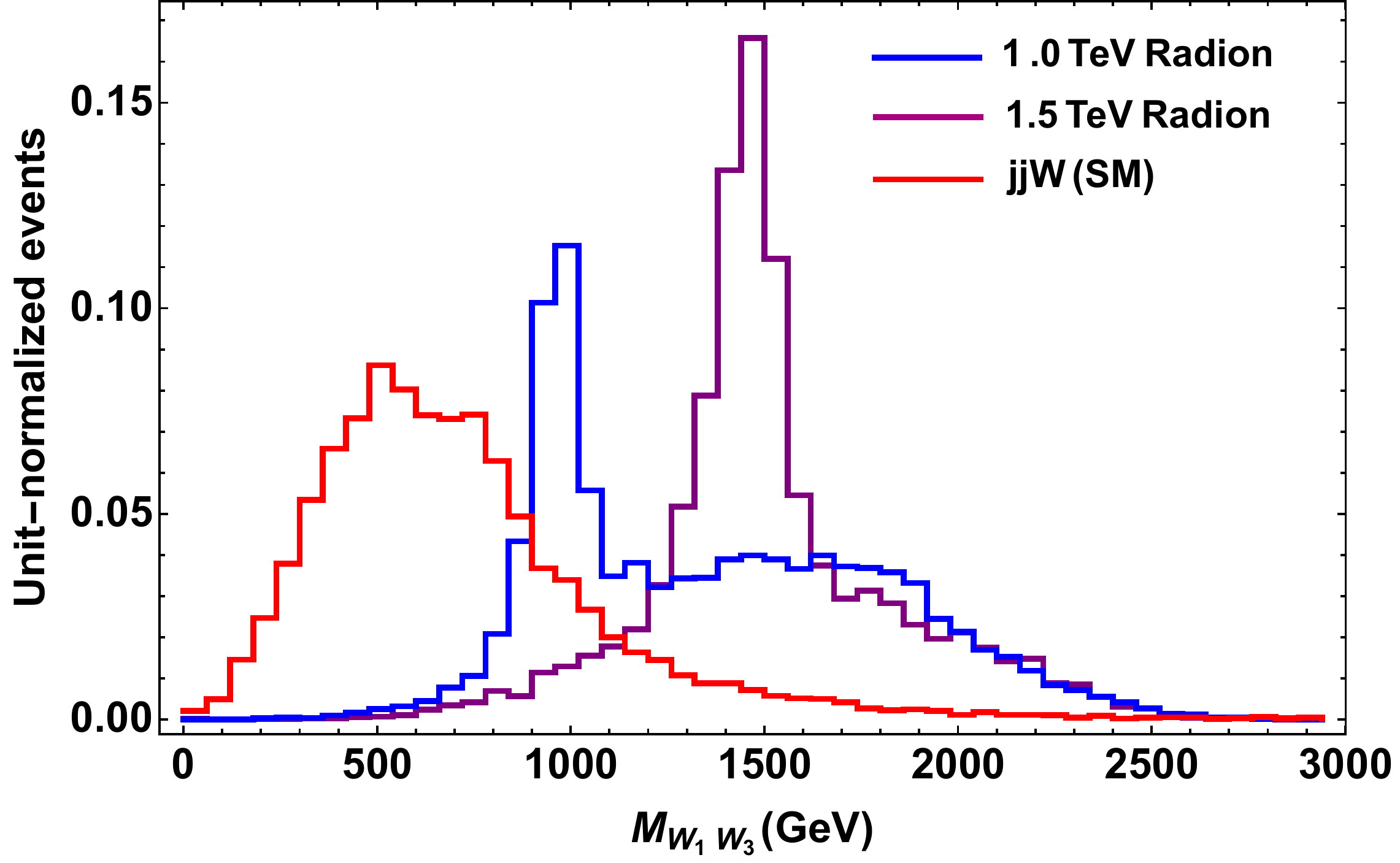}
    \includegraphics[width = 7.5 cm]{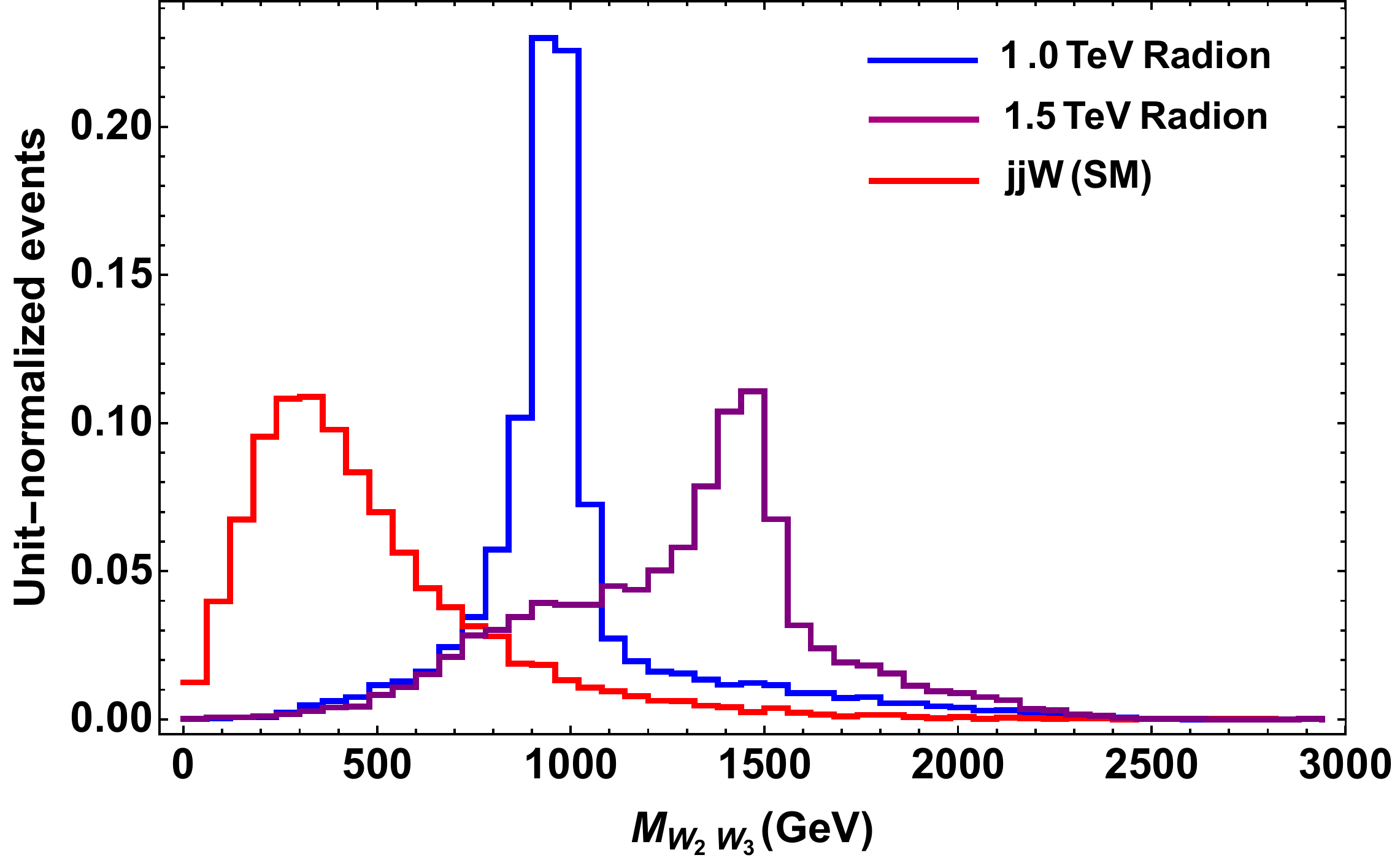}
    \includegraphics[width = 4.8 cm]{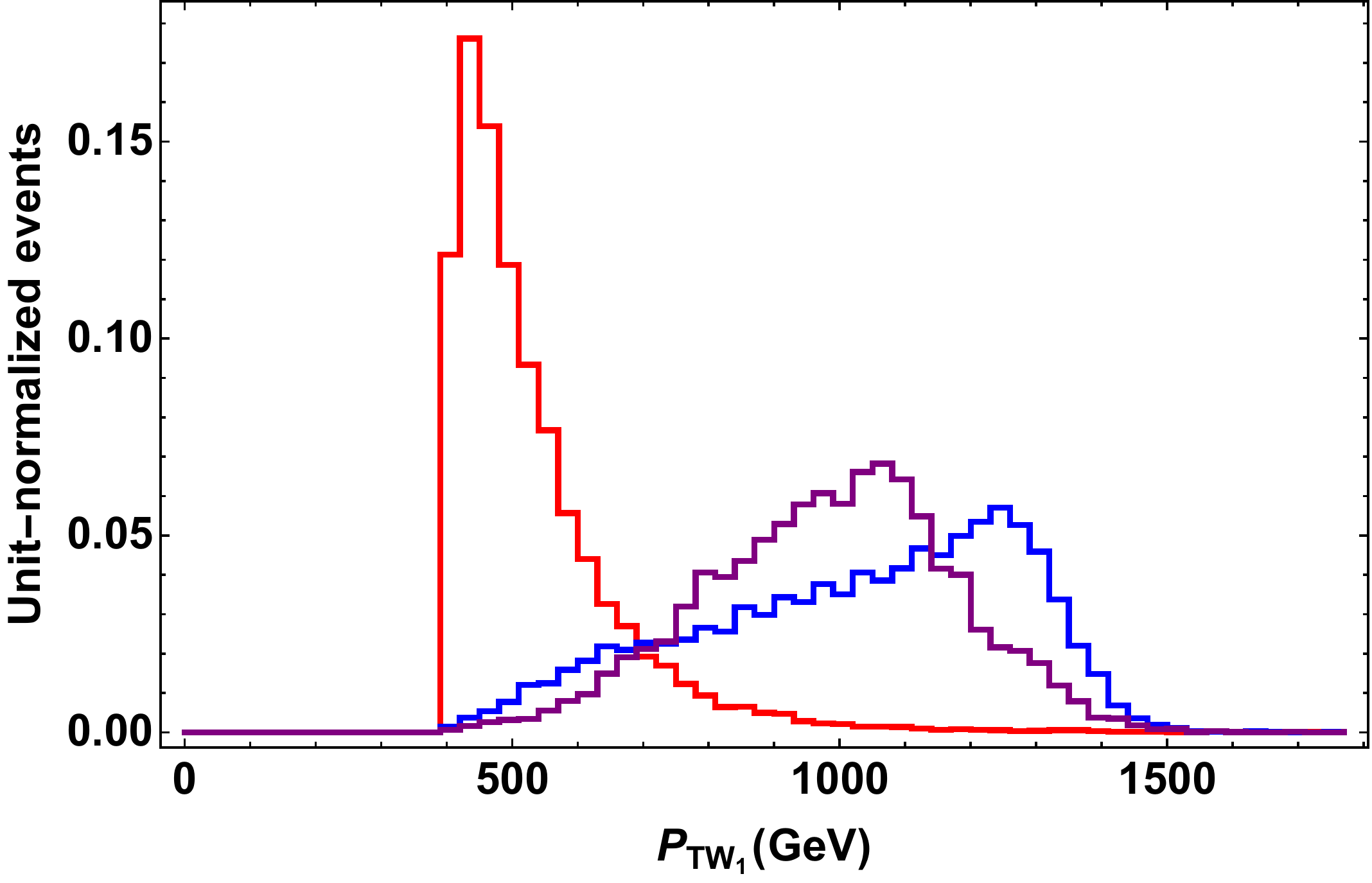}
    \includegraphics[width = 4.8 cm]{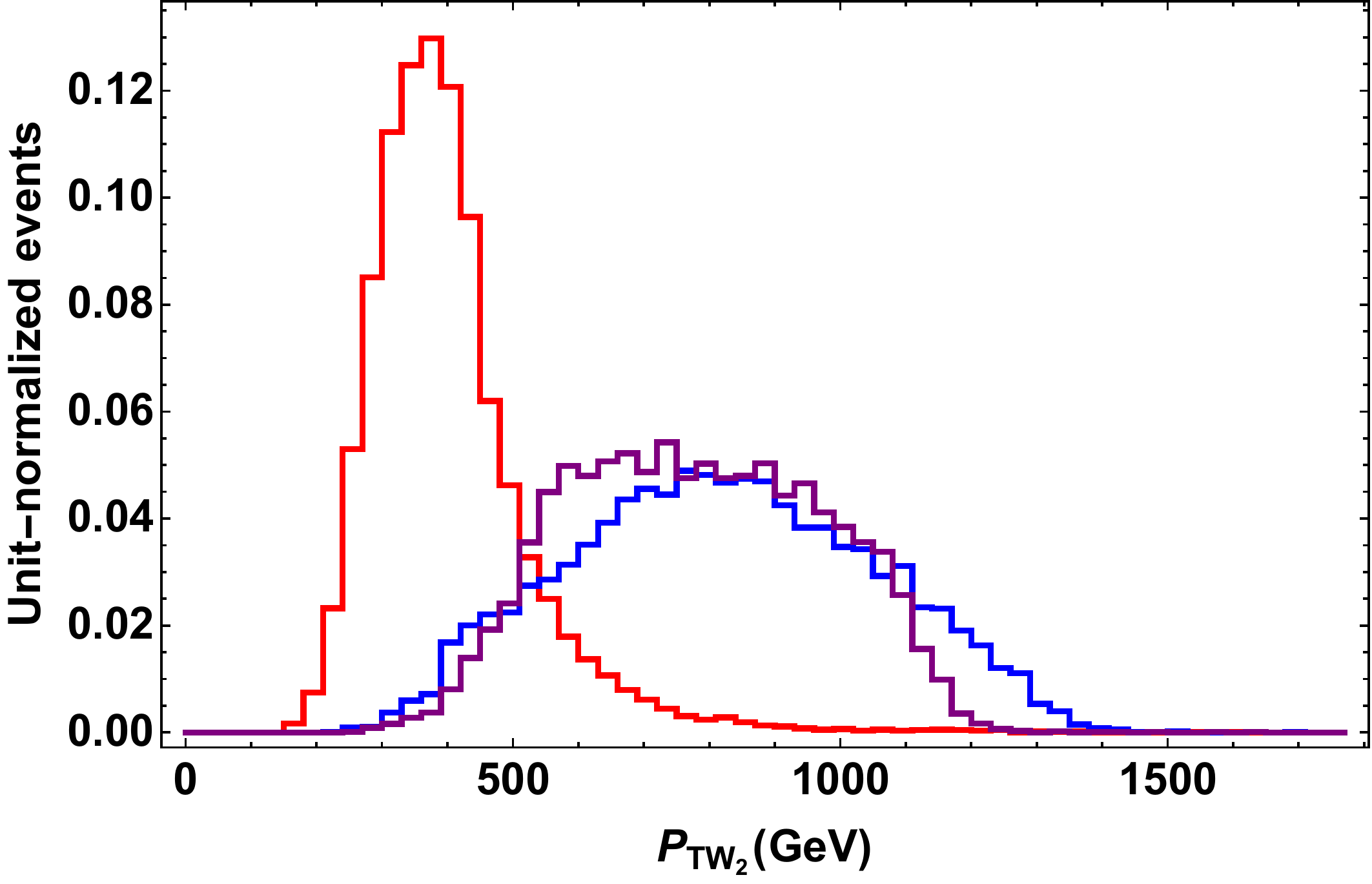}
    \includegraphics[width = 4.8 cm]{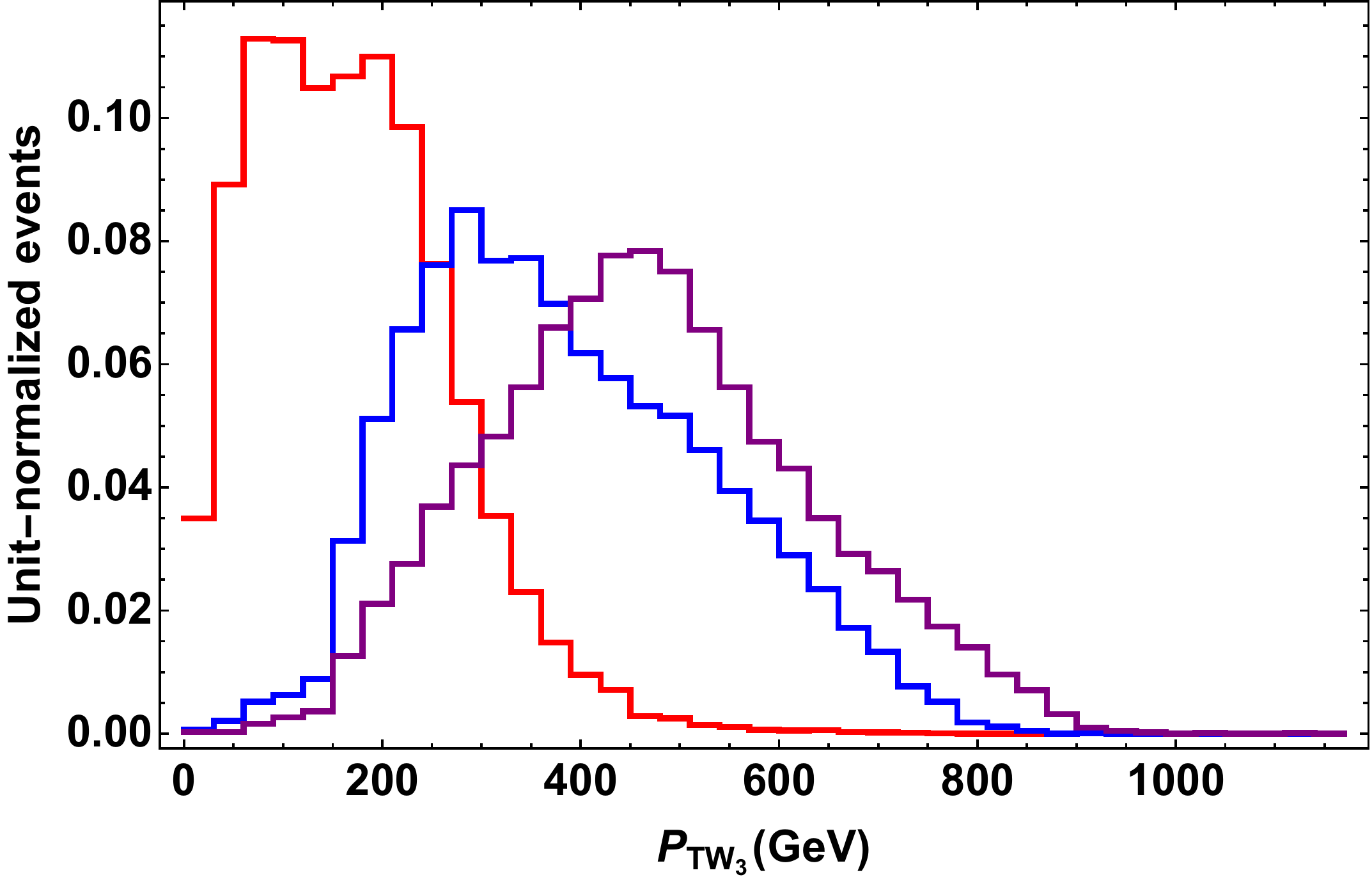}
    \includegraphics[width = 4.8 cm]{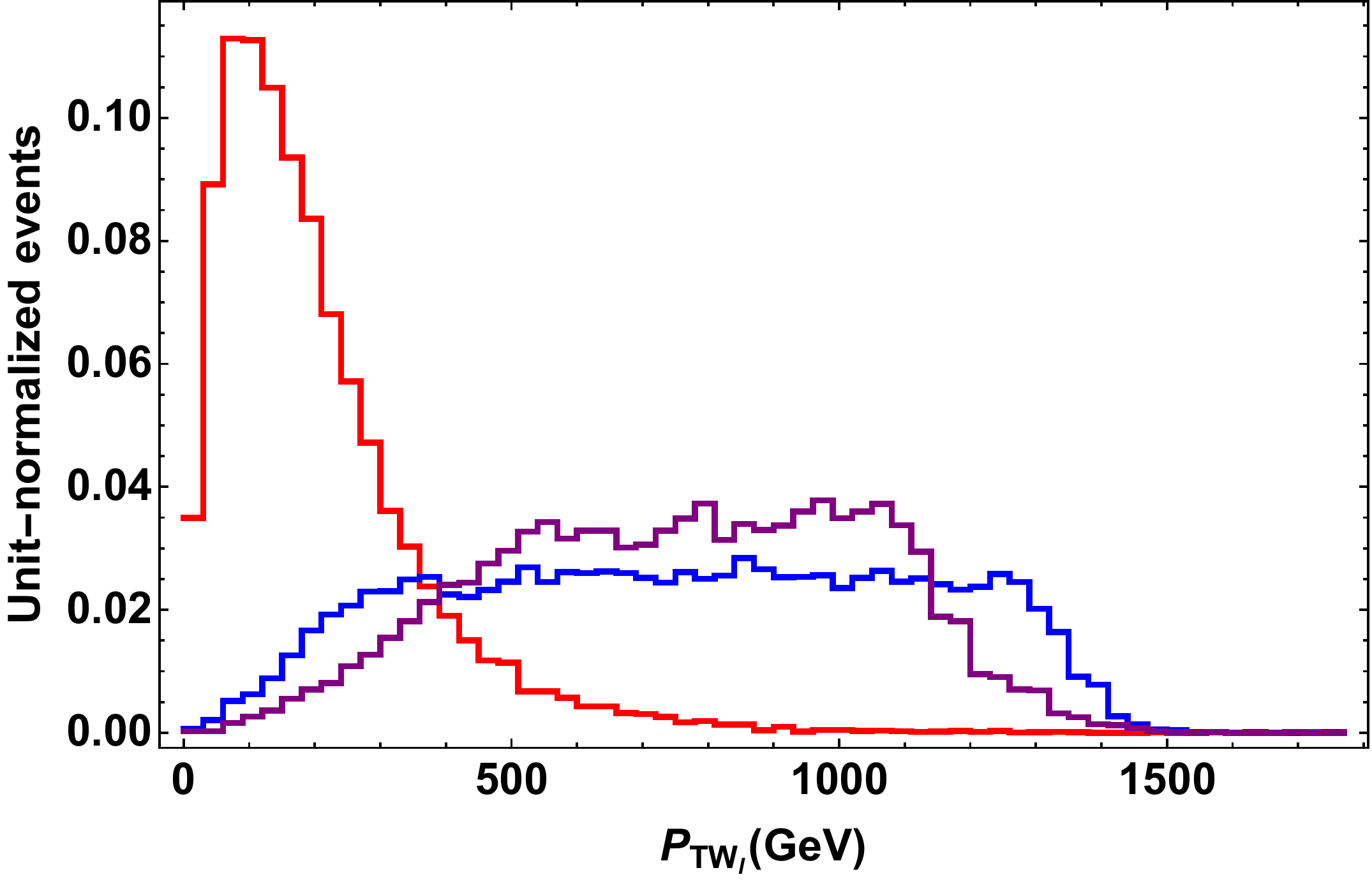}
    \includegraphics[width = 4.8 cm]{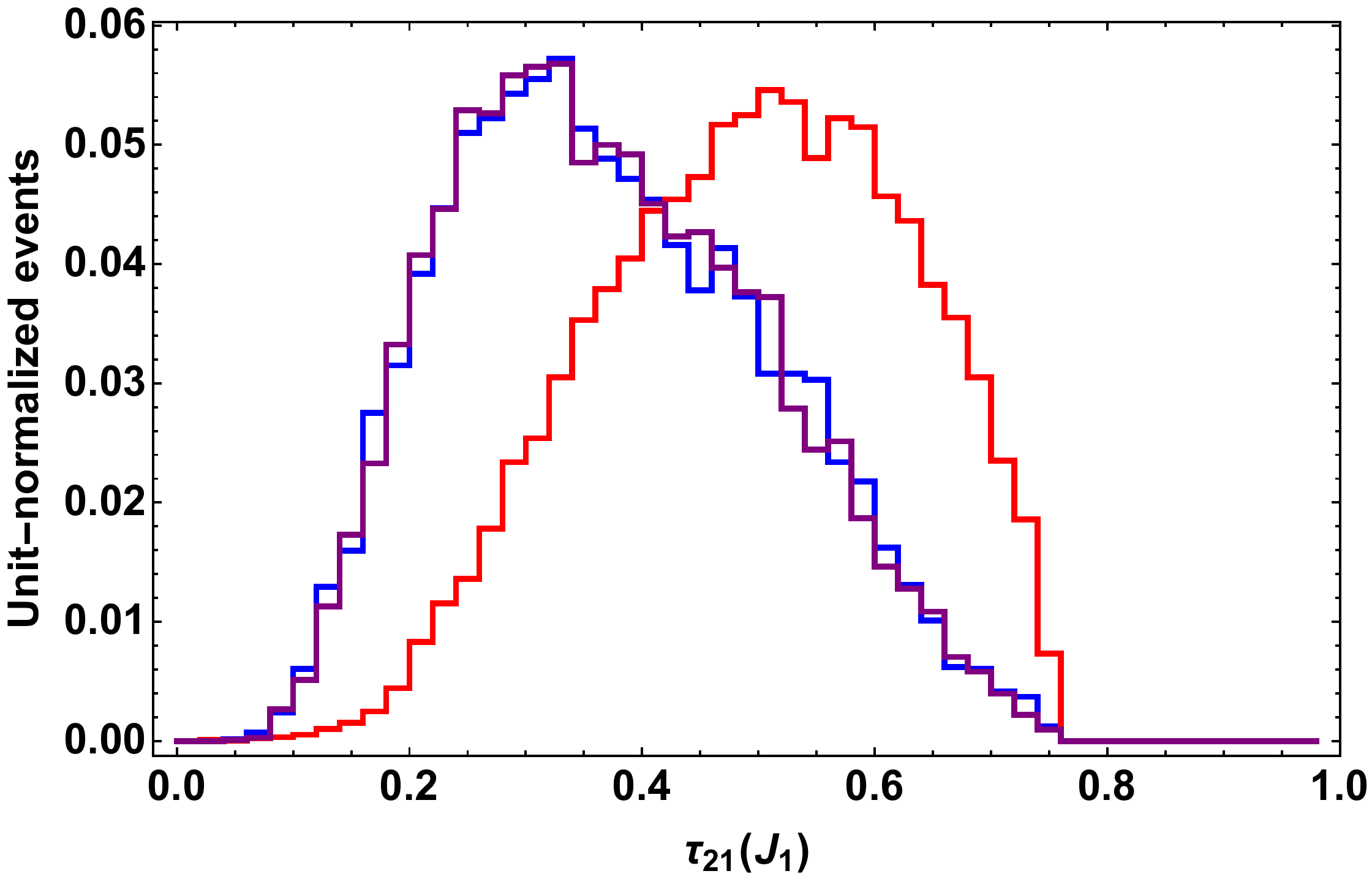}
    \includegraphics[width = 4.8 cm]{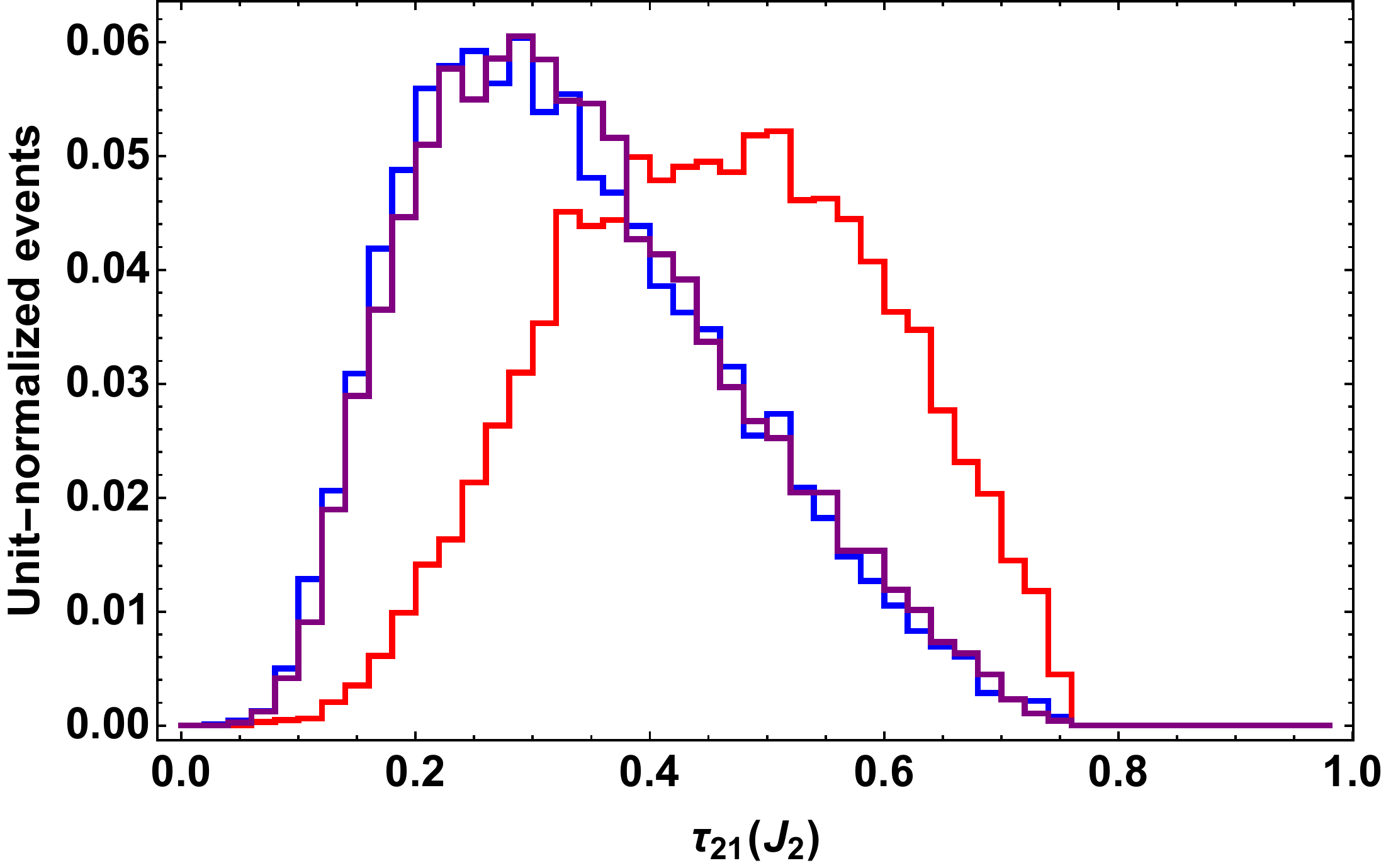}
    \caption{Distribution of kinematic variables for $W$-$WWW$-BP1($W$-$WWW$-BP2) semi-leptonic channel: $M_{WWW}$ (top row, left), $M_{W_1 W_2}$ (top row, right), $M_{W_1 W_3}$ (second row, left), $M_{W_2 W_3}$ (second row, right), $p_{T W_1}$ (third row, left), $p_{T W_2}$ (third row, middle), $p_{T W_3}$ (third row, right), $p_{T_{W_\ell}}$ (bottom row, left), $\tau_{21 J_1}$ (bottom row, middle), $\tau_{21 J_2}$ (bottom row, right) for signal with 1 TeV radion (solid blue), signal with 1.5 TeV radion (solid purple) and backgrounds (solid red). We denote $p_T$-ordered $W$'s as $W_{1,2,3}$, $W_1$ being the hardest one. 
}
\label{fig:WKK_WWW_Lep_Had_1D}
\end{figure}

We next consider the process of eq.~(\ref{eq:process_WWW}), with one $W$ decaying leptonically and the others decaying hadronically. The final state therefore consists of two fat $W$-jets and a lepton and a neutrino. As before, we impose a set of parton-level cuts for background simulation
\bea
p_{T j_1} > 400 \; \GeV, \; p_{T j_2} > 150 \; \GeV, \; M_{j_1 j_2} > 700 \; \GeV, \; \left|\eta_{\ell}\right| < 2.5, \left|\eta_j\right| < 5.
\label{eq:parton_level_cuts_WWW_semi_lep}
\eea
For pre-selection cuts, we reintroduce the same cuts at the detector level plus a harder cut on jet pseudorapidity, $|\eta_J| \leq 2.4$. 
As there is only one neutrino in the final state, we assume that the missing transverse momentum originates solely from the neutrino transverse momentum.
The strategy discussed in section~\ref{tools} allows us to reconstruct the longitudinal component of the neutrino, hence its full four momentum.
The tagging of the hadronically decaying $W$ is done in the same way as in the fully hadronic analysis. 
Figure~\ref{fig:WKK_WWW_Lep_Had_1D} shows distributions of various kinematic variables for both $W$-$WWW$-BP1 and $W$-$WWW$-BP2. We remark that the structures of the distributions, especially for two and three-body invariant masses, are similar to those in the fully hadronic channel, thus the rest of the analysis goes through a similar procedure. We report our detailed results in table~\ref{tab:W-WWW-BP1-Semi-Lep-cutflow}. 

Combining fully hadronic and semi-leptonic channels, we can get a boost in the significance.  To illustrate this point, we summarize the significances in both channels with the 1D, 2D and 3D analyses in table~\ref{tab:significance}. Also, we show the expected signal number of events and the expected background number of events, which are enclosed by parentheses, with an integrated luminosity of 300 fb$^{-1}$.
A combined significance is calculated by adding the two associated significances in quadrature. For both benchmark points, there is an improvement of significance from the 1D to 3D analyses.

\begin{table}[t]
\centering
\begin{tabular}{|c|c|c|c|c|c|c|}
\hline 
\multicolumn{7}{|c|}{$W$-$WWW$-BP1: Semi-leptonic channel } \\
\hline
-- & \multicolumn{2}{c|}{1D} &  \multicolumn{2}{c|}{2D} & \multicolumn{2}{c|}{3D} \\
\hline
Cuts & SG & $jjW$ & SG & $jjW$ & SG & $jjW$ \\
\hline \hline
Parton-level cuts & 0.13 & 10.6 & 0.13 & 10.6 & 0.13 & 10.6 \\
\hline
$N_j \geq 2, \; N_{W_\ell} \geq 1$, pre-selection cuts & 0.13 & 6.1 & 0.13 & 6.1 & 0.13 & 6.1 \\
\hline
$ p_{T, W_1} \in [600, \infty] \GeV$ & 0.12 & 1.2 & 0.12 & 1.2 & 0.12 & 1.2  \\
$ p_{T, W_2} \in [500, \infty] \GeV$ & 0.12 & 0.69 & 0.12 & 0.69 & 0.12 & 0.69 \\
$ p_{T, W_3} \in [200, \infty] \GeV$ & 0.11 & 0.33 & 0.11 & 0.33 & 0.11 & 0.33 \\
\hline
$ \tau_{21, J_1} \in [0, 0.6] $  & 0.099 & 0.22 & 0.099 & 0.22 & 0.099 & 0.22 \\
$ \tau_{21, J_2} \in [0, 0.6] $ & 0.095 & 0.18 & 0.095 & 0.18 & 0.095 & 0.18 \\
\hline
$ M_{WWW} \in [2500, 3400] \GeV$ & 0.089 & 0.039 & 0.089 & 0.039 & 0.089 & 0.039 \\
$ M_{W_1 W_2} \in [1500, 3000] \GeV$ & -- & -- & 0.088 & 0.032 & 0.088 & 0.032 \\
\hline
$ M_{W_1 W_3} \in [800, \infty] \GeV$ & -- & -- & 0.087 & 0.030 & -- & -- \\
$ M_{W_2 W_3} \in [600, 1500] \GeV$ & -- & -- & 0.081 & 0.021 & -- & -- \\
\hline
2D cut & -- & -- & -- & -- & 0.087 & 0.024 \\
\hline
$S/B$ & 2.3 & -- & 3.8 & -- & 4.6 & -- \\
$S/\sqrt{B}$ ($\mathcal{L}=300$ fb$^{-1}$) & 7.9 & -- & 9.6 & -- & 10.2 & -- \\
$S/\sqrt{S+B}$ ($\mathcal{L}=300$ fb$^{-1}$) & 4.3 & -- & 4.4 & -- & 4.5 & -- \\
\hline
\end{tabular} \\
\vspace{0.2cm}
\begin{tabular}{|c|c|c|c|c|c|c|}
\hline 
\multicolumn{7}{|c|}{$W$-$WWW$-BP2: Semi-leptonic channel } \\
\hline
-- & \multicolumn{2}{c|}{1D} &  \multicolumn{2}{c|}{2D} & \multicolumn{2}{c|}{3D} \\
\hline
Cuts & SG & $jjW$ & SG & $jjW$ & SG & $jjW$ \\
\hline \hline
Parton-level cuts & 0.11 & 10.6 & 0.11 & 10.6 & 0.11 & 10.6 \\
\hline
$N_j \geq 2, \; N_{W_\ell} \geq 1$, pre-selection cuts & 0.11 & 6.1 & 0.11 & 6.1 & 0.11 & 6.1 \\
\hline
$ p_{T, W_1} \in [700, \infty] \GeV$ & 0.097 & 0.57 & 0.097 & 0.57 & 0.097 & 0.57  \\
$ p_{T, W_2} \in [500, \infty] \GeV$ & 0.094 & 0.41 & 0.094 & 0.41 & 0.094 & 0.41 \\
$ p_{T, W_3} \in [200, \infty] \GeV$ & 0.091 & 0.23 & 0.091 & 0.23 & 0.091 & 0.23 \\
\hline
$ \tau_{21, J_1} \in [0, 0.6] $  & 0.086 & 0.16 & 0.086 & 0.16 & 0.086 & 0.16 \\
$ \tau_{21, J_2} \in [0, 0.6] $ & 0.082 & 0.12 & 0.082 & 0.12 & 0.082 & 0.12 \\
\hline
$ M_{WWW} \in [2500, 3400] \GeV$ & 0.078 & 0.032 & 0.078 & 0.032 & 0.078 & 0.032 \\
$ M_{W_1 W_2} \in [1200, 2600] \GeV$ & -- & -- & 0.077 & 0.025 & 0.077 & 0.025 \\
\hline
$ M_{W_1 W_3} \in [1200, \infty] \GeV$ & -- & -- & 0.072 & 0.018 & -- & -- \\
$ M_{W_2 W_3} \in [500, \infty] \GeV$ & -- & -- & 0.072 & 0.018 & -- & -- \\
\hline
2D cut & -- & -- & -- & -- & 0.073 & 0.016 \\
\hline
$S/B$ & 2.5 & -- & 4.1 & -- & 6.7 & -- \\
$S/\sqrt{B}$ ($\mathcal{L}=300$ fb$^{-1}$) & 7.6 & -- & 9.4 & -- & 10.5 & -- \\
$S/\sqrt{S+B}$ ($\mathcal{L}=300$ fb$^{-1}$) & 4.1 & -- & 4.2 & -- & 4.2 & -- \\
\hline
\end{tabular}
\vspace{1.0cm}
\caption{Cut flows for $W$-$WWW$-BP1/$W$-$WWW$-BP2 (upper/lower table) semi-leptonic channel and their major background in terms of cross sections (in fb). Parton-level cuts in~\eqref{eq:parton_level_cuts_WWW_semi_lep} are imposed only on the background events at the generation-level, while pre-selection cuts, which consist of the same cuts as in the parton-level cuts, are imposed on both signal and background events at the detector level after jet tagging. We refer to the main text for more detailed information on the 2D cut.}
\label{tab:W-WWW-BP1-Semi-Lep-cutflow}
\end{table} 

\begin{table}[t]
\centering
\begin{tabular}{|c|c c c|c c c|}
\hline 
\multicolumn{7}{|c|}{Significance for $WWW$ channels ($S/\sqrt{S+B}$)} \\
\hline
\multirow{2}{*}{Channels } &  \multicolumn{3}{c|}{$W$-$WWW$-BP1 } & \multicolumn{3}{c|}{$W$-$WWW$-BP2 } \\
\cline{2-7}
& 1D & 2D & 3D & 1D & 2D & 3D \\
\hline \hline
\multirow{2}{*}{Fully hadronic} &2.2  & 2.6  & 2.7  & 2.0  & 2.3 & 2.7 \\
 &(11/14) & (9.0/2.8) & (10/3.9) & (13/26) & (7.8/4.2) & (12/8.1) \\
\multirow{2}{*}{Semi-leptonic} &4.3 & 4.4 & 4.5 & 4.1 & 4.2 & 4.2  \\
& (27/12) & (24/6.3) & (26/7.2) & (23/9.6) & (22/5.4) & (22/4.8)  \\
\hline
Combined&4.8 & 5.1 & 5.3 & 4.6 &4.8 & 5.0  \\
\hline
\end{tabular}
\caption{Significances ($S/\sqrt{S+B}$) for tri-$W$ fully hadronic and semi-leptonic channels using 1D, 2D and 3D analyses. For each channel, we present two benchmark points: $W$-$WWW$-BP1($W$-$WWW$-BP2). The combined significance is obtained by adding two significances of two channels in quadrature. The values in the parentheses are the expected signal events/the expected background events with an integrated luminosity of 300 fb$^{-1}$.
\label{tab:significance}}
\end{table} 

\subsubsection{$W$ $+$ diphoton Signal}\label{subsubsec:Waa}

In this section, we consider the process where the radion from the decay of KK $W$ decays into a pair of photons.
\bea
p p \to W_{\rm KK} \to W \varphi \to W \gamma \gamma
\label{eq:process_Waa}
\eea
Although the diphoton branching ratio is much smaller than that of di-$W$, the clean nature of the associated signature and smaller SM backgrounds compensate for the smallness of the signal rate. Furthermore, unlike the case of $WWW$ final state, there is no combinatorial ambiguity as two photons come from the radion decay. Therefore, we expect that diphoton invariant mass spectrum sharply peaks at the radion mass with much less smearing, comparing with the di-$W$ case. The remaining $W$ can decay either hadronically or leptonically, both of which we consider one by one.

\subsubsection*{Hadronic $W$}
\begin{figure}
    \centering
    \includegraphics[width = 7.5 cm]{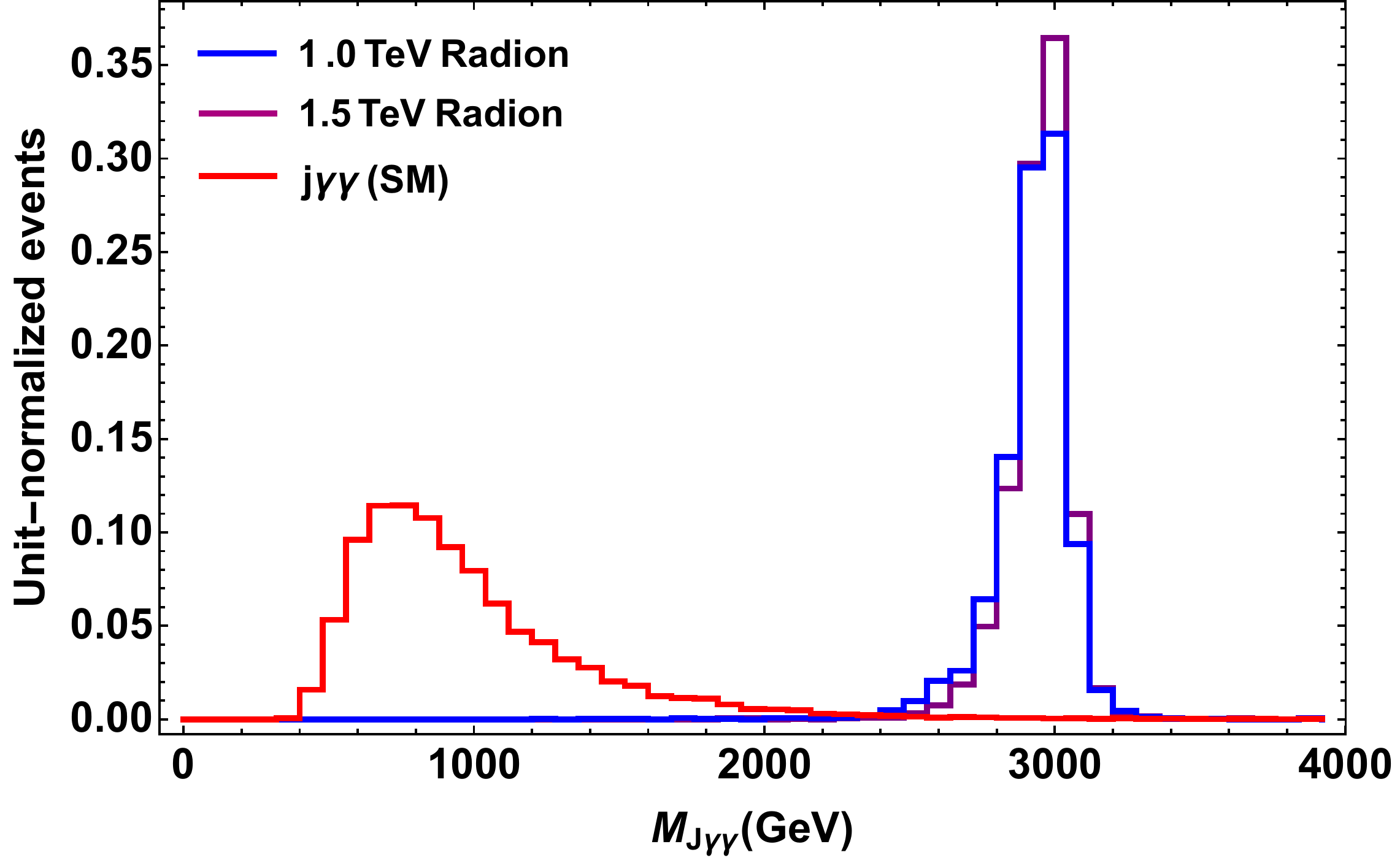}
    \includegraphics[width = 7.1 cm]{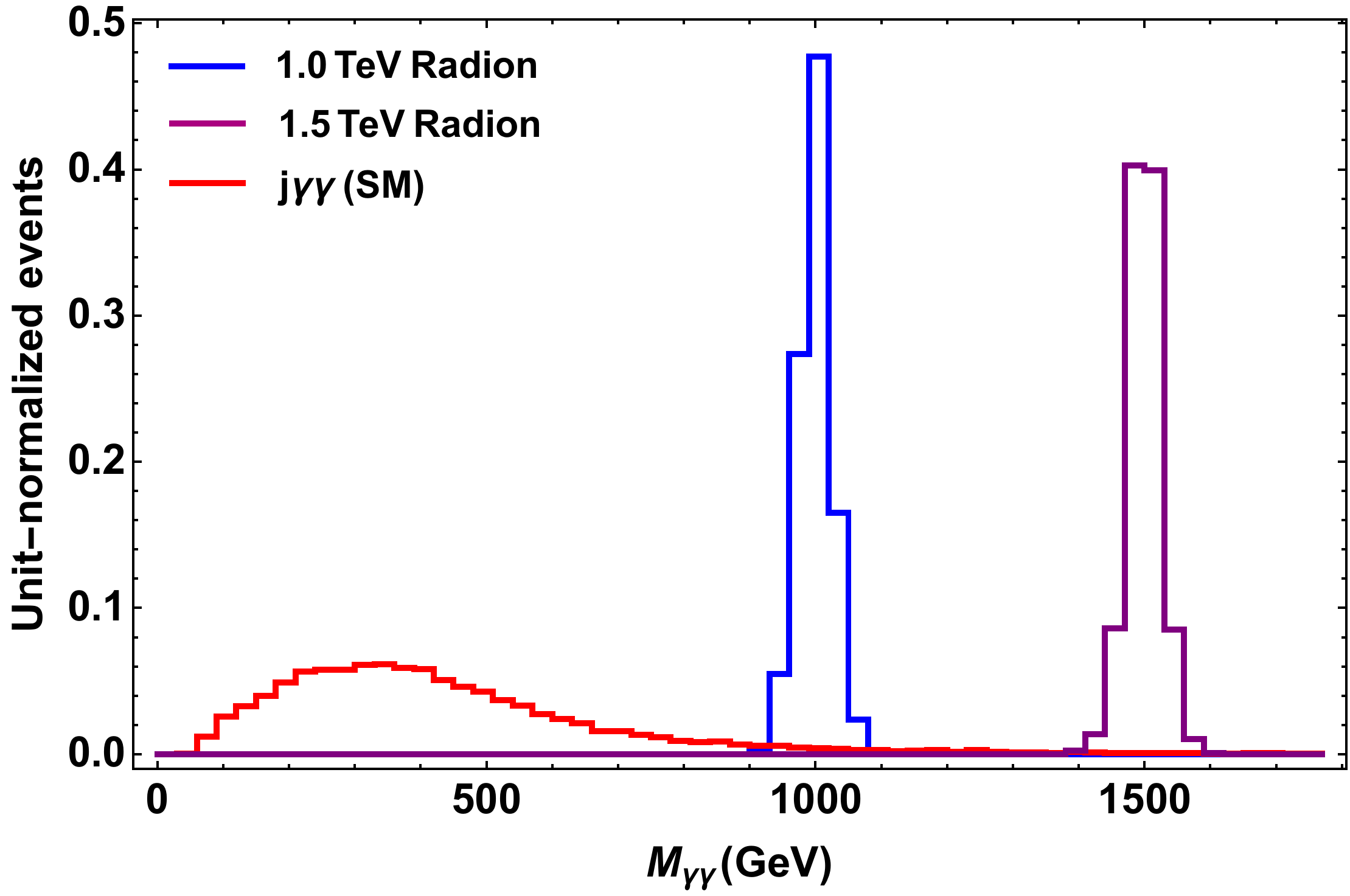}
    \includegraphics[width = 4.8 cm]{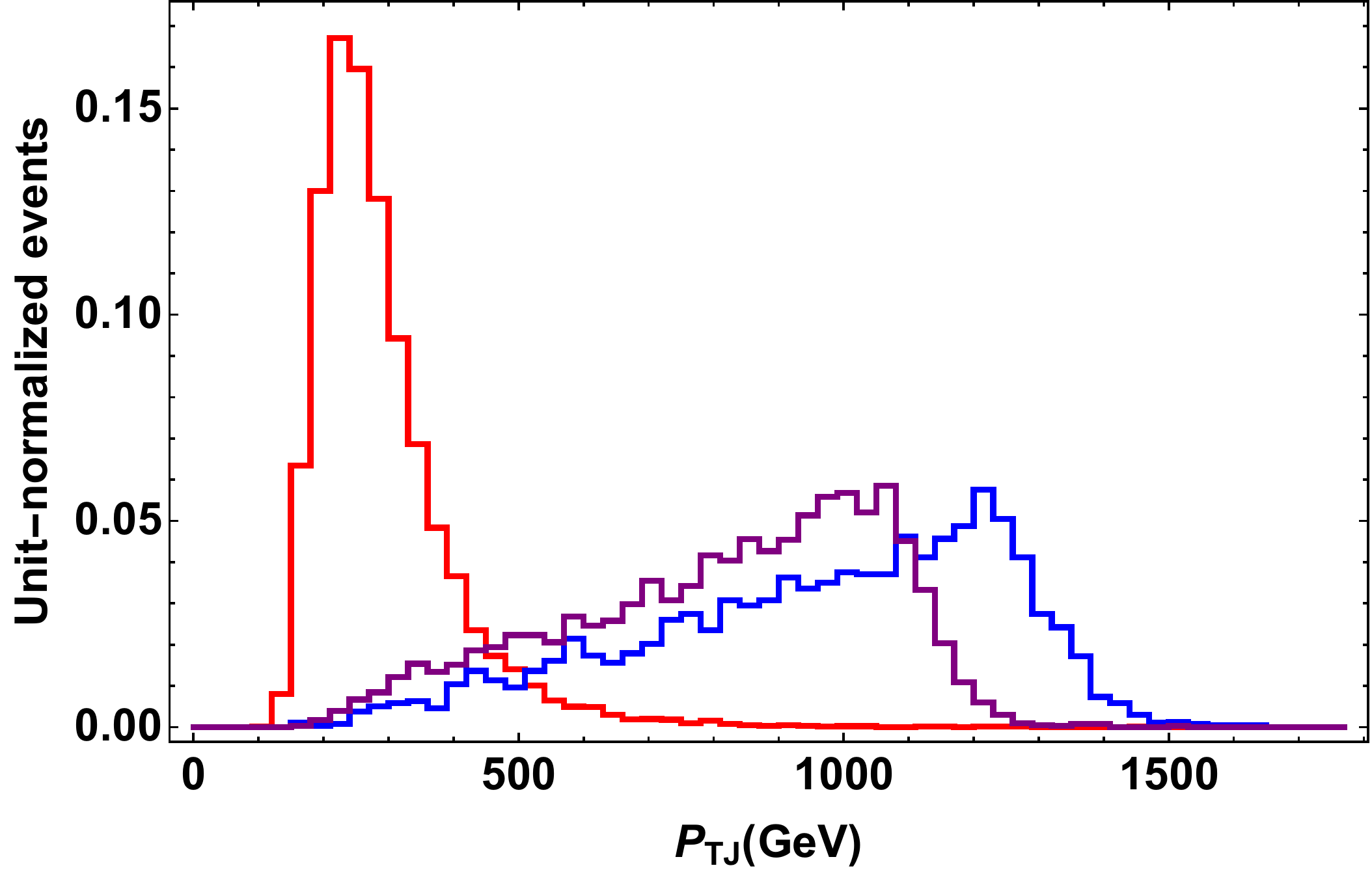}
    \includegraphics[width = 4.8 cm]{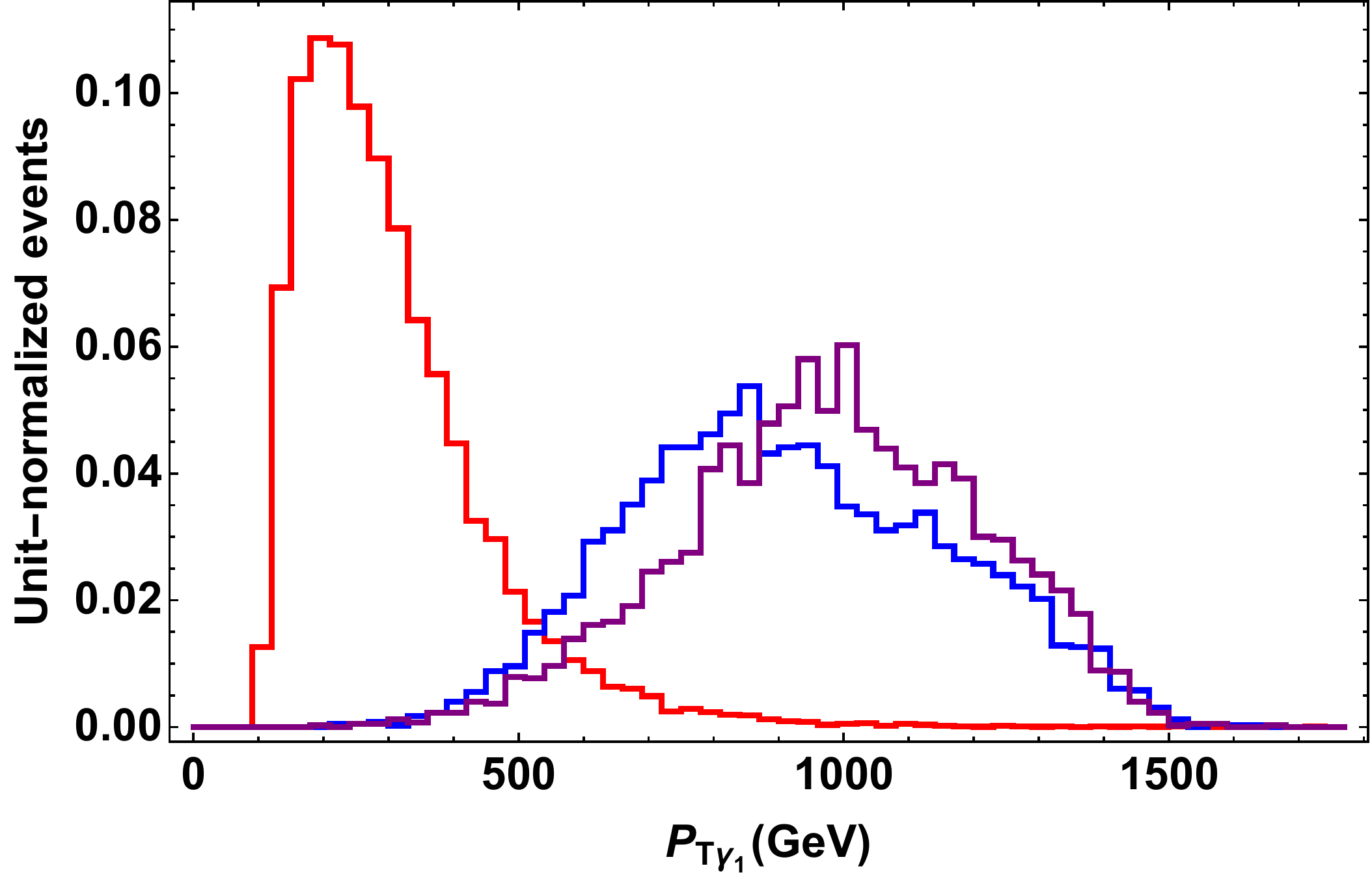}
    \includegraphics[width = 4.8 cm]{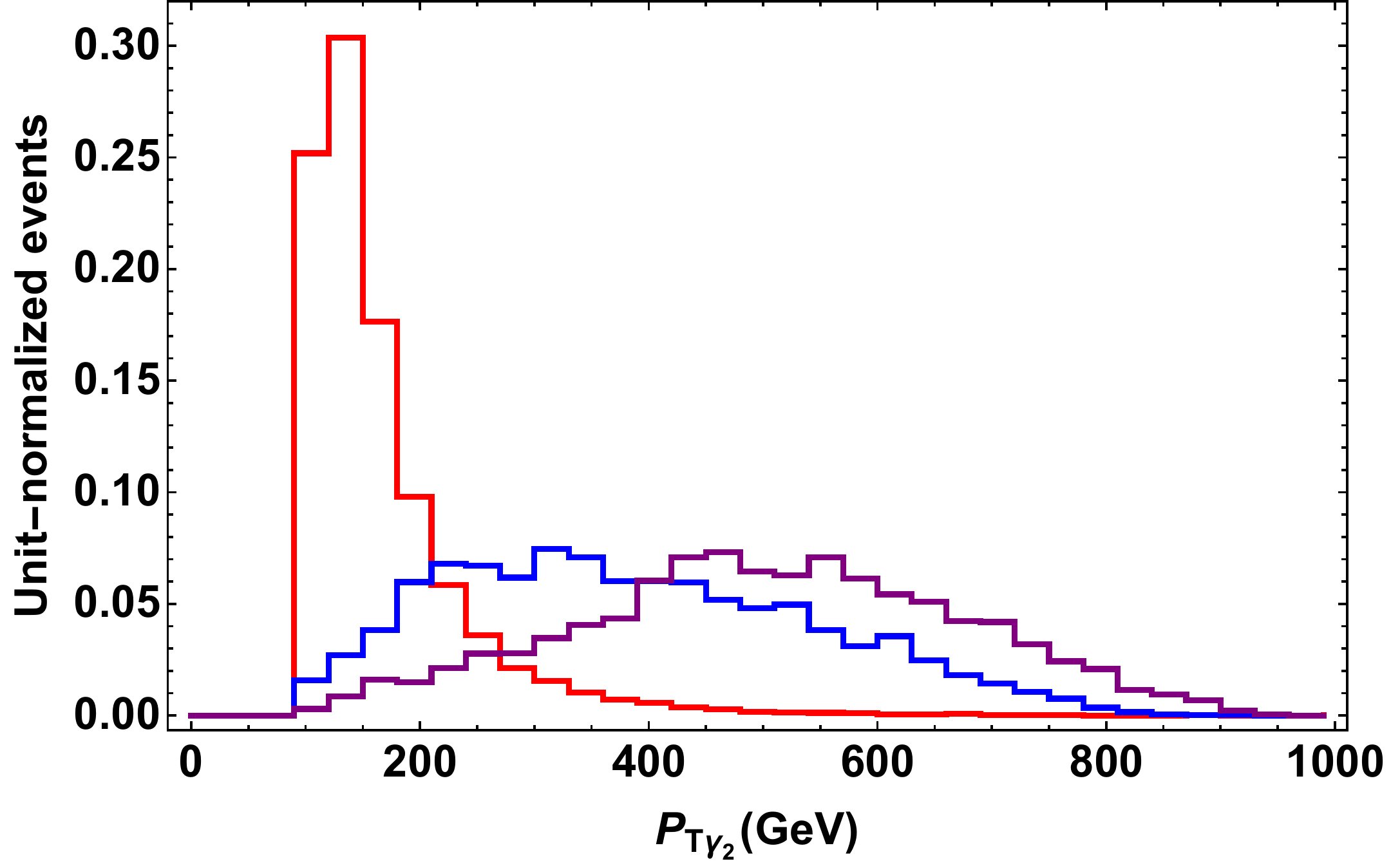}
    \caption{Distribution of kinematic variables for $W$-$W\gamma\gamma$-BP1($W$-$W\gamma\gamma$-BP2) hadronic $W$ channel: $M_{J\gamma\gamma}$ (top row, left), $M_{\gamma\gamma}$ (top row, right), $p_{T J}$ (second row, left), $p_{T \gamma_1}$ (bottom row, middle), $p_{T \gamma_2}$ (bottom row, right) for signal with 1 TeV radion (solid blue), signal with 1.5 TeV radion (solid purple) and backgrounds (solid red). We denote $p_T$-ordered $\gamma$'s as $\gamma_{1,2}$, $\gamma_1$ being the hardest one. 
}
\label{fig:WKK_Waa_had_W}
\end{figure}

The final state contains two photons and one fat $W$-jet. 
We perform the procedure of tagging $W$-jets as before. 
Given the final state, we identify $p p \to j \gamma \gamma$ and $p p \to j j \gamma$ as potential SM backgrounds. The latter can appear as background once one of the two jets is misidentified as a photon. We found the cross section for this background to be roughly $67\:\text{pb}$ after the parton level cuts, requiring us to simulate a large sample of 10 million $jj\gamma$ events using \textsc{Delphes}. After imposing the relevant cuts, we found that no Monte Carlo events survive all the cuts, allowing us to set a $95\%$ confidence limit on this cross-section of $0.02\:\text{fb}$, assuming Monte Carlo uncertainty dominates. 
Since we have a large number of signal events surviving the cuts (see table~\ref{tab:W-Waa-BP1(2)-Had-W-cutflow}), considering this background will not affect our conclusions. We therefore do not include this background any further in our analysis.
As usual, a set of parton-level cuts are applied for background simulation (but not for the signal).
\bea
p_{T j} > 100 \; \GeV, \; p_{T \gamma} > 100 \; \GeV.
\label{eq:parton_level_cuts_Waa_Had_W}
\eea
These cuts are reimposed as pre-selection cuts on the detector-level objects after jet tagging. 
We present kinematic distributions after pre-selection in figure~\ref{fig:WKK_Waa_had_W}. As anticipated, the two-body and three-body invariant mass distributions develop very sharp peaks at the radion and KK $W$ masses, respectively. These features offer excellent signal discrimination.  

\begin{table}[t]
\centering
\begin{tabular}{|c|c|c|c||c|c|c|}
\hline 
\multicolumn{7}{|c|}{$W$-$W\gamma\gamma$-BP1 and $W$-$W\gamma\gamma$-BP2: Hadronic $W$ Channel } \\
\hline
-- & \multicolumn{3}{c||}{$W$-$W\gamma\gamma$-BP1} &  \multicolumn{3}{c|}{$W$-$W\gamma\gamma$-BP2}  \\
\hline
Cuts & Ranges  & SG & $j\gamma\gamma$ & Ranges & SG & $j\gamma\gamma$ \\
\hline \hline
Parton-level cuts & -- & 0.10  & 2.3 & -- & 0.085  & 2.3  \\
\hline
$\substack{N_j \geq 1, \; N_{\gamma} \geq 2 \\ {\rm pre-selection \; cuts}}$ & -- & 0.10 & 2.3 & -- & 0.085 & 2.3  \\
\hline
$ M_{\gamma\gamma}$ (GeV) & $\in [900, 1100] $ & 0.10 & 0.068 & $\in [1400, 1600] $ & 0.085 & 0.011 \\
$ M_{J\gamma\gamma}$ (GeV) & $\in [2550, 3500] $ & 0.10 & 0.0017 & $\in [2600, 3300] $ & 0.084 & 0.0012 \\
\hline
$ p_{T, \gamma_1}$ (GeV) & $\in [150, \infty] $ & 0.10 & 0.0017 & $\in [150, \infty] $ & 0.084 & 0.0012   \\
$ p_{T, \gamma_2}$ (GeV) & $\in [150, \infty] $ & 0.096 & 0.0012 & $\in [150, \infty] $ & 0.083 & 0.0012 \\
$ p_{T, J}$ (GeV) & $\in [250, \infty] $ & 0.096 & 0.0012 & $\in [250, \infty] $ & 0.083 & 0.0012   \\
\hline
$S$ ($\mathcal{L}=300$ fb$^{-1}$) & -- & 29 & -- & -- & 25 & --  \\
$B$ ($\mathcal{L}=300$ fb$^{-1}$) & -- & -- & 0.36 & -- & -- & 0.36 \\
\hline
\end{tabular}
\begin{changemargin}{0.04in}{0.04in}
\caption{Cut flows for $W$-$W\gamma\gamma$-BP1, $W$-$W\gamma\gamma$-BP2 hadronic $W$ channel and their major background in terms of cross sections (in fb). Parton-level cuts in~\eqref{eq:parton_level_cuts_Waa_Had_W} are imposed only on the background events at the generation-level, while pre-selection cuts, which consist of the same cuts as in the parton-level cuts, are imposed on both signal and background events at the detector level after jet tagging. 
\label{tab:W-Waa-BP1(2)-Had-W-cutflow}}
\end{changemargin}
\end{table} 

Table~\ref{tab:W-Waa-BP1(2)-Had-W-cutflow} tabulates the set of analysis cuts and associated results.
We observe that even with $\mathcal{L} = 300 \; {\rm fb}^{-1}$ of statistics the expected number of background events after $M_{\gamma\gamma}$ and $M_{J\gamma\gamma}$ cuts is less than one, i.e. essentially zero, whereas the expected number of signal events is 30 (25) for $W$-$W\gamma\gamma$-BP1 ($W$-$W\gamma\gamma$-BP2). 
So, here we simply report the expected numbers of signal and background events with a luminosity of 300 ${\rm fb}^{-1}$ rather than statistical significance.
Note that we still apply $p_T$ cuts; the reason is not to achieve additional background reduction but to make sure that we impose a stronger cut than each parton-level cut, hence to stay conservative about detector smearing effects. 

\subsubsection*{Leptonic $W$}
\begin{figure}
    \centering
    \includegraphics[width = 7.5 cm]{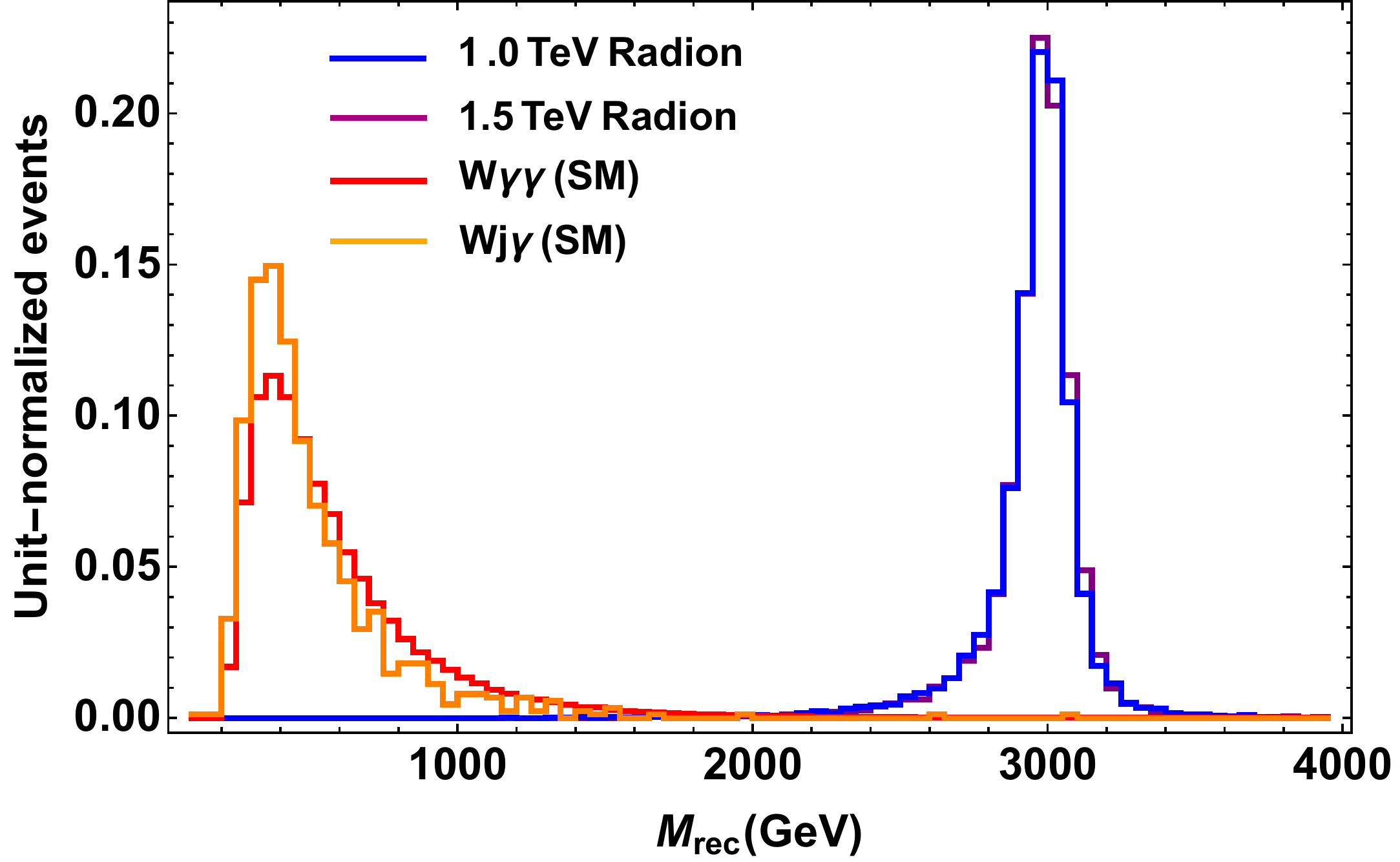}
    \includegraphics[width = 7.1 cm]{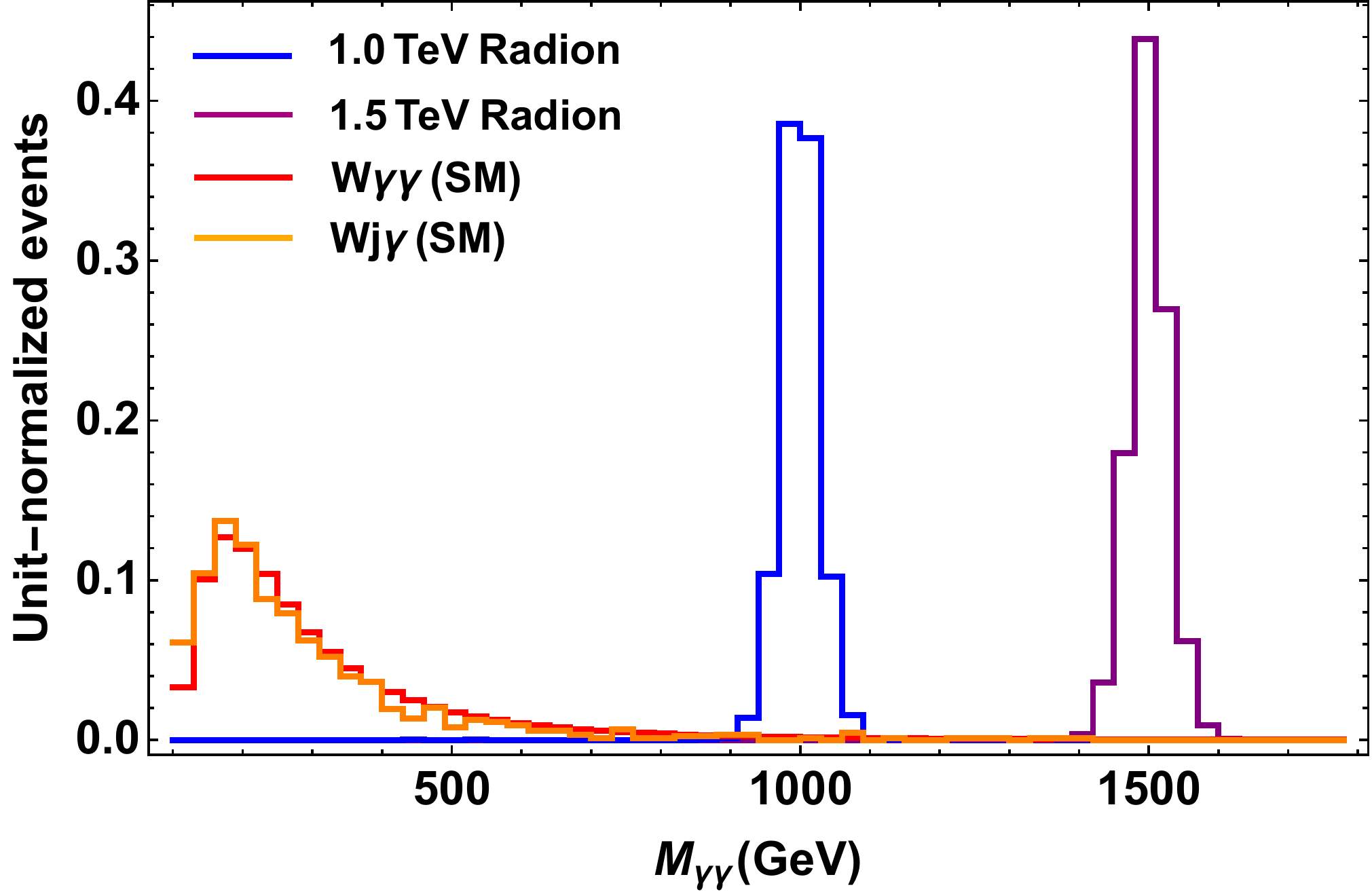}
    \includegraphics[width = 7.5 cm]{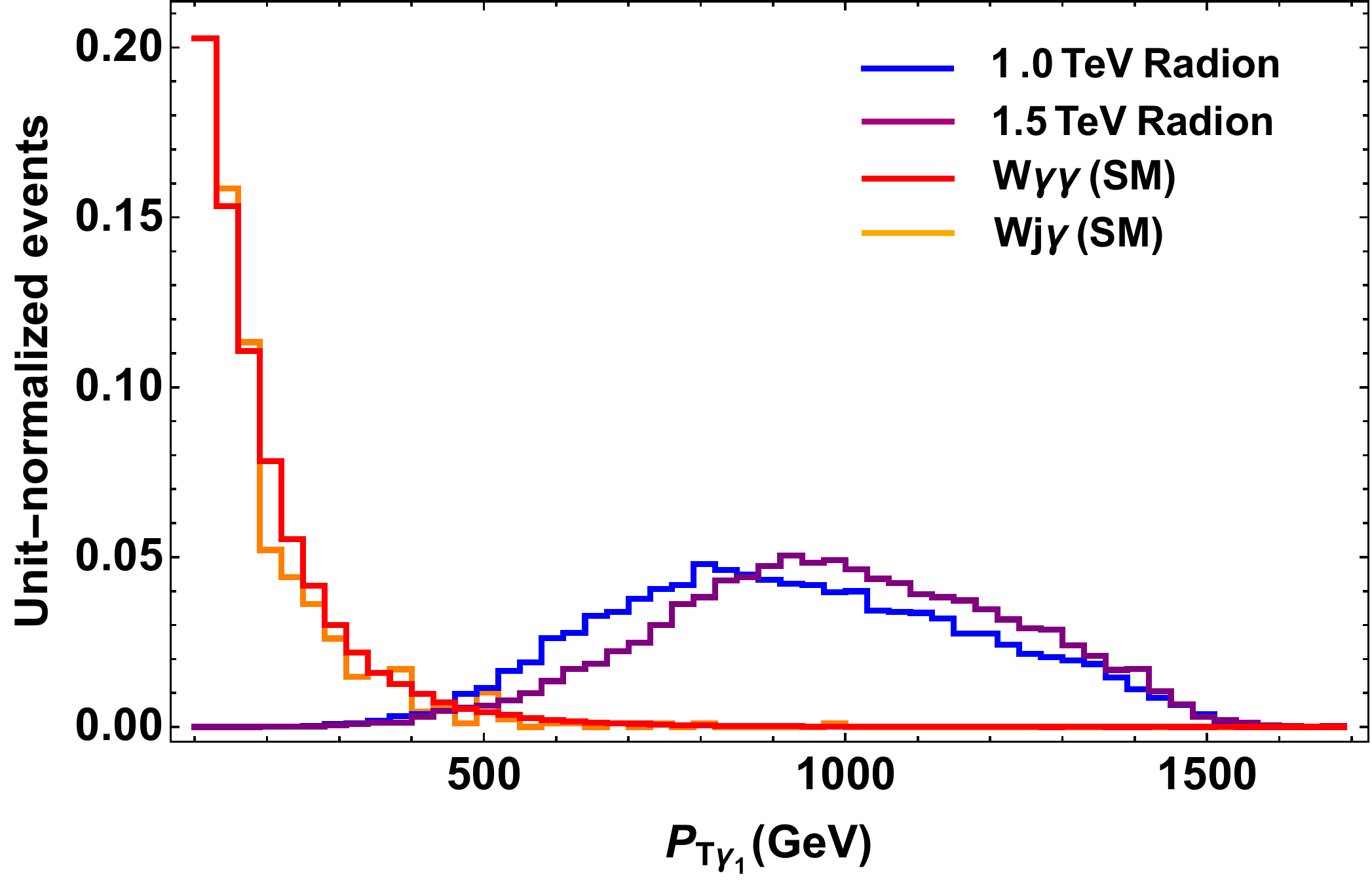}
    \includegraphics[width = 7.5 cm]{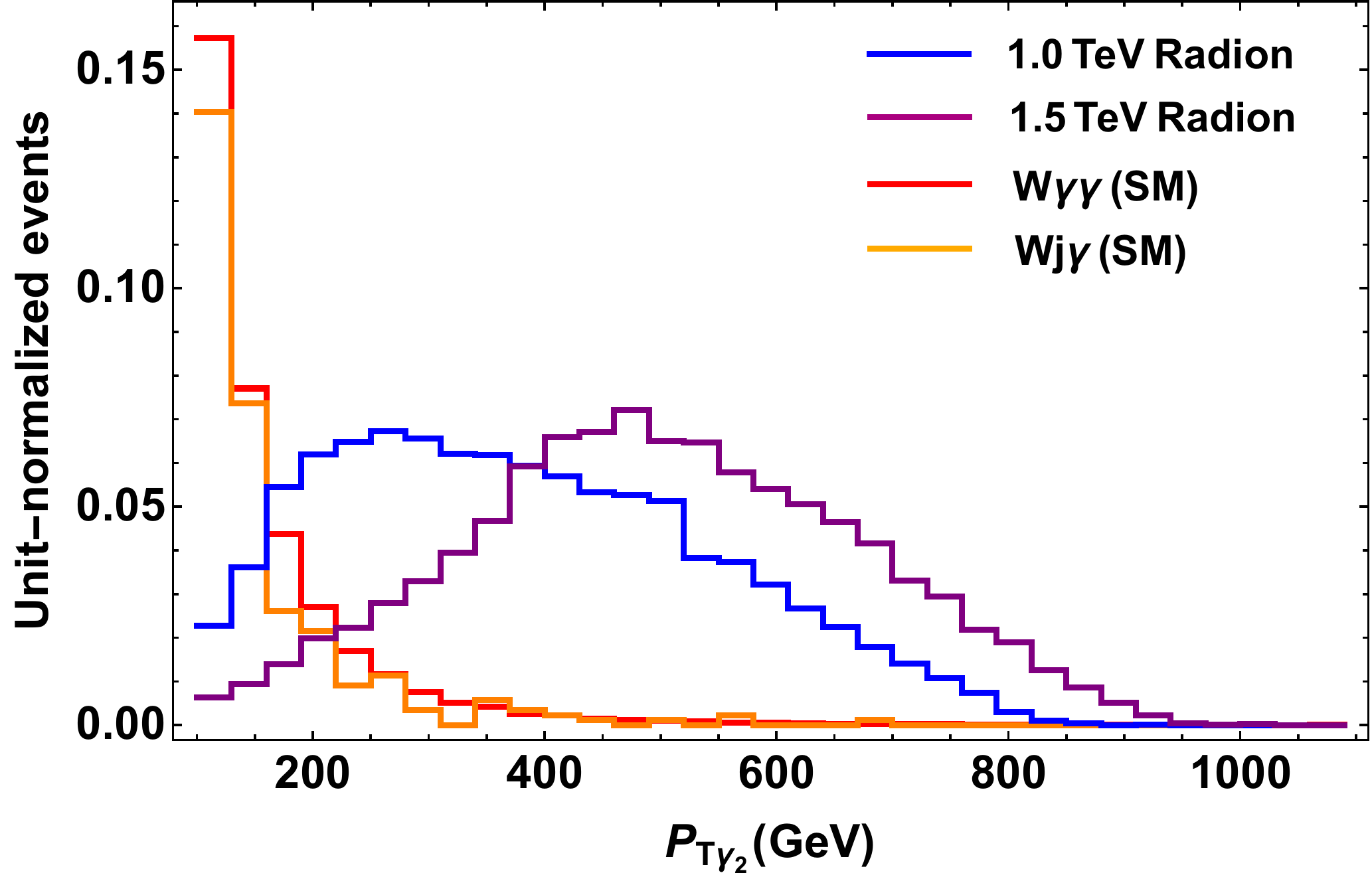}
    \caption{Distribution of kinematic variables for $W$-$W\gamma\gamma$-BP1($W$-$W\gamma\gamma$-BP2) leptonic $W$ channel: $M_{\rm rec}$ (top row, left), $M_{\gamma\gamma}$ (top row, right), $p_{T \gamma_1}$ (bottom row, left), $p_{T \gamma_2}$ (bottom row, right) for signal with 1 TeV radion (solid blue), signal with 1.5 TeV radion (solid purple), $W\gamma\gamma$ background (BK:$W\gamma\gamma$, solid red) and $Wj\gamma$ backgrounds (BK:$W j \gamma$, solid orange). We denote $p_T$-ordered $\gamma$'s as $\gamma_{1,2}$, $\gamma_1$ being the hardest one. $M_{\rm rec}$ is the invariant mass of all particles including neutrino.
}
\label{fig:WKK_Waa_lep_W}
\end{figure}

The $W$ from the direct decay of the KK $W$ now decays leptonically, and the reconstruction of the neutrino's four momentum is conducted in the same way as we described in section~\ref{tools}. We consider two relevant SM backgrounds: $p p \to W \gamma \gamma$ and $p p \to W j \gamma$ with the QCD jet faking the photon in the detector. The parton-level cuts are
\bea
p_{T, \gamma} > 50 \; \GeV, \;  p_{T, j} > 50 \; \GeV. 
\label{eq:parton_level_cuts_Waa_Lep_W}
\eea
The rest of the analysis is essentially the same as the hadronic $W$ case. In figure~\ref{fig:WKK_Waa_lep_W} we show distributions of kinematic variables, and in table~\ref{tab:W-Waa-BP1(2)-Lep-W-cutflow} we summarize the analysis cuts and results. As before, with invariant mass cuts, the signal is background-free. As long as we have enough integrated luminosity, the signal can be discovered.

\vspace{1cm}
\begin{table}[t]
\centering
\begin{tabular}{|c|c|c|c|c|}
\hline 
\multicolumn{5}{|c|}{$W$-$W\gamma\gamma$-BP1 and $W$-$W\gamma\gamma$-BP2: Leptonic $W$ Channel } \\
\hline
Cuts &  $W$-$W\gamma\gamma$-BP1& $W$-$W\gamma\gamma$-BP2 & $W\gamma\gamma$ & $Wj\gamma$\\
\hline \hline
Parton-level cuts &0.078 & 0.064 & 0.88 & 1400  \\
\hline
$N_{\ell} \geq 1, \; N_{\gamma} \geq 2$, pre-selection cuts  & 0.055 & 0.046 & 0.40 & 1.2  \\
\hline
$ M_{\gamma\gamma}\in [900, 1100] \GeV$ & 0.055 &--& 0.0048 & 0.013 \\
$ M_{\rm rec}\in [2000, 4000] \GeV$ & 0.055&-- & 0.00027 & 0.0014  \\
\hline
$ M_{\gamma\gamma}\in [1400, 1600] \GeV$ &--& 0.045 & 0.00087 & 0.0014 \\
$ M_{\rm rec}\in [2000, 4000] \GeV$ &--& 0.045 & 0.00023 & $\ll 0.0014$ \\
\hline
$S$ ($\mathcal{L}=300$ fb$^{-1}$) &  17 & 14 & -- & --  \\
$B$ ($\mathcal{L}=300$ fb$^{-1}$) &  -- & -- & 0.069 & 0.42  \\
\hline
\end{tabular}
\caption{Cut flows for $W$-$W\gamma\gamma$-BP1, $W$-$W\gamma\gamma$-BP2 leptonic $W$ channel and their major background in terms of cross sections (in fb). Parton-level cuts in~\eqref{eq:parton_level_cuts_Waa_Lep_W} are imposed only on the background events at the generation-level, while pre-selection cuts, which consist of the same cuts as in the parton-level cuts, are imposed on both signal and background events at the detector level after jet tagging. 
We obtained a statistically meaningless number with the very last cut for $Wj\gamma$, so we tabulate the expected number of background events corresponding to the cross section after the $M_{\gamma\gamma}$ cut.
\label{tab:W-Waa-BP1(2)-Lep-W-cutflow}}
\end{table} 

\subsection{Only Hypercharge in Extended Bulk}\label{subsubsec:aaa}
In this case, the production of the KK hypercharge gauge boson occurs as usual via its coupling to quarks. However, now the radion predominantly decays into $\gamma\gamma$, $Z\gamma$, and $ZZ$,
simply because its coupling to other gauge bosons are highly suppressed. 
Therefore, the prominent signal is that of a triphoton. Due to the cleanness of the collider signature, we expect signal events to be easily distinguished from SM backgrounds.
The relevant SM backgrounds are $p p \to \gamma \gamma \gamma$, $p p \to j \gamma \gamma$, and $p p \to j j \gamma$, with any jet appearing in the process misidentified as a photon in the detector. Having more jets in the process leads to larger cross section (before phase space suppression kicks in), but the combined rate of multiple jets faking photons more than compensates for the increase in cross section. 
We apply the following cuts at the parton level for backgrounds:
\bea
 \gamma \gamma \gamma :&& \; p_{T, \gamma} > 100 \; \GeV, \; M_{\gamma\gamma} > 200 \; \GeV, \\
{\rm Others:}&& \; p_{T, \gamma/j} > 100 \; \GeV, \; M_{jj/\gamma\gamma} > 200 \; \GeV, \; M_{abc} > 2500 \; \GeV ,\nonumber
\eea 
where $M_{abc}$ is the three-body invariant mass of three objects in the final state.
We then apply a set of pre-selection cuts stronger than each of above parton-level cuts. In this way, our final results will be robust even after taking into account the detector smearing for jet(s) faking photon(s). The pre-selection cuts on selected hardest three photons are
\bea
p_{T,\gamma} > 150 \; \GeV, \; M_{\gamma\gamma} > 300 \; \GeV, \; M_{\gamma\gamma\gamma} > 2700 \; \GeV.
\eea
We find that after pre-selection, with $\mathcal{L} = 300 \; {\rm fb}^{-1}$, no background events survive. 
For the  $\gamma \gamma \gamma$, $j \gamma \gamma$, $j j \gamma$ backgrounds the simulated samples correspond to effective luminosities ($\equiv$ number of simulated events $\div$ cross section) of $5 \times 10^6 \; \text{fb}^{-1}$, $2 \times 10^6 \; \text{fb}^{-1}$, and $900 \; \text{fb}^{-1}$ respectively, so we conclude that the SM backgrounds from these processes are essentially negligible. We were not able to simulate sufficient $jjj$ events to demonstrate the same for this process directly. As mentioned earlier, due to the requirement for three jets to simultaneously fake photons, we expect $jjj$ to be a subdominant background. Since both benchmark points [$B$-$\gamma\gamma\gamma$-BP1 ($B$-$\gamma\gamma\gamma$-BP2)]  predict $\mathcal{O}(30)$ events, if this scenario is realized in nature, the discovery will be made with a very spectacular diphoton and triphoton invariant mass peak, as soon as a large enough integrated luminosity is reached.  

\section{Conclusions/Outlook}
\label{conclusion}
As Run 2 of the LHC goes into full gear, an interesting and subtle change can be discerned in the research directions among certain quarters of the BSM community. There is not yet any strong evidence of new physics in the plethora of standard searches being performed at the LHC, which are mostly targeted at well motivated signatures of supersymmetry and top partners and some simple resonance topologies. Most of these searches have reached full maturity and are all set to be applied to the full luminosity and energy of the LHC, with little additional input required. This situation has prompted both theorists and experimentalists to step away from under these lampposts and explore novel channels that might be hiding new physics, and come up with strategies to unearth them.

There are broadly two partially overlapping categories of non-standard approaches. The first is centered on {\em exotic} objects as far as the detector is concerned. One example is long-lived particles undergoing displaced decays into SM particles, which might arise in supersymmetric scenarios with compressed spectra or small R-parity violating couplings or, in twin Higgs models (see, for example, ref.~\cite{Alekhin:2015byh} and references therein). Another example is boosted BSM objects resulting in multi-pronged fat jets, perhaps also with embedded leptons~\cite{Brust:2014gia,Aguilar-Saavedra:2017rzt}. Alternatively, searches can continue to encompass conventional final state objects including ordinary prompt jets, leptons, and boosted SM particles which do not require new identification techniques, but occur in {\em non-standard} topologies or combinations. This category of signals can arise even within existing frameworks by simply going to hitherto unexplored regions of parameter space, or else in minor variations of the model. 

The cascade decays considered in this paper constitute examples of the second category, in which a slight variation of the vanilla model with a warped extra-dimension leads to tri-SM final states replacing di-SM ones as the dominant decays of heavy vector resonances. The cascade decays are also present in the original model, but with very small relative rates due to the dominance of the di-SM modes. Allowing the EW gauge fields to propagate in the extended bulk 
suppresses the di-SM modes which allows for dominance of the tri-SM ones. This model is meant to serve mainly as an illustrative example and other variations are possible, for example with a different Lorentz structure of the couplings involved in the cascade. Of the existing LHC searches, the diboson (and other di-SM) searches may still be the most sensitive to the triboson cascades, but they will often fail to reconstruct any resonance in the decay cascade due to combinatoric ambiguities and they do not make use of all of the information provided by the distinctive signal. Instead we propose that the LHC experiments can search directly for these cascade decays with dedicated strategies, which will allow for reconstruction of both primary parent and secondary parent resonances. 

For the model-realization we have focussed on, we find that discovery of the signal in various triboson final states involving combinations of $W$ and $\gamma$ is possible for a $3$ TeV spin-1 particle and $1 - 1.5$ TeV (intermediate) scalar with $\mathcal O(100)$ $\hbox{fb}^{-1}$ of luminosity at the LHC. In particular, the $\gamma \gamma \gamma$ channel of the model with only hypercharge in the extended bulk is sufficiently spectacular that a discovery could be made in the $M_{\gamma \gamma \gamma}$ distribution alone with tens of signal events and zero background events, though also incorporating diphoton invariant mass distributions would help in identifying the radion. For the model with all the EW gauge fields in the bulk, the hadronic and leptonic $W \gamma \gamma$ signals are similarly striking. The $WWW$ channels are more subtle due to the combinatoric ambiguities and the non-negligible backgrounds from QCD multijet and $W$ + jets backgrounds, and due to the fact that the signal is distributed into multiple final states due to the various possible decays for the $W$'s. Focussing on the final states with zero or one lepton, we find that a simple bump hunt on the $WWW$ invariant mass  is sensitive to the signature when these two channels are combined, however sensitivity is improved by looking for a second bump in a selected $M_{WW}$ diboson invariant mass distribution. Due to the combinatoric ambiguity, most signal events actually lie in a ``+'' shape on a two-dimensional plane of two $M_{WW}$ pairings, and utilising this shape it is possible to recover additional signal strength. Combining signal channels, an observation at the 4--5$\sigma$ level seems possible for this signature.

Going beyond this specific analysis, we envisage that our work will motivate further studies in a similar spirit both within this warped framework and beyond it. In a forthcoming paper \cite{triboson_lightRad} we will consider the same warped model as above, but now with the radion being light: a few hundred GeV. Such a choice of mass is allowed in spite of LHC searches because the suppressed coupling of the radion to gluons results in a small cross section for direct radion production. When produced from the decay of an EW gauge KK mode (which are still constrained to be at least a few TeV from various searches), the radion is significantly boosted. Thus, the pair of SM EW bosons from its decay tend to be merged, creating a new object which can be called a ``boosted diboson''. Such an object requires a new dedicated algorithm for tagging, or else a general-purpose tagger designed for a variety of multi-pronged boosted object signatures~\cite{Aguilar-Saavedra:2017rzt}, otherwise it would not be spotted. This boosted diboson is accompanied by a standard boosted EW boson directly from the decay of EW gauge KK particle, just like for the case of heavy radion studied in this paper. In this way, we see simple variations can result in a combination of the two categories of new signals mentioned earlier, i.e., exotic objects {\em and} non-standard topologies. 

To summarize, there remain signatures that can arise in regions of parameter space or in plausible variations of well motivated models of new physics that could be missed in the absence of targeted search strategies. Looking forward, we hope that a combination of new search strategies both for selecting exotic objects and for identifying interesting kinematic features among collections of standard objects will broaden the LHC coverage. These ideas really constitute golden opportunities to exploit the full potential of the LHC in search for new physics, and we must be adequately prepared for any surprises on this front.

\section*{Acknowledgements}

We would like to thank Alberto Belloni, Petar Maksimovic and Nhan Tran for discussions. The work of KA, JC, PD, SH and RKM was supported in part by NSF Grant No.~PHY-1620074 and the Maryland Center for Fundamental Physics. KA, JC and SH were also supported by the Fermilab Distinguished Scholars Program and SH by NSF Grant No.~PHY-1719877. DK is supported by the Korean Research Foundation (KRF) through the CERN-Korea Fellowship program.

\appendix


\section{Same Sign Dilepton Constraints on $WWW$ Final State}
\label{SSDL}

A striking signature of $WWW$ production would be highly energetic same-sign dileptons and missing energy, in association with either a third lepton or jets, with a branching fraction of 4.6\%. With this channel it would not be possible to reconstruct the resonance masses due to the presence of two neutrinos. However due to very low Standard Model background this is still a possible discovery channel. The ATLAS search for supersymmetry in same-sign dileptons plus jets with $36 \, \text{fb}^{-1}$ of 13 TeV data \cite{Aaboud:2017dmy} requires either two same-sign leptons and at least six jets, or three leptons and at least four jets. This means that the search is not sensitive to this signal, since the hard process will produce either two leptons plus one or two jets, or three leptons and no jets. In order to verify that ISR and FSR will not produce enough energetic jets to pass the selection criteria, we used CheckMATE \cite{Drees:2013wra} to assess the sensitivity of the similar search with $3.2 \, \text{fb}^{-1}$ of data \cite{Aad:2016tuk}. The signal region SR0b5j which requires five jets is the most sensitive, with a very low efficiency of $1.2 \times 10^{-3}$ for the benchmark $W$-$WWW$-BP1.

\begin{figure}
    \centering
    \includegraphics[width = 7.5 cm]{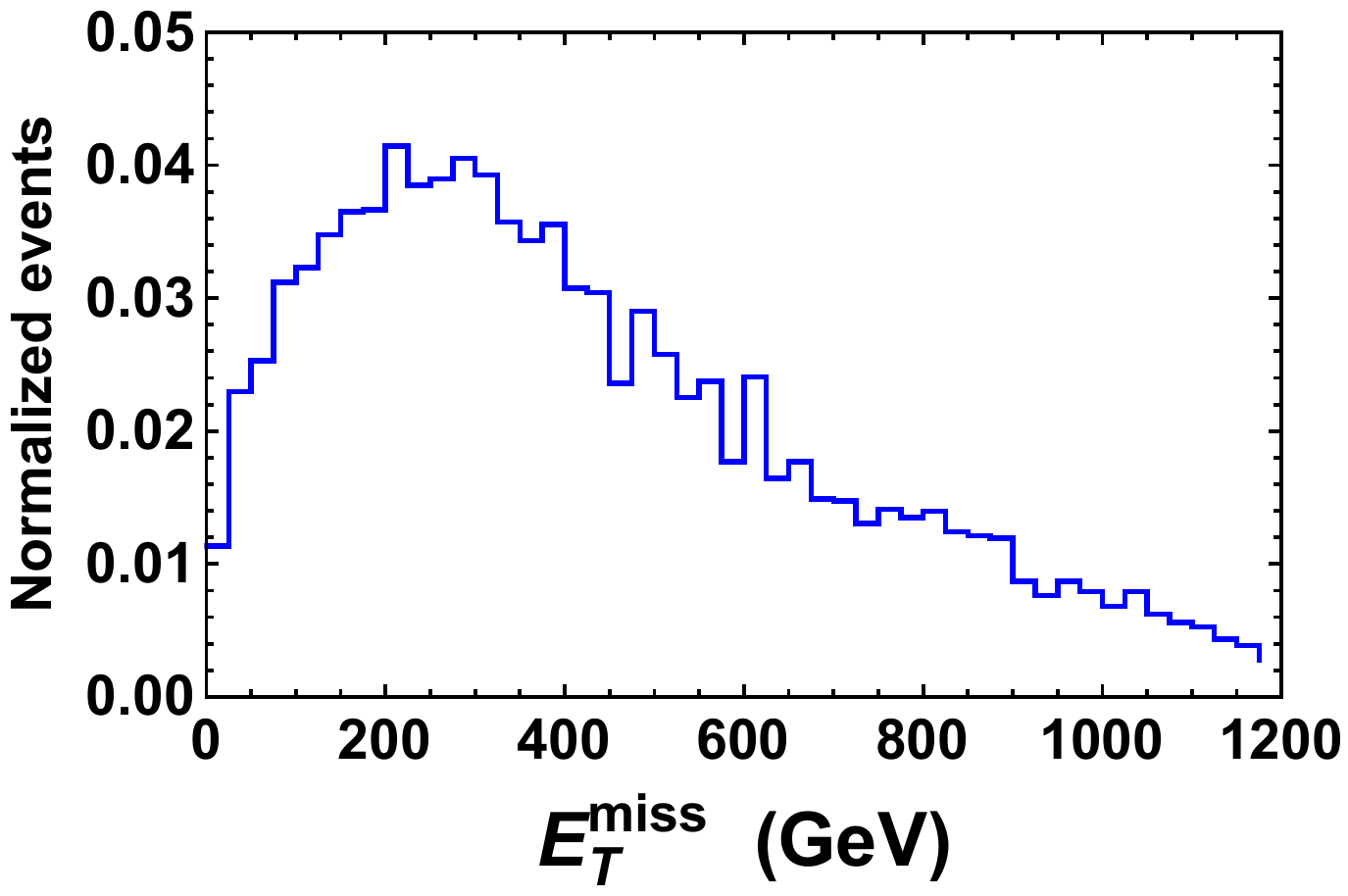}
    \caption{$E_{T}^\text{miss}$ distribution for the SSDL decay $W_{\rm KK}^\pm \to W^\pm \varphi \to W ^\pm W^\pm W^\mp \to \ell^\pm \ell^\pm \nu \nu q \bar{q}'$ with $m_\varphi = 1 \; \text{TeV}$ using benchmark point $W$-$WWW$-BP1.}
\label{fig:SSDL_MET}
\end{figure}

The CMS search for new physics in same-sign dileptons and jets \cite{CMS-PAS-SUS-16-035} is more sensitive to our signature because it has signal regions that require only at least two jets. Even if the hadronically decaying $W$ produces only a single merged jet, it is common that radiation will produce an additional jet which would pass the selection. The most sensitive signal regions will be SR44 and SR45, which require at least two jets, $E_T^\text{miss} > 500 \; \text{GeV}$, and $H_T > 300 \; \text{GeV}$, where $H_T$ is the scalar sum of the $p_T$ of jets. SR44 requires two positive leptons and SR45 requires two negative leptons, and the combined SM prediction in these bins is $4 \pm 1$ events. We assess the sensitivity of this search to the benchmark $W$-$WWW$-BP1 using a parton level analysis on the LHE file produced by \textsc{MG5@aMC}. We assume that the efficiency for reconstructing two passing jets is 100\%, and that $H_T$ is given by the sum of the quark $p_T$'s resulting from the hadronically decaying $W$. The efficiencies for leptons passing kinematic cuts are taken as 0.7 and 0.9 for electrons and muons respectively, as stated in \cite{CMS-PAS-SUS-16-035}. We find an overall efficiency of 0.16, which is driven mainly by the lepton reconstruction efficiency and the cut on $E_T^\text{miss}$ which is peaked below $500 \; \text{GeV}$ (see figure \ref{fig:SSDL_MET}). For this benchmark, this means a combined signal prediction of 0.53 events in these bins, which cannot be excluded.

Assuming that exactly the same analysis is repeated with $300 \; \text{fb}^{-1}$, we can extrapolate these results to predict 4.4 signal events on a background of 33 SM events, with estimated significance $S / \sqrt{B} = 0.8$. It is likely that the channels which we have discussed in the main text will be the discovery channels for this signature.


\section{Jet Tagging Efficiency: Fixed vs Variable jet Radius}
\label{App:JetTaggingMethods}

We have argued in the main body of the paper that for maximizing signal efficiency, one must enlarge the radius parameter $R$ used in constructing a fat jet, but not too much. In general, it is expected that the decay products from a jet with higher $p_T$ will be more collimated, so that a smaller $R$ is enough to capture all the constituents, while for smaller $p_T$, one must consider a larger $R$. In the specific case of the signal we are considering, we have 3 $W$-jets with typical $p_T$ varying from 400 to 1000 GeV, hence it is expected that a larger $R$ may give a better signal efficiency. However, the typical $\Delta R$ separation between two $W$s (before decay) sets an upper bound on how much the jet radius $R$ can be relaxed. We confirmed this behavior at parton level and detector level, and found an optimal value for $R$ which was used in our analysis.

This discussion makes it natural to consider jets with a radius $R$ that depends on the $p_T$ of the jet. Such an approach was suggested in \cite{Krohn:2009zg}, where a simple $R - p_T$ relation as
\begin{align}
R = \frac{\rho}{p_T}
\end{align}
was considered. Notice that to make up for dimensions, $\rho$ is a dimensionful quantity, and must be set by hand. One therefore has to choose such parameter $\rho$ in variable $R$ algorithm, compared with $R$ itself in fixed $R$ algorithm. Ref.~\cite{Krohn:2009zg} estimated an upper bound for $\rho$ to correctly reproduce the size of a jet,
\begin{align}
\rho \lesssim 2 p_T,
\end{align}
which is effectively the same as $R \lesssim 2$.

\begin{figure}[h]
\center

\includegraphics[width=0.45\linewidth]{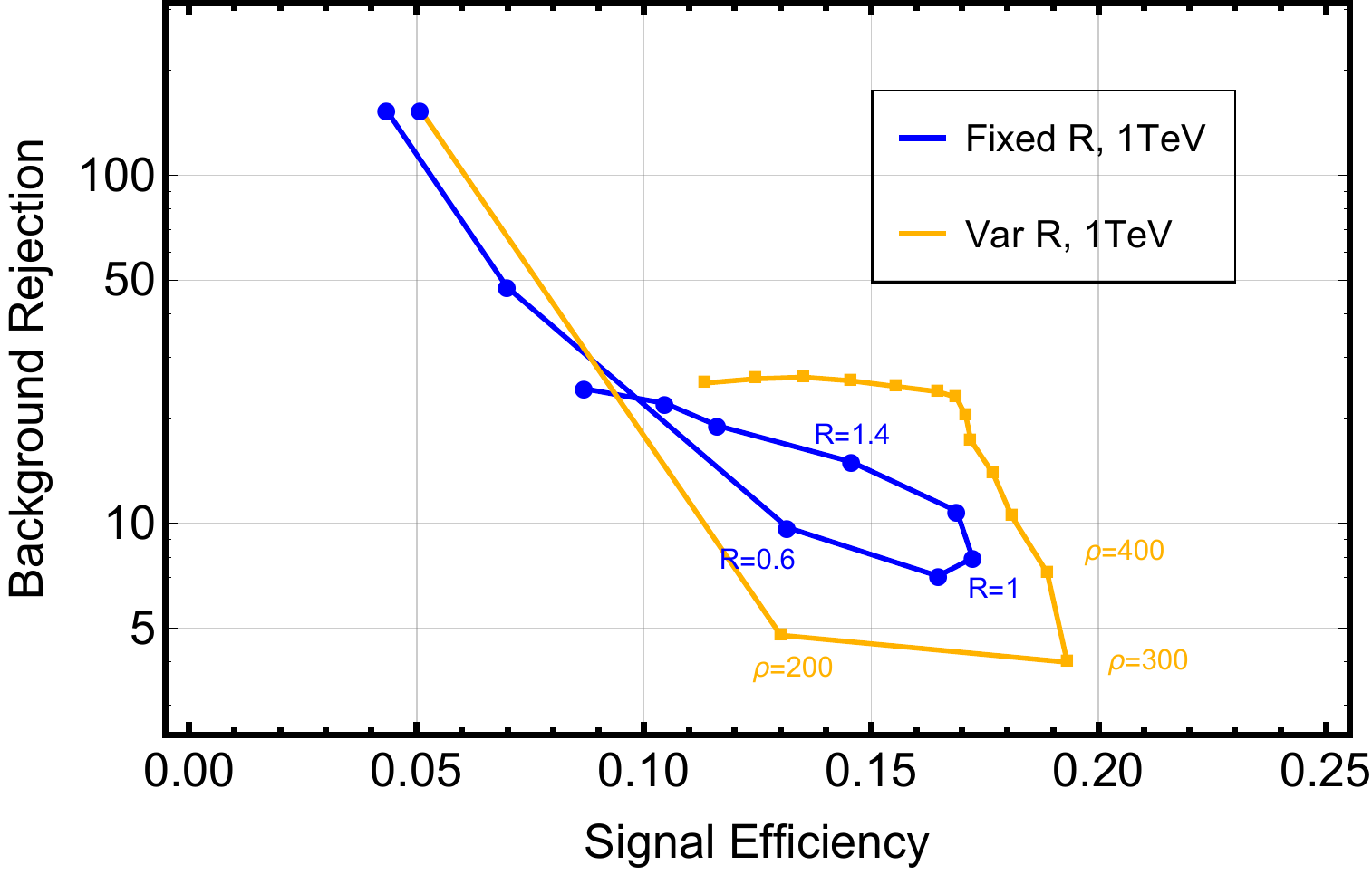}
\includegraphics[width=0.45\linewidth]{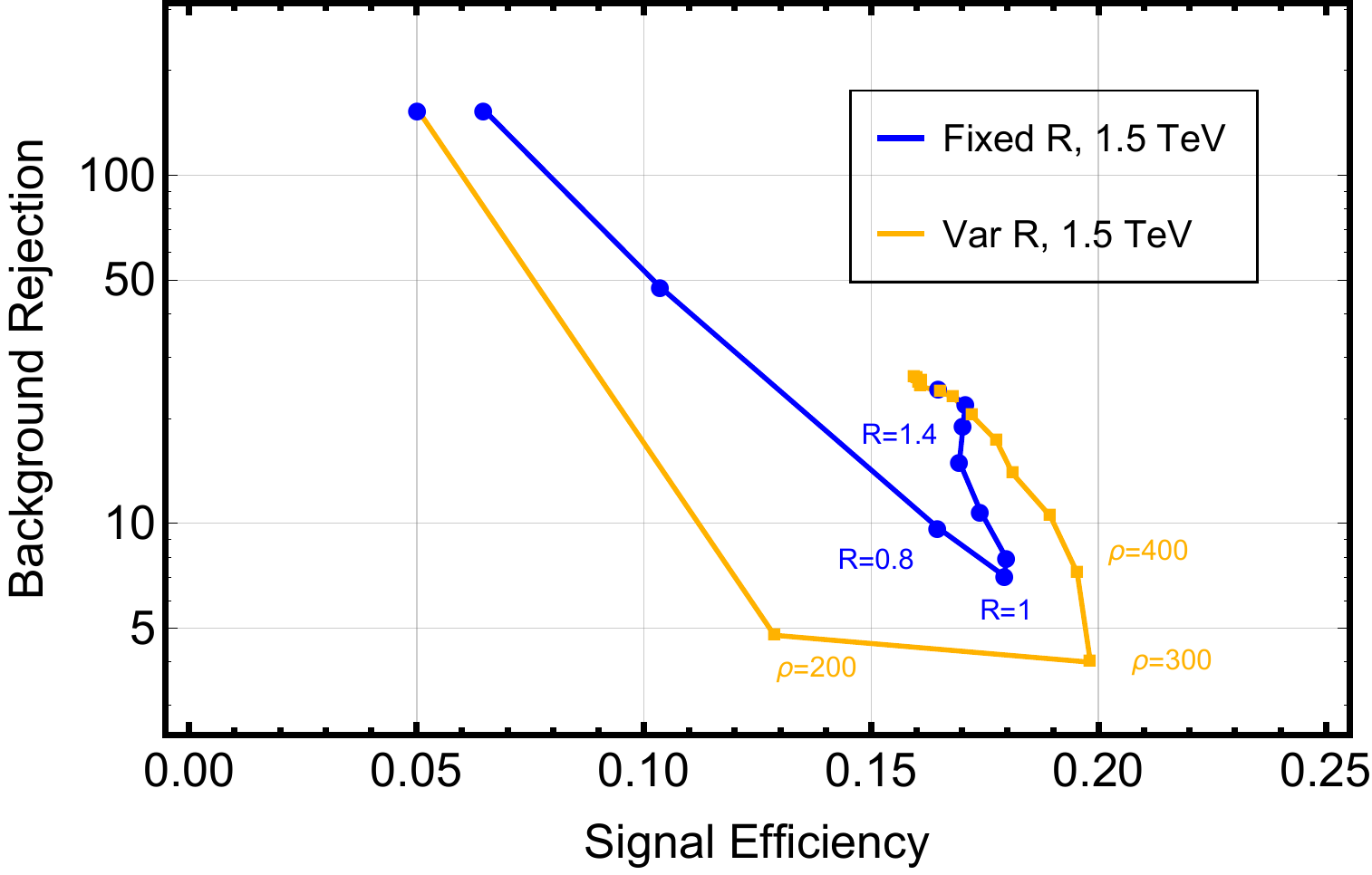}

\caption{Comparison of fixed $R$ and variable $R$ methods (blue and orange respectively), for $1$ and $1.5$ TeV radion (left and right), corresponding to benchmark points $W$-$WWW$-BP1 and $W$-$WWW$-BP2, respectively. Some representative points are labeled by the $R$ ($\rho$) values to show their value near optimal point, and the direction in which these parameters increase. }
\label{fig:roc_fixed_vs_var_jet_radius}
\end{figure}

For our signal, we can consider the performance of fixed $R$ method to ``fixed $\rho$'' method by looking at signal efficiency vs background rejection, in a ROC curve, as shown in figure~\ref{fig:roc_fixed_vs_var_jet_radius}. The event is qualified as a signal if it has three jets that are consistent from coming from a $W$ decay, as reflected in the value of their mass $m$ and substructure variable $\tau_{21}$. Different points correspond to different choice of $R$ and $\rho$ for the two algorithms. We take $R$ to vary from $0.2$ to $2.0$ in steps of $0.2$, and $\rho$ (in GeV) to vary from $100$ to $1500$ in steps of $100$. Some representative points are labeled in figure~\ref{fig:roc_fixed_vs_var_jet_radius} to show the direction in which $R$ and $\rho$ increase.

The first thing to note in the ROC curves is the double valuedness nature, in that, for a given signal efficiency, there are two possible values of background rejection and vice versa. For the fixed $R$ algorithm, as the radius increases beyond a certain optimal value, the contamination from adjacent jets reduces the signal efficiency. Further, since the background rejection increases beyond a certain point, it must be that it is harder and harder for QCD jets to look like coming from $W$, as the jet is made more and more fat (i.e. $R$ or $\rho$ is increased). The next thing to note is that the variable $R$ method gives slight increase in the maximum attainable signal efficiency, but only at a cost of reduced background rejection. However for a given signal efficiency, one can work at a higher background rejection by choosing $\rho$ wisely. Same is true if one wants to work at a fixed background rejection. While certainly interesting, such an approach is beyond the scope of the present analysis. Since the gain in overall signal efficiency is only marginal, we do not use the variable $R$ method in our analysis.

\bibliographystyle{utphys}
\bibliography{bibliography}

\end{document}